# A debiasing technique for place-based algorithmic patrol management

**Alexander Einarsson**[1], **Simen Oestmo**[2], **Lester Wollman**[2], **Duncan Purves**[3], **Ryan Jenkins**[4]
[1] Northwestern University, Department of Computer Science
[2] SoundThinking, Inc.
[3] University of Florida, Department of Philosophy
[4] California Polytechnic State University, San Luis Obispo, Department of Philosophy
aeinarsson@u.northwestern.edu, soestmo@soundthinking.com, lwollman@soundthinking.com, dpurves@ufl.edu, ryjenkin@calpoly.edu

## Abstract

In recent years, there has been a revolution in data-driven policing. With that has come scrutiny on how bias in historical data affects algorithmic decision making. In this exploratory work, we introduce a debiasing technique for place-based algorithmic patrol management systems. We show that the technique efficiently eliminates racially biased features while retaining high accuracy in the models. Finally, we provide a lengthy list of potential future research in the realm of fairness and data-driven policing which this work uncovered.

## 1 Introduction

This paper introduces a technique for eliminating racially correlated features from a place-based algorithmic patrol management system (PAPM) called ResourceRouter. The paper also evaluates the effect of removing racially correlated features on the PAPM model's accuracy.[1] PAPM is the use of computer software, often developed using machine learning, to analyze data from local law enforcement and other city agencies to assess the short-term risk of crimes across a city. In practice, these assessments are typically used to produce a patrol plan that officers can use to inform patrol decisions in their assigned area. In principle, PAPM assessments can be combined with a variety of policing tactics, but directed patrols are by far the most common tactic.

PAPM is part of a larger data-driven revolution in policing.[13] Data-driven policing is the use of big data systems to make classifications of interest to law enforcement. Big data systems make use of vast quantities of data from a variety of data sources to identify correlations or patterns in the data—often patterns that humans would otherwise overlook—to classify new instances in the domain of interest. This process of pattern recognition and classification occurs through the use of computer algorithms, which are sometimes developed through the process of machine learning, a method by which a computer system is "trained" to reason or make inferences from inputs. At its core, then, data-driven policing involves police agencies making decisions by harnessing vast quantities of data and training computer models to identify patterns in that data to produce classifications of interest to law enforcement.

[1] We use the term 'place-based algorithmic patrol management' (PAPM) instead of the more common term 'predictive policing' because the former term is more perspicuous. The term 'predictive policing' is ambiguous between a variety of technologies, each of which can be evaluated on its own technical, ethical, and legal merits.

Data-driven policing is part of a "big data" revolution across sectors, including medicine, finance, transportation, and criminal justice. Classification systems trained using machine learning techniques have recently accumulated a staggering track record of success. Consider some examples from medicine: In a trial with over 80,000 participants, algorithmically assisted breast cancer detection reduced healthcare worker workload by 44% with no increase in false positive rate and even a small improvement in breast cancer detection rate.[26] In another case, an AI model trained on 1.6 million unlabelled retinal images using self-supervised learning was found to consistently outperform "several comparison models in the diagnosis and prognosis of sight-threatening eye diseases, as well as incident prediction of complex systemic disorders such as heart failure and myocardial infarction."[52]

There is therefore a rapidly growing body of evidence to support using big data systems–especially machine learning-based models–for classification tasks. Additionally, PAPM is motivated by several specific observations about crime and policing: first, police have limited resources with which to enforce the law. Policing effectively therefore requires an efficient allocation of police resources to where they will do the most good.[39] Second, the vast majority of crime occurs in very small areas within a city.[6,11,33,41,46] Other things being equal, it is therefore more efficient for police to focus on those small, high-risk areas rather than spreading their limited resources evenly around a city. Third, a robust literature has demonstrated that crime occurrence is correlated with a variety of variables, including prior crime at a location, the presence of certain kinds of businesses, socio-economic variables, geo-spatial features, time of day, weather, and major events.[6,9] Some of these correlations are obvious–gang-related crime is more likely to occur in contested gang territory than in the affluent suburbs–but others are surprising. For example, distance to railway bridges predicts crime in some instances, and density of, and distance to, places of worship is a powerful predictor of crime in other instances. It therefore makes sense to capitalize on these correlations when allocating limited police resources. Fourth, from the perspective of public safety and police-community relations, it is preferable for police to deter crime before it occurs than to react to a crime that has already been committed.[14,47,48] So-called "proactive policing" can be paired with a variety of policing tactics, ranging in intensity from increasing patrols in high-risk areas, to addressing geographic vulnerabilities in high-risk areas, to more controversial practices like targeting high-risk people and "stop, question, and frisk."[36] The most recent large scale meta-analysis of proactive approaches to policing found that some proactive policing strategies are effective at short-term crime-reduction. About hotspots policing, a 2018 National Academy of Sciences review of the academic literature on the legal, societal, and criminological dimensions of hotspots policing says, "Evidence strongly suggests that hot spots policing strategies produce short-term crime-reduction effects without simply displacing crime into areas immediately surrounding targeted locations."[35]

PAPM is distinct from traditional hotspots policing insofar as it makes use of computer software, often developed using machine learning techniques, to automate the task of place-based crime analysis. PAPM can incorporate a much larger number of data sources than traditional hotspots policing, and PAPM is designed to produce more precise analyses of high-crime places much faster than manual crime analysis. Because PAPM is a new policing innovation, there are insufficient academic studies to support a clear conclusion about the general efficacy of PAPM in crime-reduction.[31,35] Some studies have found statistically significant decreases in crime when police operations are directed using computer software, compared with traditional hotspots techniques.[32,38] Other studies have found no evidence of greater crime reduction when police used the software-driven crime modeling, compared to traditional crime modeling techniques.[18] Meta-analyses of the efficacy of PAPM techniques are inconclusive but also likely outdated.[31,35] For example, a study published after the most recent reviews found "substantial benefits for [reducing] property crime" when marked patrol cars were given dedicated assignments to crime areas predicted by the PAPM software Hunchlab.[2,38]

The potential of PAPM technologies, and their predictive policing predecessors, to replicate or introduce racial bias in policing has been a central focus of recent academic literature.[7,8,15,16,22,28] Two seminal articles demonstrated the potential for racial bias in PAPM systems. In a 2016 article, Kristian Lum and William Isaac demonstrated that a PAPM system developed to forecast drug crime on the basis of arrest data can lead to a "feedback" loop of escalating police attention in Black communities. The first problem with arrest data, as Lum and Isaac describe it, is that, "Because [arrest] data is collected as a by-product of police activity, predictions made on the basis of patterns learned from this data do not pertain to future instances of crime on the whole. They pertain to future instances of crime that become known to police."[28] When police spend more time in Black communities looking for drugs, they make more drug arrests in those communities. A PAPM system that uses arrest data to forecast crime will therefore recommend that police go to those same communities to look for drugs. As police are deployed in greater numbers, they will make even more drug arrests, confirming prior predictions, and this arrest information will be fed back into the PAPM system to update its forecasts. In this way, a PAPM system can lead to a feedback loop of intensifying police attention directed to minority communities. In 2018, Ensign et al. prove a similar result.[15] They also propose mitigating feedback loops by filtering input data from a location by the probability that the PAPM forecast recommended sending a police patrol to that location.

---

[2] Hunchlab, the software, has since been acquired by SoundThinking (formerly ShotSpotter). The Hunchlab software was the foundation of ResourceRouter, SoundThinking's PAPM system.

The debiasing techniques described by Lum and Issac, as well as Ensign et al. are tailored to address feedback loops caused by PAPM systems that are trained on arrest data. As discussed above, arrests are but one of a variety of sources of information used by contemporary PAPM systems to forecast crime. When a PAPM system includes features like seasonality, time of month, day of week, time of day, holidays, upcoming events, weather, and locations of liquor establishments, data about which is not collected as a by-product of police activity, it is unclear whether the mitigation techniques proposed by Ensign et al apply.[15] And yet each of those features might demonstrate problematic correlations with race and ethnicity—e.g. we might expect that liquor stores and pawn shops act as ecological attractors for crime *and* that they are concentrated in minority neighborhoods. It is therefore desirable to have a method for mitigating all instances of racial bias in a PAPM system, regardless of the data source.

It has repeatedly been shown that it is impossible to maximize both fairness and accuracy for all fairness metrics used to assess algorithmic classification systems.[4,23,42] For example, in 2021 Jabri found that PAPM systems "decrease serious property and violent crime" but also found disproportionate racial impacts in both arrests for violent crimes and lower-level offenses.[21] As such, addressing algorithmic fairness in an application area where accuracy is paramount must confront the tradeoff between fairness and accuracy.

Previous work has found that addressing racial bias in PAPM systems might compromise accuracy. In 2018, Mohler et al. developed a method to achieve demographic parity using point process models of crime.[34] This method adds "a penalty term to the likelihood function that encourages the amount of police patrol received by each of 1,…,M demographic groups to be proportional to the representation of that group in the total population."[34] One of their key findings is that "patrol rates can be matched to population demographics, but at a cost to the accuracy of the algorithm and a lowering of the maximum crime rate reduction possible by predictive policing." One central challenge to addressing racial bias in PAPM systems is thus to achieve an acceptable balance between fairness and accuracy.[3,10] Demographic parity is also the fairness measure in terms of which criticisms of PAPM are typically raised. For example, no critical papers that we are aware of argue that PAPM systems have different *error rates* for different racial or ethnic groups. The present paper follows recent work in this area by assuming that demographic parity is a suitable measure of fairness to use to assess PAPM systems, insofar as our goal is to increase equity between demographics while minimizing costs to accuracy. Further work on achieving strict, mathematical demographic parity is left for future work.

This tradeoff provides the motivation for the present study, the objective of which is to illustrate a method to eliminate racially correlated features in a real-world PAPM application while preserving accuracy.[3]

[3] This work is inspired in part by the pioneering research of Ibo van de Poel, Jeroen van den Hoven, and others, in incorporating values into the process of technology design and negotiating conflicts in those values as they arise.[44]

We apply this technique to remove racially correlated features to SoundThinking's ResourceRouter PAPM system (**Figure 1**). The SoundThinking PAPM system is designed to direct officers to precise patrol areas within their normal area of assignment for short, focused periods of time. ResourceRouter provides officers with an interface that displays current directed patrols for their particular shift and jurisdiction. It also provides officers with a list of recent crime events in the patrol area being visited so that officers have awareness of recent activities and potential patterns or trends.

ResourceRouter includes five features designed to limit problems with enforcement bias from traditional patrolling methods: (1) It bases modeling on crime types that are in the majority of cases reported by citizens or victims themselves (part I crimes).[4] (2) It uses crime theory/pattern-agnostic modeling. This means that the model gives equal weight to multiple theories or patterns of crime to assess risk of crime.[38] (3) It utilizes intelligent patrol metering to measure the amount of time an officer spends in a directed patrol area and to minimize occurrences of over-policing. Each directed patrol area has a visible 15-minute timer and patrol meter to keep track of time and number of visits to an area. A separate function blocks out visited boxes for a set time so that the officer must patrol other boxes first before the current patrol can be visited again.[21] This is in line with Koper's results in his 1995 paper, which found that patrolling areas in 10-15 minute intervals results in the highest crime reduction benefit and can last for hours.[24,50] (4) It utilizes weighting of crime to prioritize more harmful violent crime over non-violent crimes.[29] And (5) it has reporting features that facilitate transparency and accountability for reporting and community relations. Reporting helps the police prioritize oversight and accountability. ResourceRouter logs patrol activities including time, place, and tactics used. Agencies can generate reports to show what areas officers visited during a shift, what tactics were employed, and how much time was spent in each area. These reports provide insights into officer activities during their uncommitted time and provide a level of oversight to command staff that can be fed into future assignments and strategies.[5]

[4] This distinction is drawn from the FBI's Uniform Crime Reporting (UCR) Program, which distinguishes between Part I and Part II crimes.[43]

[5] This description of ResourceRouter's reporting features is based on insight from SoundThinking's Data Science project team.

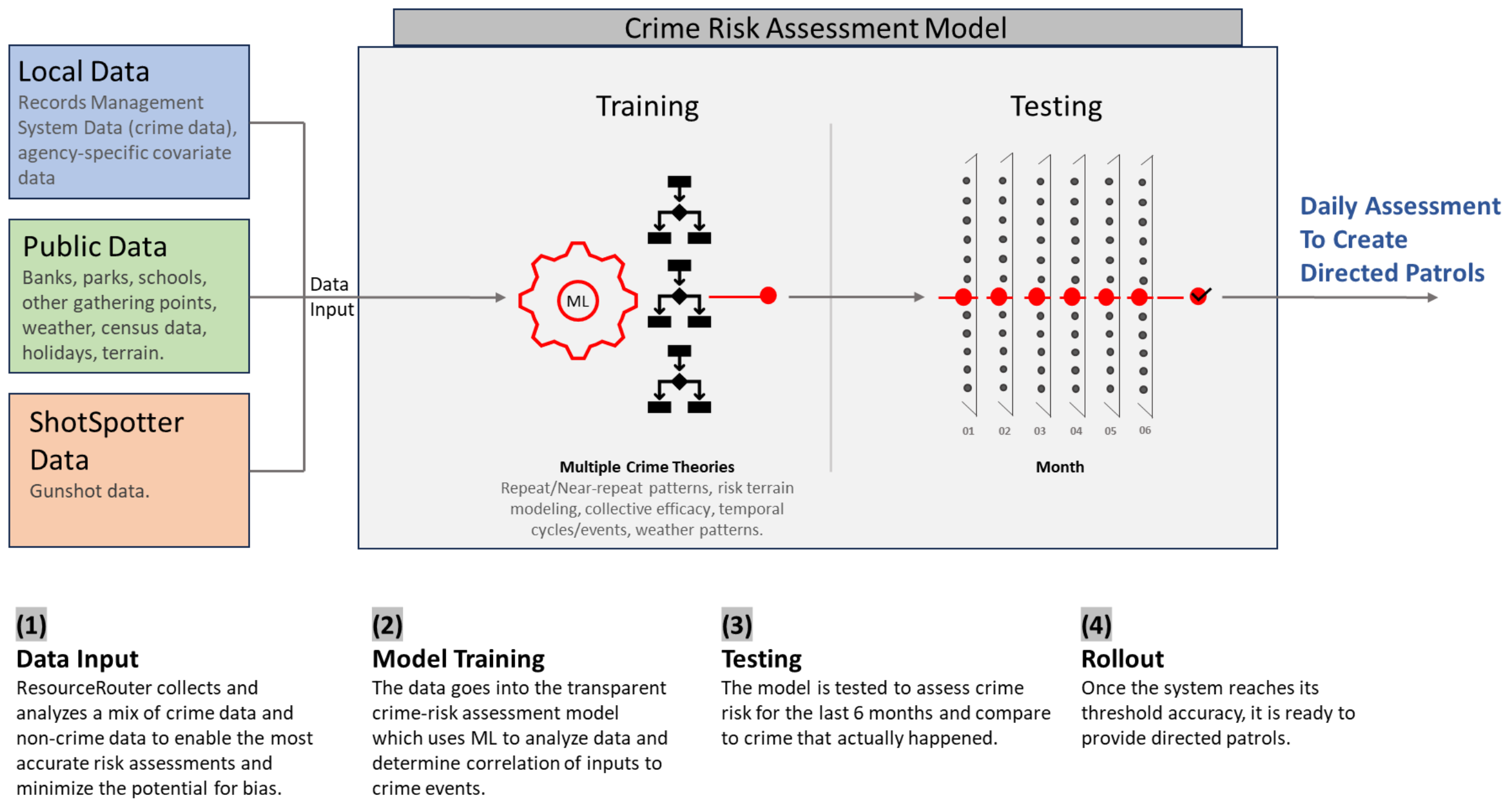

**Fig. 1 |** General Diagram of SoundThinking ResourceRouter PAPM System.

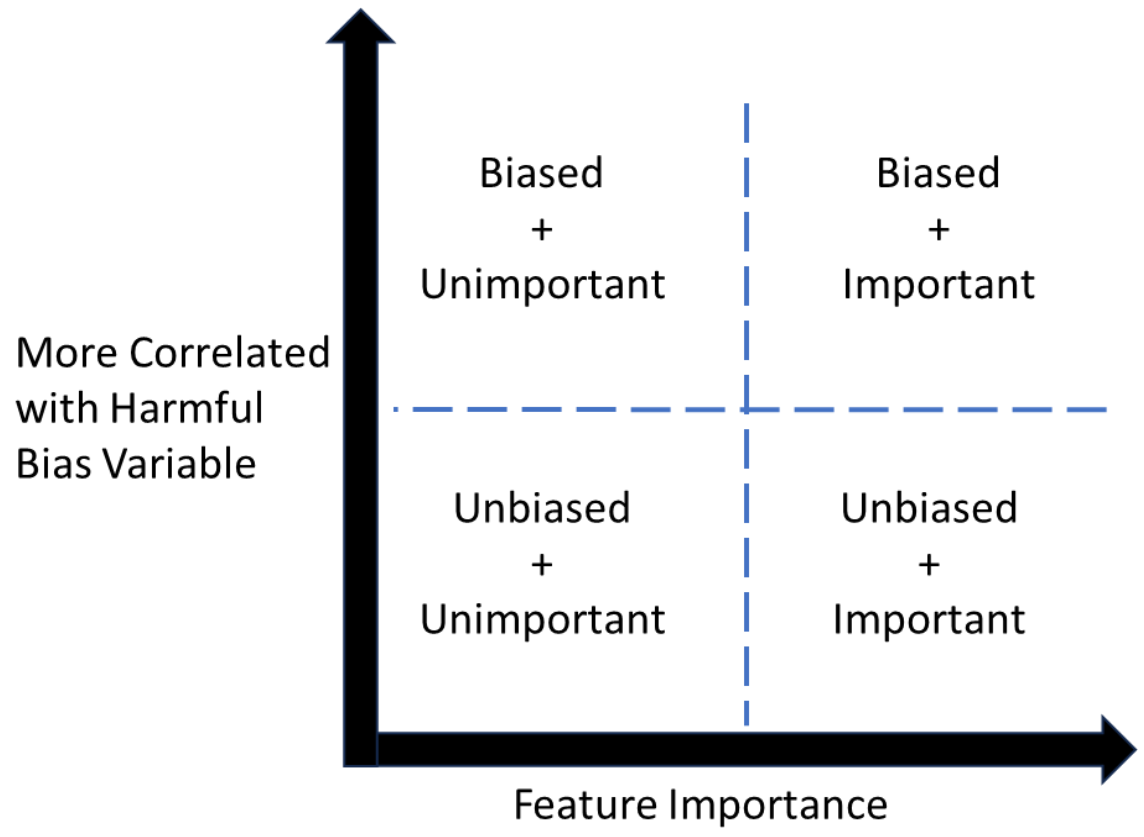

**Fig. 2 |** Conceptual zones of relationships between feature prediction importance and correlation to a harmful bias variable.

The debiasing method introduced here aims to minimize the influence on prediction of variables that are moderately to strongly associated with racial or ethnic features. Because variables with low feature importance (left quadrants in **Figure 2**) already have minimal influence on the predictions, this work prioritizes removing variables with high feature importance (top right quadrant in **Figure 2**).

The debiasing method (see methods sections for a more in depth description), for the set of all features used to train a model to predict the location of crime, involves the following steps:
(1) measure the relative gain to model prediction for each feature;
(2) measure the association between each feature and protected features of individuals, with particular focus on race or ethnicity;
(3) remove all features with a medium to high correlation with race/ethnicity and also low gain using a static cut-off;
(4) eliminate all features that fall above a defined acceptable threshold of relative gain to model prediction vs. association with protected class;

(5) retrain the model on the remaining features
(6) repeat the preceding two steps until all features with high association with protected class have been eliminated or until the model performance decreases below a predetermined value.[6]

'Gain' can be viewed as a measure of how much each feature contributes to minimizing entropy in the decision trees in the XGBoost model, and is therefore a good feature importance metric. Our decision to remove features with high 'Gain' may be somewhat controversial. Although the main purpose of this study is not to create an interpretable AI system, feature sparsity and decision trees both promote certain notions of interpretability.[27] If the model is driven by few features with high gain, its process will be more interpretable by the user. If those features have relatively low racial bias, that would both indicate that the model as a whole is less biased, and help developers show stakeholders that their model is less prone to biased decisions. As such, our goal with the method is to ensure that the model is driven by features with low racial or ethnic bias, and that features with moderately high racial or ethnic bias are either cut from the model or are relatively unimportant for the model predictions.

**Figure 3** depicts an example of the debiasing technique applied to a simplified toy crime model. In actuality, redundancy, which is imperative for success in using this debiasing method, would require the model to be trained using many more features.

**Figure 3, graph 1** illustrates that after training the model with all features, many of the features provide little information gain and high racial bias (top left corner, above threshold line). Eliminating these low-gain, high-bias features is justified as a precautionary measure. These features do little work in driving model predictions, but they are potentially damaging if they become high information gain features in future training runs. Therefore, these features are removed from the training set as a preliminary step. The last feature removed has the second highest information gain, but potentially problematically high association with racial bias. The algorithm has then removed 6 features from the training set, and is retrained on the remaining 14.

**Figure 3, graph 2** illustrates that retraining the model increases the information gain from the remaining features, which can be seen in the graph as features shifting to the right. Here, the result is that one feature moves above the threshold and is therefore removed from the training set. The model is retrained using the remaining 13 features.

[6] This raises a concerning possibility: that each time features are removed from consideration and the model is retrained, the location of each remaining feature on the plot will change. For each feature, its normalized importance is relative to all of the features in the model. Thus, by removing some features, the relative importance of each remaining feature will change. One early worry was that this debiasing method would generate an endless cycle, as it were, whereby developers would remove some features, and then re-train the model to find that more features were now above the acceptable threshold, remove them, and so on. Upon further inspection, though, this problem is unlikely, as each feature would likely shift to the right rather than to the left. Features would certainly not shift up because their racial correlation would not change during a new training cycle. At any rate, in our testing, this problem was not encountered in any of the (real-world) production models used.

Finally, **Figure 3, graph 3** illustrates that after retraining the model a third time, no individual feature's gain is great enough for it to fall above the threshold line. Thus a stop condition is met and the algorithm finishes running.

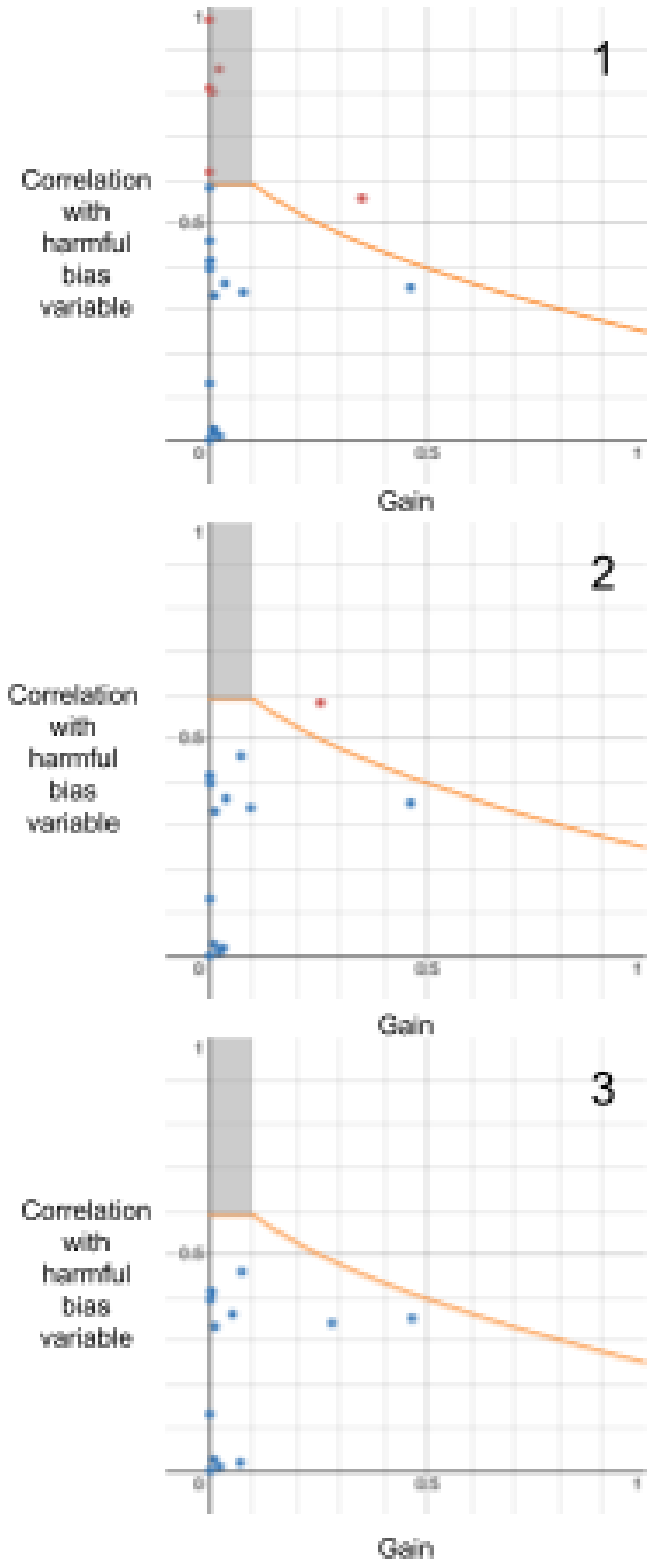

**Fig. 3 |** Algorithm running until the stop condition is met.

We applied this method to the production models in SoundThinking's ResourceRouter PAPM system across four law enforcement agencies. These are listed as Agencies A, B, C, and D, in the interest of anonymity. The present paper shares the results of testing the efficacy of this method in eliminating correlations between predictive features and race/ethnicity. We also discuss the conceptual and technical issues we encountered.

# 2 Results

Eliminating all variables which demonstrated potentially problematic racial/ethnic bias—that is, features that fell above the threshold for acceptable racial/ethnic correlation—did not significantly degrade the performance of the model. In all four cities, the accuracy of the crime models for multiple types of crime did not change significantly (**Figure 4**). The standard error of these results across 10 model repetitions also shows little variance in the Area Under Curve (AUC) metric (**Supplementary Tables 1-4**).

The following variables were most frequently eliminated due to co-correlation with race or ethnicity above a 0.5 correlation threshold regardless of gain value across multiple crime models: percent of population below poverty level when correlated with Blackalone variable, percent of rented houses when correlated with Whitealone and non-white variables, and median household income when correlated with Hispanic variable (**Supplementary Table 5**).

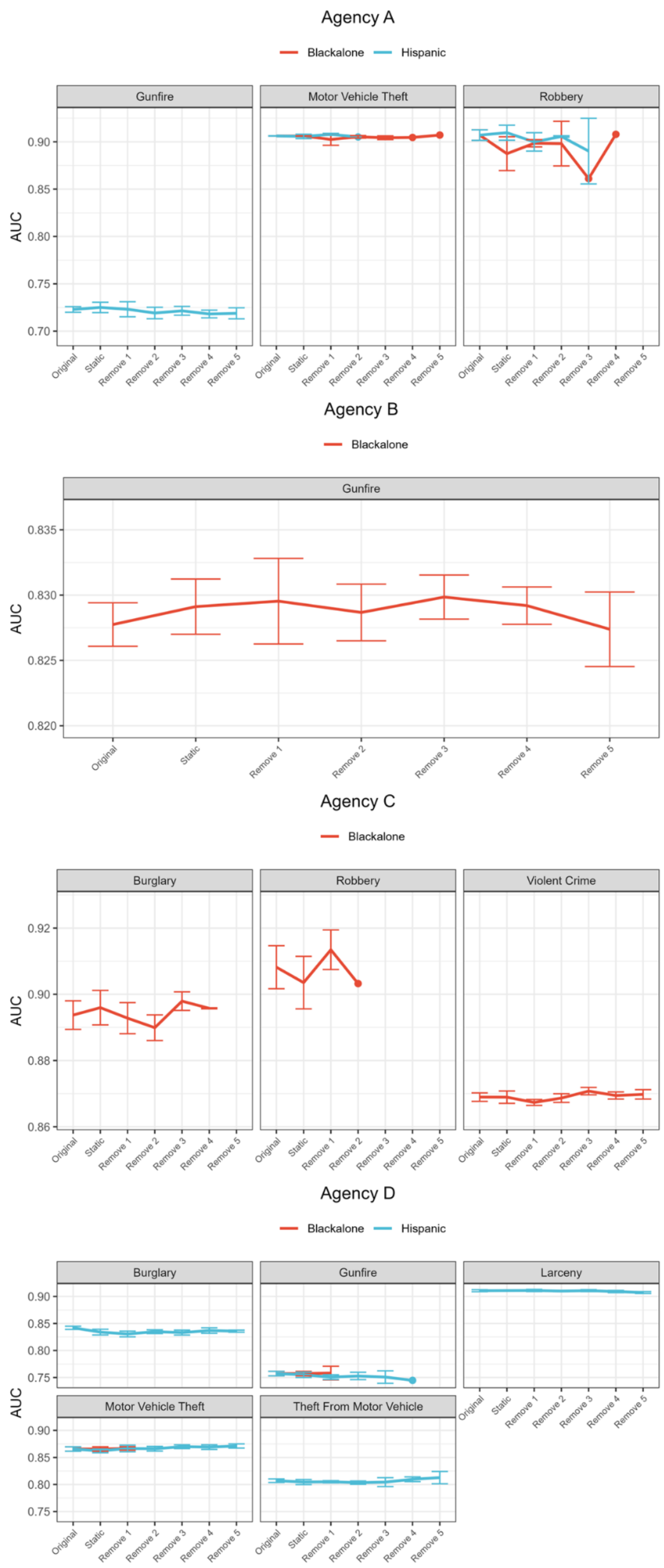

**Fig. 4 |** Change in AUC with different number of features removed for different crimes. The solid lines represent the mean across model repetitions. Vertical lines with horizontal bars represent the 95% Confidence Interval (CI) on the mean. The solid dots represent model runs with only one repetition.

**Figure 5** (also see **Supplementary Figures 1-11**) illustrates a dynamic realization of what was described in **Figure 3** for two different crime models in two different cities. The history of the gain values of selected features at all 10 model repetitions are shown from the original step (model run with all features) to the static cutoff, then to each of the feature elimination steps until the feature is eliminated, or the performance of the model sufficiently degrades, or there are no more features to be eliminated, or after five elimination steps.[7]

As the graphs demonstrate, features often show a sharp increase in gain after other related features have been eliminated. This increase in gain causes these remaining features to pass the bias tolerance threshold. They are therefore eliminated. The process continues, eliminating features with a combination of high gain and high correlation to race variables.

[7] In the interest of conserving compute resources, we chose a limit of five elimination steps. More than five elimination steps can in principle be performed, but of the 2,804 model runs conducted for this study, only 3.9% (114) of those completed five rounds of feature elimination before one of the other stop conditions was met, which suggests that five elimination steps is sufficient for the great majority of cases.

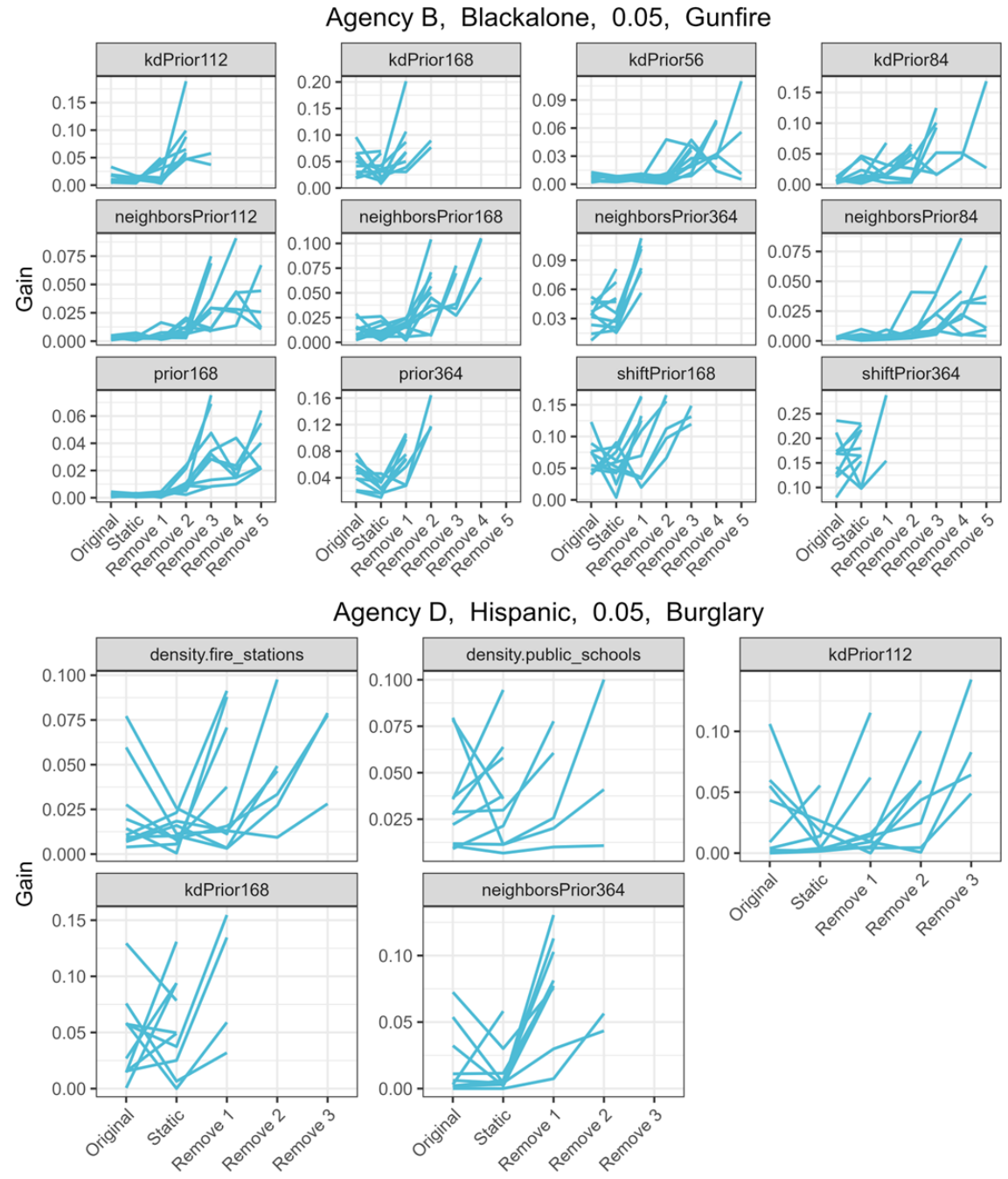

**Fig. 5 |** Difference in gain for features until elimination. Each line is a model training run.

# 3 Related work

Like the debiasing techniques analyzed by Skeem and Lowenkamp, this work focuses on algorithmic bias mitigation for AI applications, and explores an accuracy against fairness tradeoff.[42] As with Skeem and Lowenkamp's preferred methods from that work, our bias mitigation algorithm allows access to race/ethnicity in order to mitigate bias. It may be viewed as a type of proxy elimination method, although the purpose of this is not to eliminate features with high racial correlation, but rather to minimize the number of features with high racial correlation which also have high information gain for the model. This work differs from their work as it is an exploration of using a specific debiasing method via proxy elimination applied on an existing PAPM system, rather than a viability comparison of different debiasing methods for criminal justice data.

Another popular method is debiasing through preprocessing of variables to make them statistically independent of the problematic variables, explored in the field of criminal justice by both Skeem and Lowenkamp and later Aliverti et al.[1,42] Preprocessing is a viable and effective method, but retaining features in their original state through the entire process of training to prediction promotes explainability by demonstrating which features drive prediction as well as the extent of correlation between those features and race. This provides valuable information for PAPM developers, users, and other stakeholders that is not provided by preprocessing methods.

The bias mitigation method described here also differs from the method described in Ensign et al. The authors propose a method for addressing feedback loops that seeks specifically to reduce the impact of officer behavior on crime data by filtering input data–primarily arrest data–from a location by the probability that the PAPM forecast recommended sending a police patrol to that location.[15] The method of proxy elimination proposed in the present paper can be applied to a broader array of model features, including, for instance, environmental features for which officer bias in data collection is not a central concern. With that difference in mind, the method proposed here is compatible with the method described in Ensign et al. Features can be eliminated based on their correlation with protected characteristics, and, at the same time, new input data can be filtered based on the probability that the data was collected as a result of a PAPM forecast.

# 4 Discussion

The methods section below discusses “harmful racial bias,” without defining the term. As expected, preliminary experiments found that the correlation between model features and race and ethnicity depends on the demographics of the city in which the PAPM system operates and the crime(s) being modeled. Whether a model feature like the location of churches correlates with race or ethnicity depends on how those categories are conceptualized. The American Community Survey, which furnished the data used here to measure race and ethnicity, allows developers to analyze the composition of a neighborhood in terms of its Hispanic population, Black or minority population, white or non-white population, and so on. The appropriateness of choosing one of these representations over another is open for debate and will differ between localities.

Which racial or ethnic group is the appropriate focus of bias mitigation efforts, in turn, is a function of differences in the prevalence of different racial and ethnic groups as well as a variety of social and historical factors, such as redlining targeting racial minorities in U.S. cities. In cities in the Southwestern U.S., Hispanic people might be the ethnic group of greatest interest, while in the Southeastern U.S. we might have the greatest a priori concern about racial bias against Black people. Choosing appropriate race and ethnicity variables to evaluate racial bias will thus require knowledge about the historical and current demographics in cities, social and historical facts about those demographic groups, as well as the crime model used on part of the developers.

As can be seen in the supplemental material, the features that are removed by the mitigation technique vary between agencies, crime, and racial variables (see **Supplementary Figures 3** and **10** for agencies A and D, Motor Vehicle Theft on nonwhite bias). This reveals that it is likely impossible to know *a priori* which features of a model will demonstrate problematic correlations with race or ethnicity in a given community. Instead, performing the kind of debiasing method illustrated here will be necessary to uncover those correlations. In our experiments, we used four different race or ethnicity variables (see **Supplementary Table 6**).

## Future work

Because this is an exploratory study on debiasing methods for PAPM systems, this work suggests multiple avenues for future research.

The present study, as well as the ResourceRouter PAPM system under examination, use calls for service as a proxy for crimes committed. While we consider this a suitable proxy, and certainly an improvement over arrest data, we also recognize that calls for service are not a perfect proxy because calls for service are affected by factors such as trust or confidence in police. We encourage ideas for alternative proxies for crimes committed that are well suited for testing the effectiveness of these models, as work on improving their fairness are futile without effective testing.

Before starting the experiments, we explored multiple different types of equations for our threshold function, all designed to fulfill the requirement that high bias features should be removed ,especially if they also have high information gain for the model. For the experimental stage, we used a single equation while altering the bias tolerance line for different crimes.

In this debiasing method the primary stop condition is the threshold function: when there are no longer any features that fall above the threshold function, the feature elimination and retraining process ends. The method includes two secondary stop conditions: the maximum AUC loss allowed, and the maximum allowed number of retraining runs (in our case five). The first stop condition is designed to preserve the accuracy of the model. The second stop condition is key to preserving limited resources when retraining the model, a resource-intensive activity.

A fourth stop condition we considered including was to set the number of allowed features to be cut before the algorithm started, instead of cutting features until none fall above the threshold line.

These stop conditions all put more or less responsibility on the developers in the process. In this proposed method, the developers must choose not only a good general threshold function, but also constants in the function that result in neither too few or too many features being cut.

If, instead, the maximum AUC loss allowed and the number of features to be dropped are predetermined, the values in the threshold functions that are held constant in this work could be made variable, and those values could be adjusted as part of the algorithm until either the AUC or number of features cut stop condition is met. This would reduce the developers' responsibility to predict the perfect threshold function, as the algorithm would find suitable values.

Furthermore, it is likely that other threshold functions than the logarithm-based one used in the present study better suit the goal of this debiasing method. This should also be explored in future research.

The rationale for setting different thresholds for different crimes stems from the idea of cost of crime, as operationalized by the RAND Corporation, but the costs of higher- and lower- cost crimes differ by orders of magnitude, and can't therefore feasibly be implemented in determining constants.[18] In this work, we attempted to incorporate cost of crime in the different bias tolerance thresholds (see the Methods section), but we recognize that this may not be satisfactory. More work is needed to determine how the concept of cost of crime should be incorporated into this and other debiasing methods.

One key takeaway from this work is the importance of redundancy in the set of features the model trains on. Redundancy in a model is key to eliminating racial/ethnic correlations while preserving predictive power. A variable is redundant just in case its predictive value is replicated by some other variable(s) already included in the data set. Without redundant variables, we could not eliminate variables without sacrificing predictive accuracy.

Redundant features will often be proxies for other features. Using a definition from Kraemer et al (2001), a proxy is a correlate of a strongly predictive risk factor that also seems to be a risk factor for the same outcome—but the only connection between the correlate and the outcome is the strong risk factor correlated with both.[25] In the case of crime risk, some demographic variables such as household income or property values are likely to be redundant in virtue of being proxies for some other risk factor—perhaps social mobility—that is itself the true driver of crime. When PAPM models demonstrate redundancy in this way, it is in principle possible to eliminate a variety of demographic variables from a model without losing accuracy. This is a plausible explanation for the model's continued strong performance even after features that demonstrate a problematic correlation with race are removed: the redundancy in the original set of features allowed for the information yield from features eliminated for their problematic racial bias to be replaced with other information from one or more other features which had similarly strong predictive strength, but which lacked the problematic racial bias.

This work leans heavily on lower-bias features replacing much of the information gain when higher-bias features are removed. Performing a qualitative study exploring the relationship between lower bias features that eventually replace the information gain from the removed higher bias features could be very valuable. If it is possible to know *a priori* which lower bias features might eventually supplant the removed higher bias features, that would limit the number of times the model has to be retrained, saving time and money involved in retraining a large model multiple times. A qualitative study on the association between lower-bias replacing features with the higher-bias replaced features could have substantial impact in this area of research.

It is, moreover, unclear how many features are needed for this "redundancy requirement" to be fulfilled. Insight into this question would both allow for better understanding of what redundancy does for debiasing and potentially lower the bar of entry for developing these types of algorithms, if the number of redundant features required is much smaller than the approximately 150 features used by ResourceRouter.

The role of redundancy in explaining the success of our feature removal technique raises two conceptual issues in need of further investigation.

First, earlier work has indicated that removing problematic features in datasets with high redundancy may result in the models' recoding other features to replace the removed ones.[3] We believe that our method of removing a large number of high bias features, as well as iteratively removing influential high bias features, reduces this risk. One explanation of our key result–that it is possible to eliminate high-bias features without compromising model accuracy–is therefore that some *further* feature persists in the model that both encodes the same information as the eliminated features and dominates in driving the model assessments. One might worry, then, that the model is sending police patrols to the same places, that this distribution is racially unequal, even after the feature elimination process, and that this explains why model accuracy is preserved. For example, calls for service might encode the same information as some subset of eliminated features, and calls for service might demonstrate strong correlations with race.

This possibility raises a second conceptual issue in need of further investigation: what makes for legally or ethically problematic correlations between facially neutral variables and race? It is possible that some facially neutral variables will be both (a) very strong predictors of crime and (b) correlated strongly with race. As noted above, one motivation for adopting PAPM in the first place is that crime is often concentrated in a very small area, or sets of areas, within a city. Cities are also often *de facto* segregated by race.[5,17] Therefore, it is unsurprising that some ethnic groups will be overrepresented in these small areas where crime is concentrated. Suppose that the best available evidence of criminal activities at a place was the totality of crime reporting from available sources, including crime scene reports, arrest reports, and calls for service data. In this world, given limitations in evidence, using the best evidence available to predict crime might produce crime predictions that “overrepresent” members of one ethnic group. But this would not appear to be in any obvious way discriminatory, if there is in fact a correlation between race and the base rate of crime at a place, and assuming that predicting crime is a legitimate law enforcement activity. In this situation, it would be a mistake to eliminate crime data from the model simply because of its correlation with race or ethnicity. A core question when evaluating PAPM technologies for racial impact, then, is whether there is independent evidence that some ethnic groups are concentrated in high-crime areas within a city of deployment. If they are, then this fact must be taken into account when deciding whether to eliminate facially-neutral features that correlate with race from a model.

Finally, as noted in the introduction, there are limited rigorous academic studies of either the efficacy or racial bias of PAPM systems. Furthermore, it is difficult to draw general conclusions about PAPM by comparing studies due to differences between them.[35] PAPM technologies continually evolve, and the policing tactics deployed in response to PAPM risk assessments vary significantly across studies. For example, contemporary PAPM systems like ResourceRouter provide new directed patrols for every shift, whereas the Hunt et al (2014) study examined a PAPM system that forecasted high-risk areas to inform police strategy for an entire month.[20] These are very different time horizons from which to formulate an operational response.

Even if the theory underlying PAPM rests on a firm foundation, randomized controlled trials (RCTs) would offer stronger evidence. More RCTs are needed, even if the conclusions are limited to a specific technology, in a specific city, using specific law enforcement tastics. Further avenues of study include: the effects of PAPM on the public’s attitudes toward police, especially trust and legitimacy; the effects of PAPM on equality in law enforcement (measured by equality of stops or arrests); and the effects of PAPM on collective efficacy and community perceptions of crime and criminality.

# 5 Methods

The SoundThinking ResourceRouter PAPM System uses an extreme gradient boosting (XGB) machine learning model (see **Supplementary Table 7** for hyperparameters and values used).[12] This model is trained on a diverse set of features (**Supplementary Tables 8-12**) each of which represents different crime theories/patterns such as baseline crime levels, near repeat patterns, risk terrain modeling, collective efficacy, temporal cycles – seasonality, recurring temporal events, and weather.[19,30,37,40,45] For the purpose of this paper a set of additional American community survey (ACS) data points from the census.gov API was turned into features, and a set of original ACS data points were engineered to be presented both as density and percent features, and added to the standard model training feature-set used by the SoundThinking PAPM in its live production system (see **Supplementary Tables 8-12** for feature names). This was done to increase the chance that more features would be affected by the method presented in this study.

This model was used to train and test crime models from four cities representing four different regions of the country with four different demographic profiles (**Supplementary Table 13**). A set of city and crime model combinations were sampled for this study to investigate the variance of outcome in both prediction accuracy and removal of features given a threshold of bias tolerance. In total 24 city and crime model combinations using different bias tolerance thresholds were run 10 times each (except for two city-crime combinations, **Supplementary Tables 1-4** and **Supplementary Figures 12-20**). The models were run against our different harmful bias variables: Whitealone; Nonwhite; Hispanic; Blackalone (See **Supplementary Table 6** for details) to address the potential effect that different demographics have on the relationship between modeling features and harmful bias variables.

Census data was extracted from the census.gov API prior to modeling. For each city, the 2021 five-year average for a set of census data in the form of geographical information systems (GIS) shapefiles on a census-tract level from the ACS dataset was downloaded and compiled into one shapefile with multiple features (see **supplementary material Appendix A** for R script to download the data via API). Once this shapefile was compiled it was used to create raster-based .tif files that SoundThinking's APM system can use to generate features with (see **Supplementary Table 8** for features and **supplementary material Appendix B** for R script to change ACS shapefile to useful features).

For a given model, training data was first extracted from a database that houses the formated training and test data. This data is feature-engineered from raw data into column format where each column is a feature and each row is a combination of a time slice (in our case a police shift) and a location on a map (the jurisdiction of the police department across a city is divided into 250 meter by 250 meter raster cells). This combo is called a chronon and it is the unit of analysis for the machine learning model.

To test the accuracy of a model a hold-out test set consisting of the last 6 months of data was applied to the trained model object. We used the Area Under The Curve (AUC) metric to assess how accurate the model was at predicting crime or no crime in each chronon in the test set. The AUC method is a ranking method. It gives you the probability that the model ranks a random chronon with an actual crime more highly than a random chronon without a crime. The AUC is presented as a percentage meaning that it represents how often, for two given cells and shifts -- one containing a crime and one not -- the model can predict which examples contain the crime. For example, if the model was guessing randomly then it would score 0.50 (a 50/50 chance of selecting between the two examples). If this value is 0.90, it means that when the model is asked to pick from the two examples, it is able to select the correct example 90% of the time. Values can range from zero to one; values closer to one are better.

The following steps were taken for a given model and crime combination. First the full model using all variables was trained. Secondly, correlations are run between chosen harmful bias variables and all the model variables. The Spearman's Rank Correlation method was used to measure the strength and direction between variables.[51] Once completed, a static cut-off filter was applied to any variable with gain above 0.1 and co-correlation with harmful bias variable above 0.5. This was followed by retraining of the model without those variables dropped by the static cut-off. Then an iteration and elimination function was started that applied the chosen bias tolerance threshold line given the crime model. In total, 2804 model training runs were conducted, including original training (n=872), static cut-off round (n=872), and elimination rounds (1=322, 2=270, 3=188, 4=166, 5=114).

A set of features were not considered in the static cut-off and in the iteration and elimination function in every model run because they had a correlation value of near-zero due to being time-based variables (**Supplementary Table 14**). It is important to note that these features were still part of the model training but could not be cut by the threshold function.

The threshold line used in our experiments is defined by the following equation:

$$\frac{log(2+10*Gain*corr+20*Gain*corr^3)-0.69}{2}$$

Which generalizes to:

$$\frac{log(A+B*Gain*corr+C*Gain*corr^3)-D}{E}$$

Using that threshold line, any variable above the threshold line was cut. **Figure 6** shows where the bias tolerance threshold lines fall on the graph.

All three threshold lines are superimposed on the graph to show their difference in shape and depth along the axis. The gray zone is the static cut-off area.

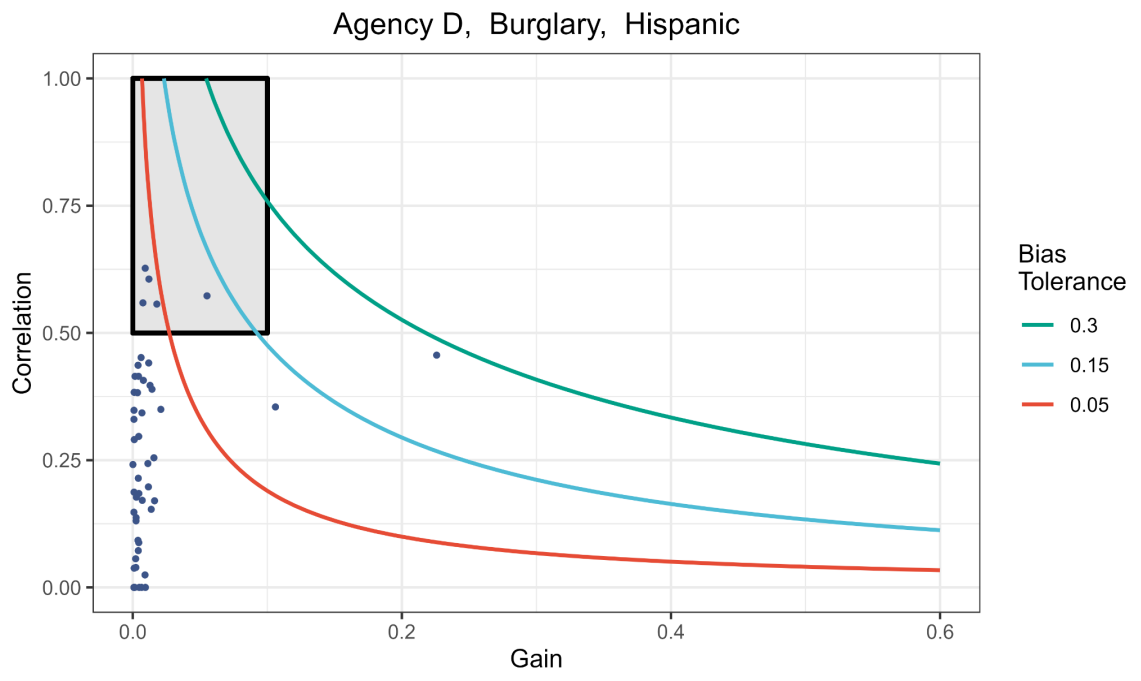

**Fig. 6 |** Bias tolerance threshold examples.

After the first iteration, having removed variables that surpassed the bias tolerance threshold, the model was re-trained again to see if any new variables surpassed the threshold line. If yes, repeat (in this study the repetitions were stopped after five rounds) until either no more variables surpass the threshold line or the AUC metric for model accuracy falls by more than 10% (see **supplementary material Appendix C, D,** and **E** for R scripts to retrieve data from an example database, run the model procedure described above, and produce graphs).

## Acknowledgements

Simen Oestmo and Lester Wollman would like to thank Scott Lamkin for his invaluable help with the threshold equation developed for this study.

## Supplementary materials

[Appendix A - Download ACS data R Notebook]
[Appendix B - Manipulate ACS data R Notebook]
[Appendix C - Machine Learning Functions Code R Script]
[Appendix D - Run Experiment Iterations Code R Notebook]
[Appendix E - Data Analysis Code R Notebook]
[Supplementary Figures]
[Supplementary Tables]

## Appendix A - Download ACS data R Notebook

````
---
title: "Grab ACS data (variables and geometry) - A de-biasing technique for place-based algorithmic patrol management - Einarsson et al."
output: html_notebook
---

Code to retrieve ACS data for census.gov API

```{r}
library(dplyr)
library(tidycensus)
library(sf)
library(tigris)  # for getting Census block group geometry
```

```{r}
# Set the API key
options(tigris_use_cache = TRUE) # This line will use local cache to speed up the process
#Insert you census.gov API key: - after this you can remove the API key from this script as it is stored.
census_api_key("", overwrite = TRUE, install = TRUE)
#Reload R environment:
readRenviron("~/.Renviron")
#Check if the API key is ready to use:
Sys.getenv("CENSUS_API_KEY")
```

```{r}
# Get the variable names and labels for the ACS 2021 data-set
acs_vars <- load_variables(2021, "acs5")

# retrieve a full list of ACS variables by block group
acs_blockgroup <- acs_vars %>%
  filter(geography=='block group')

# list of ACS variables we want to retrieve.
variables <- c("B01003_001",
"B01002_001",
"B15003_016",
"B20002_001",
"B19013_001",
"B17010_002",
"B19001_002",
"B25001_001",
"B25002_003",
"B25002_002",
````
````

```
"B25003_003",
"B25064_001",
"B25046_001",
"B03002_012",
"B02001_001",
"B02001_002",
"B02001_003",
"B02001_004",
"B02001_005",
"B02001_006",
"B02001_007",
"B02001_008",
"B02001_009",
"B02001_010",
"B23025_005",
"B23025_002"
)

#create subset of the block groups= list with just the variables we want.
ACS_Vars_sub <- acs_blockgroup %>%
  filter(name %in% variables)

# for logging purposes write out the list of ACS variables you are interested
in by block group
write.csv(ACS_Vars_sub,"acs_vars.csv")
```

```{r}
# Set state and county FIPS codes to download
state_fips <- ""  #  <--- set your own state fips code
county_fips <- ""  #  <--- set your own county fips code

# Get Census block group geometry for the state and county FIPS you set
bg_geo_all <- block_groups(state_fips, county_fips, cb = TRUE)

#Drop columns that is not needed from bg_geo data frame (only two that are
needed: GEOID and geometry)
bg_geo = subset(bg_geo_all, select = c("GEOID", "geometry"))
```

```{r}
# Define a dictionary to rename the columns of each variable
cols <- c(
  "GEOID" = "GEOID",
  "NAME" = "NAME",
  "B01003_001E" = "poptotal",
  "B01002_001E" = "medage",
  "B15003_016E" = "unedu",
  "B20002_001E" = "medinc",
```

```
  "B19013_001E" = "medhhinc",
  "B17010_002E" = "poppov",
  "B19001_002E" = "hhnoinc",
  "B25001_001E" = "houses",
  "B25002_003E" = "vachouse",
  "B25002_002E" = "occhouse",
  "B25003_003E" = "rentunit",
  "B25064_001E" = "medrent",
  "B25046_001E" = "vehicles",
  "B03002_012E" = "hispanic",
  "B02001_001E" = "race",
  "B02001_002E" = "whitealone",
  "B02001_003E" = "blackalone",
  "B02001_004E" = "amindnalone",
  "B02001_005E" = "asianalone",
  "B02001_006E" = "hawaiialone",
  "B02001_007E" = "otherracealone",
  "B02001_008E" = "tworaces",
  "B02001_009E" = "tworacesincone",
  "B02001_010E" = "threeraces",
  "B23025_005E" = "unempl",
  "B23025_002E" = "laborforce")

# Create table with Census data for the state and county you have set
bg_data <- tidycensus::get_acs(
  geography = "block group",
  variables = variables,
  state = state_fips,
  county = county_fips,
  survey = "acs5",
  year = 2021,
  output = "wide"
)

#convert to tibble for easier manipulation
bg_data2 <- as_tibble(bg_data)

# Identify columns to keep, and make new table
bg_data3 <- bg_data2 %>% dplyr::select(-ends_with("M"))

# Rename the columns using the dictionary
names(bg_data3) <- cols[names(bg_data3)]
```

```{r}
# Merge the block group variable data with the block group geographic data
bg_sf <- st_as_sf(bg_geo) %>%
  left_join(bg_data3, by = "GEOID")
```

````
```{r}
# Write the data to a shapefile - this file will be used in the ProcessACSVariables R Notebook next.
st_write(bg_sf, ".shp")
```
````

## Appendix B - Manipulate ACS data R Notebook

````
---
title: " Process an ACS shapefile - A de-biasing technique for place-based algorithmic patrol management - Einarsson et al."
output: html_notebook
---

Code to take the ACS shapefile with multiple data variables retrieved with the GetCensusData R notebook and turn the shapefile into raster tiffs that is useful for the ML model. These raster tiffs shows either percentage of variable in a raster cell, density of a variable in a raster cell, or mean/median of a variable in a raster cell.

```{r echo=FALSE}
# Install Libraries:
````
````

````
library(optparse)
library(raster)
library(spatstat)
library(sp)
library(rgeos)
library(rgdal)
library(maptools)
library(caret)
library(foreach)
library(doParallel)
library(dplyr)
library(ipred)
```

```{r}
####
# PARSE AND PRINT FUNCTIONS
####
option_list = list(
  make_option(c("--cores"), type="integer", default=4,
              help="Set number of CPU cores to use [default %default]",
              metavar="number"),
  make_option(c("--verbose"), type="integer", default=2,
              help="Adjust level of status messages [default %default]",
              metavar="number")
)
parser = OptionParser(usage = "%prog [options] extent shapefile",
option_list=option_list)
arguments = parse_args(parser, positional_arguments=TRUE)

# extract the remaining options
opt = arguments$options

kVerbose = opt$verbose

PrintStatus = function (visible.at, ...) {
  # Outputs status messages for logging purposes
  #
  # Args:
  #   visible.at: works with kVerbose constant to determine what to print
  #   ...: any number of other variables
  if(visible.at <= kVerbose) print(paste0("Status --  ", ...))
}
```

```{r}
####
# SETUP PARAMETERS for parallel cores if needed
####
````

````
# set number of cores to use
cores = 8
kParallelCores = cores

kRasterSplitFactor = 4

# enable multithreaded environment
if(kParallelCores > 1) {
  registerDoParallel(kParallelCores)
}
```

```{r}
# set file paths
# extent raster file for agency/city - set your own path - Take the shapefile of a jurisdiction and create a raster map out of it. Take that raster file and use it here.
kExtentFile <- file.path("")

# shapefile of county/state/country - set our own path - this file is generated by the "GetCensusData" R Notebook
kShapeFile <- file.path("")
```

```{r}
####
# MAIN
####

# read in files
PrintStatus(1, "Reading extent file...")
extent = raster(kExtentFile)
print(extent)

# create mask for adaptive density function
extent.fine = disaggregate(extent, fact=kRasterSplitFactor)

# create output directory
output.dir = file.path(dirname(kShapeFile), basename(kExtentFile))
dir.create(output.dir, recursive=TRUE)

PrintStatus(1, "Reading shapefile...")

shape = readOGR(dirname(kShapeFile),gsub("\\.shp", "", basename(kShapeFile)))
```

```{r}
# recalculate area in projected units
````

````
PrintStatus(1, "Reprojecting shapefile to extent projection...")
shape.projected = spTransform(shape, CRS(proj4string(extent)))
shape.projected$AREA = gArea(shape.projected, byid=TRUE)

# add centroid coordinates
shape.data.coordinates = coordinates(shape.projected)
shape.data.coordinates = data.frame(shape.data.coordinates)
names(shape.data.coordinates) = c("coord_x", "coord_y")
shape.data.original = cbind(shape.projected@data, shape.data.coordinates)
shape.data.original$LOGRECNO = shape.data.original$GEOID =
shape.data.original$GEOID_STRP = NULL
shape.data.original$PERIMETER = shape.data.original$CENTROID_X =
shape.data.original$CENTROID_Y = NULL
```

```{r}
# remove columns with zero variance or nearly so
shape.data.original.filtered = shape.data.original[,
-nearZeroVar(shape.data.original)]
shape.data.original.filtered = shape.data.original.filtered %>%
mutate_each(list(as.numeric))
```

```{r}
# fix NA values
preProcessor = preProcess(shape.data.original.filtered, method=c("bagImpute"),
verbose = TRUE)
shape.data.original.filtered.noNA = predict(preProcessor,
shape.data.original.filtered)
```

```{r}
# calculate useful variables
# density variables
shape.data.processed =
data.frame(house.density=(shape.data.original.filtered.noNA$houses /
shape.data.original.filtered.noNA$AREA))
shape.data.processed$population.density =
shape.data.original.filtered.noNA$poptotl /
shape.data.original.filtered.noNA$AREA
shape.data.processed$vehicles.density =
shape.data.original.filtered.noNA$vehicls /
shape.data.original.filtered.noNA$AREA
shape.data.processed$vacanthouses.density =
shape.data.original.filtered.noNA$vachous /
shape.data.original.filtered.noNA$AREA
shape.data.processed$rentedhouses.density =
shape.data.original.filtered.noNA$rentunt /
shape.data.original.filtered.noNA$AREA
````

````
shape.data.processed$unemployment.density =
shape.data.original.filtered.noNA$unempl /
shape.data.original.filtered.noNA$AREA
shape.data.processed$popbelowpovertylevel.density =
shape.data.original.filtered.noNA$poppov /
shape.data.original.filtered.noNA$AREA
shape.data.processed$hhnoincome.density =
shape.data.original.filtered.noNA$hhnoinc /
shape.data.original.filtered.noNA$AREA
shape.data.processed$belowhsedu.density =
shape.data.original.filtered.noNA$unedu /
shape.data.original.filtered.noNA$AREA
```

```{r}
# median and mean variables
shape.data.processed$age.median = shape.data.original.filtered.noNA$medage
shape.data.processed$hhincome.median =
shape.data.original.filtered.noNA$medhhnc
shape.data.processed$rent.median = shape.data.original.filtered.noNA$medrent
shape.data.processed$hhsize.mean = pmin(10,
shape.data.original.filtered.noNA$poptotl /
(shape.data.original.filtered.noNA$houses + 1))
```

```{r}
# percent variables
shape.data.processed$hhnoincome.percent =
shape.data.original.filtered.noNA$hhnoinc /
(shape.data.original.filtered.noNA$houses + 1)
shape.data.processed$vacanthouses.percent =
shape.data.original.filtered.noNA$vachous /
(shape.data.original.filtered.noNA$houses + 1)
shape.data.processed$rentedhouses.percent =
shape.data.original.filtered.noNA$rentunt /
(shape.data.original.filtered.noNA$occhous + 1)
shape.data.processed$unemployment.percent =
shape.data.original.filtered.noNA$unempl /
(shape.data.original.filtered.noNA$labrfrc + 1)
shape.data.processed$belowhsedu.percent =
shape.data.original.filtered.noNA$unedu /
(shape.data.original.filtered.noNA$poptotl + 1)
shape.data.processed$popbelowpovertylevel.percent =
shape.data.original.filtered.noNA$poppov /
(shape.data.original.filtered.noNA$poptotl + 1)
```

```{r}
# race/ethnicity variables
````

```
shape.data.processed$whitealone.percentage =
shape.data.original.filtered.noNA$whiteln /
shape.data.original.filtered.noNA$race
shape.data.processed$blackalone.percentage =
shape.data.original.filtered.noNA$blackln /
shape.data.original.filtered.noNA$race
shape.data.processed$hispanic.percentage =
shape.data.original.filtered.noNA$hispanc /
shape.data.original.filtered.noNA$race
shape.data.processed$nonwhite.percentage =
rowSums(cbind(shape.data.original.filtered.noNA$blackln,

shape.data.original.filtered.noNA$amndnln,

shape.data.original.filtered.noNA$asianln,

shape.data.original.filtered.noNA$hawailn,

shape.data.original.filtered.noNA$othrrcl,

shape.data.original.filtered.noNA$tworacs,

shape.data.original.filtered.noNA$tworcsnc,

shape.data.original.filtered.noNA$thercs)) /
shape.data.original.filtered.noNA$race
```

```{r}
# z score the values to put them all on the same scale
shape.data.processed = data.frame(scale(shape.data.processed))
```

```{r}
# copy new values back to create a spatial object
shape.projected.processed = shape.projected
shape.projected.processed@data = shape.data.processed

#function to generate a raster tiff for each variable.
PrintStatus(1, "Rasterizing the polygons...")
for(c in 1:ncol(shape.data.processed)) {
  c.name = names(shape.data.processed)[c]
  PrintStatus(1, "Calculating raster for variable: ", c.name, "...")

      c.rasterized.fine = rasterize(shape.projected.processed, extent.fine,
field=c.name)
            c.rasterized = aggregate(c.rasterized.fine, fun=mean,
fact=kRasterSplitFactor)
```

````
    output.file = file.path(output.dir, paste0("acs.", c.name, "_", gsub("\\.shp", "", basename(kShapeFile))))
  writeRaster(c.rasterized, output.file, format="GTiff", overwrite=TRUE)
}
```

```{r}
PrintStatus(1, "Complete")
```
````
````

## Appendix C - Machine Learning Functions Code R Script

```
# Functions used to run the model- A de-biasing technique for place-based algorithmic patrol management - Einarsson et al.

################################################################################
#######
# dat.raw is the inputted data that consists of a list with 5 named elements:
#
# raw.x: input training data with a large number of columns.
# The first 4 columns are datetime, cellid, row, and col.  The rest are the
# input features.
#
# raw.y: output training data (the labels) with 6 columns:
# The first 4 columns are datetime, cellid, row, and col. The 5th is count (the
# number of crime events in the cell), and the 6th is presence (whether a crime
# happened or not)
#
# raw.weightsfile: weights file for the training data with 5 columns
# The first 4 columns are datetime, cellid, row, and col. The 5th is the cell weight
#
# test.input: input test data (same columns as raw.x)
#
# test.output: output test data (same columns as raw.y)
################################################################################
#######

# Create the training and validation dataframes
MakeTrainVal <- function(raw.x, raw.y, raw.weightsfile, kLimitObs, limitVal, kDropRowCol) {
  if(nrow(raw.y) - kLimitObs < 100000) {
    # set limit obs to 2/3 of the data
    kLimitObs = round(nrow(raw.y) * 0.666)
  }

   trainIndex = createDataPartition(raw.y$count, p = (kLimitObs / nrow(raw.y)), list=FALSE, times=1)

  raw.y.train = raw.y[trainIndex, ]
  raw.x.train = raw.x[trainIndex, ]
  raw.y.validate = raw.y[-trainIndex,]
  raw.x.validate = raw.x[-trainIndex,]

  raw.weightsfile.train = raw.weightsfile[trainIndex, ]
  raw.weightsfile.validate = raw.weightsfile[-trainIndex, ]
  raw.weightsfile = NULL

  # Limit validation data to increase speed
```

```
  if(!is.na(limitVal)){
    indxLim <- sample(nrow(raw.x.validate), limitVal)
    raw.x.validate <- raw.x.validate[indxLim, ]
    raw.y.validate <- raw.y.validate[indxLim, ]
    raw.weightsfile.validate <- raw.weightsfile.validate[indxLim, ]
  }

  train.output.presence.onezero = raw.y.train$presence
  train.output.count = raw.y.train$count

  validate.output.presence.onezero = raw.y.validate$presence
  validate.output.count = raw.y.validate$count

   # Drop the first 3 columns (datetime, cellid, row, col)?  If no, just drop
datetime and cellid
  if(kDropRowCol == TRUE) {
    train.input = raw.x.train[,-c(1:4)]
    validate.input = raw.x.validate[,-c(1:4)]

  } else {
    train.input = raw.x.train[,-c(1:2)]
    validate.input = raw.x.validate[,-c(1:2)]
  }

   list(train.input = train.input, validate.input = validate.input, raw.y.train
= raw.y.train,
                  raw.y.validate  =  raw.y.validate,  raw.weightsfile.train  =
raw.weightsfile.train,
                        raw.weightsfile.validate   =   raw.weightsfile.validate,
train.output.presence.onezero = train.output.presence.onezero,
                                 train.output.count    =    train.output.count,
validate.output.presence.onezero = validate.output.presence.onezero,
          validate.output.count = validate.output.count, dtcid = raw.x.train[,
1:2])
}


# Functions used to generate exponential weights that weight the recent data
more than the past data.
# Used in the GenerateWeights function
ExpTemporalWeightHalfLife = function(indexes, halflife=28*2, zerobelow=0.01) {
  indexes.max = max(indexes)
  weights = 2^(- (indexes.max - indexes) / halflife)
  return(ifelse(weights >= zerobelow, weights, 0))
}

ExpTemporalWeightEndAt = function(indexes, endat=0.01) {
  indexes.max = max(indexes)
  indexes.min = min(indexes)
```

```
  halflife = - (indexes.max - indexes.min) / log2(endat)
  return(ExpTemporalWeightHalfLife(indexes, halflife, 0))
}

# Generate weights dataframes.  Can optionally weight the recent data more than
the past data ("exponential" weighting)
GenerateWeights <- function(train.input,
                            validate.input,
                            raw.y.train,
                            raw.y.validate,
                            raw.weightsfile.train,
                            raw.weightsfile.validate,
                            kWeightStrategy) {
  if(kWeightStrategy == "constant") {
    train.weights = rep.int(1, nrow(train.input))
    validate.weights = rep.int(1, nrow(validate.input))

  } else if(kWeightStrategy == "exponential") {
    train.weights = ExpTemporalWeightEndAt(raw.y.train$datetime, 0.1)
    validate.weights = ExpTemporalWeightEndAt(raw.y.validate$datetime, 0.1)
  }

  train.weights = train.weights * raw.weightsfile.train$weight
  validate.weights = validate.weights * raw.weightsfile.validate$weight

  list(train.weights = train.weights, validate.weights = validate.weights)
}

# Makes the train and test tables into the types of objects required by
xgboost.
# "r" is the ratio of total size of the training data to the sum of the
presence
# values (total number of rows that had crime events).  It is used for the
# scale_pos_weight parameter in xgboost.
MakeXGBData <- function(dat.tv, wts) {
  Dtrain <- xgb.DMatrix(as.matrix(dat.tv$train.input),
                                                                   label   =
as.numeric(dat.tv$train.output.presence.onezero),
                        weight = wts$train.weights)
  Dtest <- xgb.DMatrix(as.matrix(dat.tv$validate.input),
                                                                   label   =
as.numeric(dat.tv$validate.output.presence.onezero))
            r      <-      length(dat.tv$train.output.presence.onezero)      /
sum(dat.tv$train.output.presence.onezero == 1)
  list(Dtrain = Dtrain, Dtest = Dtest, r = r)
}

####################
# Modeling functions
```

```
####################

# Function to perform xgboost with optional Bayes optimization tuning
xgb_bayesopt <- function(Dtrain,
                         Dtest,
                         r,
                         param.fixed.default,
                         param.opt.default,
                         nrounds,
                         nthread,
                         early_stopping_rounds,
                         verbose = 0,
                         init_points = 10,
                         n_iter = 8,
                         acq = "ucb",
                         kappa = 4,
                         eps = - 0.1) {

  # Function to create scores for inputting into BayesOptimization.  Runs
xgboost with inputted hyperparameters
  xgb_bayes <- function(eta, max_depth, subsample, colsample_bytree) {
    paramlist <- list(eta = eta,
                      max_depth = max_depth,
                      subsample = subsample,
                      colsample_bytree = colsample_bytree)  # Hyperparameters
to be tuned
    paramlist <- c(paramlist, param.fixed.default)   # All hyperparameters
needed for xgboost, default plus those to be optimized
    res <- xgb.train(data = Dtrain,
                     nrounds = nrounds,
                     verbose = verbose,
                     nthread = nthread,
                     early_stopping_rounds = early_stopping_rounds,
                     weight = wts$train.weights,
                     scale_pos_weight = r,
                     objective = "binary:logistic",
                     eval_metric = "auc",
                     maximize = TRUE,
                     param = paramlist,
                     watchlist = list(train = Dtrain, test = Dtest))
     list(Score = res$best_score, Pred = predict(res, Dtest))   # Output best
metric score from early stopping and predictions
  }

  # Find best parameters using BayesianOptimization function using xgb_bayes
function above
  # If error, use default values
  res <- tryCatch(BayesianOptimization(
    xgb_bayes,
```

```
    bounds = list(max_depth = c(1L, 13L),
                  eta = c(.02, .3),
                  subsample = c(0.5, 1),
                  colsample_bytree = c(.2, 1)),
    init_grid_dt = NULL, init_points = init_points, n_iter = n_iter,
    acq = "ucb", kappa = kappa, eps = eps,
    verbose = TRUE
  ), error = function(e) list(Best_par = param.opt.default))

  res$Best_Par
}

# Run xgboost using either default parameters (useTuner = 0) or tuned
parameters from above (useTuner = 1)
# First run cross-validation to pick optimal number of iterations, then run on
full training data
BuildXGBTuned <- function(Dtrain,
                          Dtest,
                          r,
                          param.fixed.default,
                          param.opt.default,
                          wts = wts,
                          nrounds = 500,
                          nfold = 4,
                          nthread = 8,
                          early_stopping_rounds = 10,
                          useTuner = 1) {
  if (useTuner == 1) {
    cat("Tuning XGB model using Bayesian optimization ...")
    res.opt <- xgb_bayesopt(Dtrain,
                            Dtest,
                            r,
                            param.fixed.default,
                            param.opt.default,
                            nrounds = nrounds,
                            nthread = nthread,
                            early_stopping_rounds = early_stopping_rounds,
                            verbose= 2)
  } else res.opt <- param.opt.default
  paramlist <- c(param.fixed.default, res.opt)
  res.cv <- xgb.cv(data = Dtrain,
                   nrounds = nrounds,
                   verbose = 2,
                   nfold = nfold,
                   nthread = nthread,
                   early_stopping_rounds = early_stopping_rounds,
                   weight = wts$train.weights,
                   scale_pos_weight = r,
                   objective = "binary:logistic",
```

```
                    eval_metric = "auc",
                    maximize = TRUE,
                    param = paramlist)
  xgb.train(data = Dtrain,
            nrounds = res.cv$best_iteration,
            verbose = 2,
            nthread = nthread,
            weight = wts$train.weights,
            scale_pos_weight = r,
            objective = "binary:logistic",
            eval_metric = "auc",
            maximize = TRUE,
            watchlist = list(train = Dtrain, test = Dtest),
            param = paramlist)
}

Predictions.xgb <- function(model, test.input) {
  predictions <- predict(model, as.matrix(test.input))
  list(predictions = predictions)
}

# Bootstrap calculations of AUC
BootAUC = function(actual, predictions) {
  ci = ci.auc(actual, predictions, conf.level=0.95)
  return(data.frame(LowCI=ci[1], Metric=ci[2], HighCI=ci[3]))
}

# Run BootAUC.  This function allows other types of metrics as well, if needed
AllMetrics <- function(counts, predictions) {
  bootauc <- BootAUC(as.numeric(counts > 0), predictions)
  return(list(BootAUC = bootauc))
}

# Run all functions in order
runModel <- function(dat.raw, kWeightStrategy, useTuner) {
  dat.tv <- MakeTrainVal(dat.raw$raw.x,
                         dat.raw$raw.y,
                         dat.raw$raw.weightsfile,
                         kLimitObs,
                         limitVal,
                         kDropRowCol)
  wts <- GenerateWeights(dat.tv$train.input,
                         dat.tv$validate.input,
                         dat.tv$raw.y.train,
                         dat.tv$raw.y.validate,
                         dat.tv$raw.weightsfile.train,
                         dat.tv$raw.weightsfile.validate,
                         kWeightStrategy)
  xgbdata <- MakeXGBData(dat.tv, wts)
```

```
  model <- BuildXGBTuned(xgbdata$Dtrain,
                         xgbdata$Dtest,
                         xgbdata$r,
                         param.opt.default,
                         param.fixed.default,
                         wts = wts,
                         nrounds = 500,
                         nfold = 4,
                         nthread = 12,
                         early_stopping_rounds = 10,
                         useTuner = useTuner)
  predictions <- Predictions.xgb(model, dat.raw$test.input[, -(1:2)])
  metrics <- AllMetrics(dat.raw$test.output$count, predictions$prediction)
  gc()  # Garbage collection
  list(auc = metrics$BootAUC, model = model)
}
```

## Appendix D - Run Experiment Iterations Code R Notebook

````
---
title: "Run iterations - A de-biasing technique for place-based algorithmic
patrol management - Einarsson et al."
output: html_notebook
---

Code used to run the model with de-biasing technique and output several tables
The inputs for each run are run_id (job), agency, crimetype, threshold, reps
(number of repetitions), and dat.raw (described in the "Functions" script).

The code is written with the assumption that it will be looped over many
combinations of agencies, crime types, and thresholds. The output tables are
exported to csv and can be appended to existing csv files. Descriptions of the
output tables are given later on.

### Libraries
```{r}
library(DBI)
library(xgboost)
library(dbplyr)
library(ROCR)
library(purrr)
library(data.table)
library(caret)
library(mgcv)
library(boot)
library(pROC)
library(rBayesianOptimization)
library(dplyr)
library(tidyr)
```

**Put in the paths to where the function scripts are located and to where the
resultant csv files will be saved**
```{r}
## codepath <- < Folder for the scripts >
## path <- < Folder for the results >
```

### Source the script functions to run the model
```{r}
source(file.path(codepath, "Functions.R"))
```

### Initial values
```{r}
````
````

````
racevars <- c("acs.blackalone.percentage_ACS2021", "acs.hispanic.percentage_ACS2021", "acs.nonwhite.percentage_ACS2021", "acs.whitealone.percentage_ACS2021")  # Race features in the model

kParallelCores <- 20  # Number of cores for parallelization
corcutoff.static <- 0.5  # Minimum correlation for the static cutoff run
gaincutoff.static <- 0.1 # Minimum gain for the static cutoff run.  Features with correlation > corcutoff and gain < gaincutoff will be removed
maxiter <- 5  # Maximum number of allowed iterations for iteration runs so it doesn't run forever
auc.maxdroppct <- 10  # Percent drop-off in AUC above which we stop the iterations.

# Don't change these
kLimitObs <- 2000000
kDropRowCol <- FALSE
kStochasticHyperParams <- FALSE
kNTrees <- 300
limitVal <- NA # Limit validation data to increase speed

# Default hyperparameters for xgboost.  param.fixed are fixed, param.opt can be optimized by tuning
param.fixed.default <- list(max_delta_step = 2, min_child_weight = 6, gamma = 2, lambda = 4, alpha = 4)
param.opt.default <- list(eta = 0.1, max_depth = 5, subsample = 0.86, colsample_bytree = 0.42)
```

## Which combinations of the model versions to run? These models compete and the best one is chosen
### The options are:
### -   constant vs. exponential weighting of recent dates
### -   use autotuner vs. fixed hyperparameters
```{r}
# modelcombos <- expand_grid(kWeightStrategy = c("constant", "exponential"), useTuner = 0:1)  # This version uses 4 combos
modelcombos <- expand_grid(kWeightStrategy = c("constant"), useTuner = 0)  # This version uses only 1 to increase speed
```

### The function to calculation the threshold score from the gain and correlation
```{r}
calcBiasScore <- function(correlation, Gain) (log(2+(10*Gain*correlation)+(20*Gain*(correlation**3)))-0.69)/2
```

### Function to remove features at each step for a model run. This function will be run for each race variable after the correlations are calculated and
````
````

the winning model version is chosen for a given agency, crime type, and threshold

```{r}
ModelRemoveVars <- function(cors, dat, model.chosen, job, agency, crimetype, threshold) {

  # modelsummary is a list of lists that contains information about each model run:
  # racevar, iteration name, model info (runid, crime type, weight strategy, usetuner, auc),
  # variable importance, and variables removed
  # Append summary of model output at each run to this variable
  racevar <- cors$RaceVar[1]
  print(paste("Race variable =", racevar))
  features <- names(dat$test.input)  # Get list of feature names
  modelsummary <- list()

  # Calculate variable importance at beginning ("Original") and append to model summary
  varimp <- xgb.importance(names(dat$test.input)[-(1:2)], model.chosen$model)[, c("Feature", "Gain")]
  modelsummary <- c(info = list(bind_cols(racevar = racevar,
                                          iteration = "Original",
                                          model.chosen$info[, 1:6])),
                    vars.removed = NA,
                    varimp = list(varimp))

  print("Removing high correlation, low gain features")

  # Static cutoff.  Remove any features that meet the criteria.  Removed features = vars.remove
  vars.remove <- cors %>%
    inner_join(varimp, "Feature") %>%
    filter(abs(correlation) >= corcutoff.static & Gain < gaincutoff.static) %>%
    pull(Feature)

  features <- setdiff(features, vars.remove)    # Remaining features

  # If features are removed, run model (model.reduced) on data with remaining
  # features.  Calculate variable importance and append all to modelsummary.
  # Otherwise, append same results as original
  if (length(vars.remove) > 0) {
    dat$raw.x <- select(dat$raw.x, all_of(features))
    dat$test.input <- select(dat$test.input, !all_of(vars.remove))
    model.reduced <- runModel(dat,
                              kWeightStrategy = model.chosen$info$weightstrategy,
                              model.chosen$info$useTuner)
```

```
                   varimp     =     xgb.importance(names(dat$test.input)[-(1:2)],
model.reduced$model)[, c("Feature", "Gain")]
    modelsummary.reduced <- c(info = list(data.frame(racevar = racevar,
                                                        iteration = "Remove high
cor low gain",
                                                     run_id = job,
                                                     agency = agency,
                                                     crime_model = crimetype,
                                                             weightstrategy =
model.chosen$info$weightstrategy,
                                                                   useTuner =
model.chosen$info$useTuner,
                                                                        auc =
model.reduced$auc$Metric)),
                              vars.removed = list(vars.remove),
                              varimp = list(varimp))
    } else {
    modelsummary.reduced <- c(info = list(bind_cols(racevar = racevar,
                                                        iteration = "Remove high
cor low gain",
                                                    model.chosen$info[, 1:6])),
                              vars.removed = NA,
                              varimp = list(varimp))
    }
  modelsummary <- c(list(modelsummary), list(modelsummary.reduced))

  # Iterate the model, removing features above threshold.  Stop if no features
  # are removed or AUC drops too much.
  for (i in 1:maxiter) {

    print(paste("Removing high bias score features, iteration", i))
    auc.last <- modelsummary.reduced$info$auc  # Starting AUC

       #  Calculate  bias  score  and  remove  features  above  threshold.   Removed
features = vars.remove
    vars.remove <- cors %>%
      inner_join(varimp, "Feature") %>%
      mutate(biasscore = calcBiasScore(correlation, Gain)) %>%
      filter(biasscore > threshold) %>%
      pull(Feature)

    if(length(vars.remove) == 0) break

    print(vars.remove)

    features <- setdiff(features, vars.remove)    # Remaining features

    # Run model (mode.reduced) on data with remaining features.  Calculate
    # variable importance and append all to modelsummary.
```

```
    dat$raw.x <- select(dat$raw.x, all_of(features))
    dat$test.input <- select(dat$test.input, !all_of(vars.remove))
    model.reduced <- runModel(dat,
                                                              kWeightStrategy  =
model.chosen$info$weightstrategy,
                              model.chosen$info$useTuner)
                   varimp    =    xgb.importance(names(dat$test.input)[-(1:2)],
model.reduced$model)[, c("Feature", "Gain")]
    modelsummary.reduced <- c(info = list(data.frame(run_id = job,
                                                     agency = agency,
                                                     crime_model = crimetype,
                                                     racevar = racevar,
                                                       iteration = paste("Remove
high gain", i),
                                                                weightstrategy =
model.chosen$info$weightstrategy,
                                                                      useTuner =
model.chosen$info$useTuner,
                                                                           auc =
model.reduced$auc$Metric)),
                              vars.removed = list(vars.remove),
                              varimp = list(varimp))
    modelsummary <- c(modelsummary, list(modelsummary.reduced))

    gc()

    # If AUC drops more than 10% from start, break
      if  (modelsummary.reduced$info$auc  <  auc.last  *  (1  -  auc.maxdroppct/100))
break
  }

  gc()

  modelsummary
}
```

### Main function to run all the code for each agency/crime type/threshold combination. Inputs (from the runs table) are run_id, agency, crimetype, threshold, and \# repetitions. The function calculates the correlations to all the race variables, runs the model on all requested model types, picks the winner, runs ModelRemoveVars on all race variables, outputs modelresult (same as modelsummary above, a list of lists with all important results). It uses modelresult to output 4 summary tables that are appended to csv files.

### The 4 tables are:
### -   *variablesremoved:* variables removed at all steps
### -   *varimpsummary:* variable importances at all steps

### - *modelresultsummary:* summary of model results at all steps, including AUC
### - *correlations.out:* correlation of all features to each race variable

```{r}
RunModelWithMitigation <- function(dat.raw, job, agency, crimetype, threshold, reps) {

      # Calculate correlations between features and race variables
      print("Calculating correlations")
            correlations <- cor(select(dat.raw$raw.x, !all_of(c(c("datetime", "cellid"), racevars))),
                          select(dat.raw$raw.x, all_of(racevars)))
      correlations <- as.data.frame(as.table(correlations))
      names(correlations) <- c("Feature", "RaceVar", "correlation")

      # Remove race variables from the features used to train the model
      dat.raw$raw.x <- select(dat.raw$raw.x, !all_of(racevars))
      dat.raw$test.input <- select(dat.raw$test.input, !all_of(racevars))

      # Loop over repetitions
      for (i in 1:reps) {
        print(paste("Repetition = ", i))

         # Run the model on all model variations in the modelcombos table, save in
        # list, remove failed runs from list
        print("Picking winning model")
        model4 <- pmap(modelcombos[1, ], function(kWeightStrategy, useTuner) {
          possibly(\(kWeightStrategy, useTuner) {
            modelout <- runModel(dat.raw, kWeightStrategy, useTuner)
            list(weightstrategy = kWeightStrategy,
                 useTuner = useTuner,
                 auc = modelout$auc$Metric,
                 model = modelout$model)
          }, otherwise = NULL)(kWeightStrategy, useTuner)
        }) %>%
          compact()

        # Pick model with highest AUC
        model.chosen <- model4 %>%
          imap_dfr(~ data.frame(run_id = job,
                                agency = agency,
                                crime_model = crimetype,
                                weightstrategy = .x$weightstrategy,
                                useTuner = .x$useTuner,
                                auc = .x$auc, i = .y)) %>%
          arrange(desc(auc)) %>%
          slice_head(n = 1)
```

```
        model.chosen <- c(info = list(model.chosen), model = map(model4, ~.x$model)[model.chosen$i[1]])

        # Run ModelRemoveVars on all race variables using the chosen model
        modelresult <- correlations %>%
          split(.$RaceVar) %>%
          map(~ ModelRemoveVars(.x,
                                dat.raw,
                                model.chosen,
                                job,
                                agency,
                                crimetype,
                                threshold = threshold)) %>%
          reduce(c)

        # Create summary dataframes by pulling out and combining different elements of modelresult
        modelresultsummary <- modelresult %>% map_dfr(~ .x$info)
        variablesremoved <- modelresult %>% map_dfr(~ bind_cols(racevar = .x$info$racevar,
                                                                iteration = .x$info$iteration,
                                                                data.frame(features_removed = .x$vars.removed)))
        varimpsummary <- modelresult %>% map_dfr(~ bind_cols(racevar = .x$info$racevar,
                                                             iteration = .x$info$iteration,
                                                             rank = 1:nrow(.x$varimp),
                                                             .x$varimp))

        # Append dataframes to csv files.  doAppend checks if files exist.  If they
        # do, add row of column names.  Before starting, make sure all output files
        # are erased or moved, or they will be appended to.
        doAppend <- file.exists(file.path(path, "varimpsummary.csv"))

        variablesremoved <- bind_cols(rep = i,
                                      run_id = job,
                                      agency = agency,
                                      crime_model = crimetype,
                                      threshold = threshold,
                                      variablesremoved)
        varimpsummary <- bind_cols(rep = i,
                                   run_id = job,
                                   agency = agency,
                                   crime_model = crimetype,
```

```
                             threshold = threshold,
                             varimpsummary)
  modelresultsummary <- bind_cols(rep = i,
                                  threshold = threshold,
                                  modelresultsummary)

  write.table(variablesremoved,
              file.path(path, "variablesremoved.csv"),
              sep = ",",
              row.names = FALSE,
              col.names = !doAppend,
              quote = FALSE,
              na = "",
              append = doAppend)
  write.table(varimpsummary,
              file.path(path, "varimpsummary.csv"),
              sep = ",",
              row.names = FALSE,
              col.names = !doAppend,
              quote = FALSE,
              na = "",
              append = doAppend)
  write.table(modelresultsummary,
              file.path(path, "modelresultsummary.csv"),
              sep = ",",
              row.names = FALSE,
              col.names = !doAppend,
              quote = FALSE,
              na = "",
              append = doAppend)

}

doAppend.corr <- file.exists(file.path(path, "correlations.csv"))

correlations.out <- bind_cols(run_id = job,
                              agency = agency,
                              crime_model = crimetype,
                              correlations)
write.table(correlations.out,
            file.path(path, "correlations.csv"),
            sep = ",",
            row.names = FALSE,
            col.names = !doAppend.corr,
            quote = FALSE,
            na = "",
            append = doAppend.corr)

modelresult
```

}
    ```

### Run RunModelWithMitigation once (example). Can loop and run through many sets of data by mapping, each time extracting another dat.raw. 4 csv tables will be created at each run and appended to existing tables. modelout will also be outputted which is a list of lists as described above in case more analysis is needed. Each element in model is one run of the model.

```{r}
modelout <- RunModelWithMitigation(dat.raw, job = "123456", agency = "Agency A", crimetype = "Burglary", .05, 10)
```
```

**Appendix E - Data Analysis Code R Notebook**

```
---
title: "Analysis - A de-biasing technique for place-based algorithmic patrol
management - Einarsson et al."
output: html_notebook
---

### Libraries
```{r}
library(scales)
library(ggplot2)
library(dplyr)
library(DBI)
library(purrr)
library(stringr)
library(ggsci)  # Library to provide color palettes for scientific publications
```

### Paths to where the csv files from the model runs are located (need to enter
path)
```{r}
## dirresults <- < Path to results csv files >
```

### Some code for making graphs look better
```{r}
theme_set(theme_bw() + theme(plot.title = element_text(hjust = 0.5)))

# Recode names of iteration steps to make them shorter for graphs
newlabels <- c("Original" = "Original", "Remove high cor low gain" = "Static",
"Remove high gain 1" = "Remove 1",
               "Remove high gain 2" = "Remove 2", "Remove high gain 3" = "Remove
3","Remove high gain 4" = "Remove 4", "Remove high gain 5" = "Remove 5")
```

### Read in csv files in results directory and make into data frames. Make the
race variables more readable and add in repetition numbers for all tables
(except for correlations)
```{r}
filenames <- list.files(dirresults, pattern = ".*\\.csv")

for (fn in filenames) assign(gsub(".csv$", "", fn),
read.csv(file.path(dirresults, fn)))

# Make the race variables more readable and add in repetition numbers for all
tables (no rep numbers for correlations)
correlations <- correlations %>%
  mutate(RaceVar = str_to_sentence(str_split_i(RaceVar, "\\.", 2)))
```
```

````
variablesremoved <- variablesremoved %>%
  mutate(racevar = str_to_sentence(str_split_i(racevar, "\\.", 2))) %>%
   mutate(rep = cumsum(iteration == "Original"), .by = c(run_id, crime_model,
racevar, threshold))

modelresultsummary <- modelresultsummary %>%
  mutate(racevar = str_to_sentence(str_split_i(racevar, "\\.", 2))) %>%
   mutate(rep = cumsum(iteration == "Original"), .by = c(run_id, crime_model,
racevar, threshold))

varimpsummary <- varimpsummary %>%
  mutate(racevar = str_to_sentence(str_split_i(racevar, "\\.", 2))) %>%
     mutate(rep = rep(1:length(rle(rep)$values), rle(rep)$lengths), .by =
c(run_id, crime_model, threshold))
```

### Variation of AUC of original run
```{r}
modelresultsummary %>%
  filter(iteration == "Original") %>%
  group_by(run_id) %>%
   mutate(run = factor(cur_group_id())) %>%  # So the runs #'s are factors, not
integers
   ggplot(aes(x = auc, y = forcats::fct_rev(factor(agency)))) +    # Reverse the
order of the agency names for plotting purposes
    geom_boxplot(fill = "blue") +
    facet_wrap(~ crime_model) +
    labs(x = "AUC", y = NULL)
```

### How many runs had no features removed at all? Which ones? Just for
information
```{r}
modelresultsummary %>%
  group_by(run_id, crime_model, threshold) %>%
  mutate(removed = grepl("high gain", iteration)) %>%
   summarize(anyremoved = any(removed)) %>%  # Were features removed in any of
the repetitions?
  ungroup() %>%
  count(anyremoved)
```

### Plot of number of steps where features were removed for each variable
(racevar, threshold, crime, agency)
```{r}
niterationssummary <- modelresultsummary %>%
  mutate(removed = grepl("high gain", iteration)) %>%
````

```
   summarize(sumremoved = factor(sum(removed), levels = as.character(0:5)), .by
= c(run_id,

crime_model,

agency,

threshold,

racevar,

rep))

roundcolors <- setNames(pal_npg()(n_distinct(niterationssummary$sumremoved)),
                        unique(niterationssummary$sumremoved))  # Nice colors

# Function to make plots of the number of features removed for any of the
variables, where var = (racevar, threshold, crime, agency)
plotNiterDist <- function(var, ylabel) {
  ggplot(niterationssummary, aes(y = factor({{var}}), fill = sumremoved)) +
     geom_bar(position = position_dodge2(preserve = "single", reverse = TRUE),
width = 0.7) +
    scale_fill_manual(name = "# Steps", values = roundcolors) +
    labs(x = "Number of Feature Removal Steps", y = ylabel)
}

plotNiterDist(racevar, "Race Variable")
plotNiterDist(threshold, "Threshold Value")
plotNiterDist(crime_model, "Crime")
plotNiterDist(agency, "Agency")
```

### AUC values at each step, by crime, with a separate plot for each agency. Show AUC at each repetition plus the mean over all repetitions

```{r}
auccurves <- modelresultsummary %>%
  filter(racevar %in% c("Blackalone", "Hispanic") & threshold == 0.05) %>%
  group_by(agency, crime_model, racevar) %>%
  filter(n_distinct(iteration) > 2) %>%
  add_count(agency, iteration, racevar, crime_model, name = "onerep") %>%
  mutate(onerep = onerep == 1)

auccurves %>%
    group_by(agency) %>%
    group_map(
         ~ ggplot(.x, aes(x = iteration, y = auc, color = racevar, group =
interaction(crime_model, racevar))) +
        geom_line(stat = "summary", fun = "mean", linewidth = 1) +
        geom_point(aes(alpha = onerep), size = 2, show.legend = FALSE) +
```

````
          geom_errorbar(stat = "summary", fun.data = "mean_se", fun.args =
list(mult = 1.96), width = 0.6) +
        scale_x_discrete(labels = newlabels) +
        scale_alpha_manual(values = c(0, 1)) +
        scale_color_npg(name = NULL) +
        facet_wrap(~ crime_model) +
        labs(title = .y$agency[1], x = NULL, y = "AUC") +
        theme_bw() +
        theme(plot.title = element_text(hjust = .5),
              axis.text.x = element_text(angle = 45, hjust = 1, size = 7),
              legend.position = "top")
    )
```

## Change of gain values/ranks over iterations. Show the history of all reps
### Keep only the top 12 features in each agency/racevar/threshold/crime group.
Don't include static cutoff in count
### Require at least one rep with 3 steps for each feature
```{r}
varimpsummary2 <- variablesremoved %>%
  filter(grepl("high gain", iteration)) %>%
     count(run_id, agency, crime_model, racevar, threshold, Feature =
features_removed, name = "nremoved") %>%
   slice_max(nremoved, n = 12, by = c(run_id, agency, racevar, threshold)) %>%
# Top 12 by how many times they have a feature removed
   inner_join(varimpsummary, by = join_by(run_id, agency, crime_model, racevar,
threshold, Feature)) %>%   # Inner join back to original dataframe
   mutate(Feature = gsub("_ACS2021$", "", Feature)) %>%   # Lop off irritating
"_ACS2012" suffix
   filter(n() >= 2, .by = c(run_id, agency, crime_model, racevar, threshold,
iteration, Feature)) %>%   # Require at least 2 repetitions
    filter(n_distinct(iteration) >= 3, .by = c(run_id, agency, crime_model,
racevar, threshold, Feature))   # Require at least 3 feature removal steps in
some rep

# Unique colors for each crime type
crimecolors <- setNames(pal_npg()(n_distinct(varimpsummary2$crime_model)),
unique(varimpsummary2$crime_model))

# Each crime, all reps on the same plot
varimpsummary2 %>%
  group_by(agency, threshold, racevar, crime_model) %>%
  group_map(
    ~ ggplot(.x, aes(x = iteration, y = Gain, group = rep)) +
      geom_line(color = unname(crimecolors[1]), linewidth = 0.7) +
      facet_wrap(~ Feature, scales = "free_y") +
      scale_x_discrete(labels = newlabels) +
      labs(title = paste0(.y$agency[1],
                          ", ",
````

````
                          .y$racevar[1],
                          ",  ",
                          .y$threshold[1],
                          ",  ",
                          .y$crime_model),
           x = NULL) +
      theme(axis.text.x = element_text(angle = 45, hjust = 1))
  )
```

### Summary of results in a table (number of reps, standard deviation of
auc's). For original results only. Use shortened names of iteration.
```{r}
(summarytable.mean <- modelresultsummary %>%
  filter(iteration == "Original") %>%
  group_by(agency, crime_model, threshold, racevar) %>%
  summarize(nreps = n()) %>%
  inner_join({
    modelresultsummary %>%
      mutate(iteration = factor(iteration),
                                   iteration   =   forcats::fct_recode(iteration,
!!!setNames(names(newlabels),

newlabels))) %>%
      group_by(agency, crime_model, threshold, racevar, iteration) %>%
      summarize(AUCmean = mean(auc)) %>%
       pivot_wider(names_from = iteration, values_from = AUCmean)   # One column
per iteration step
  }, by = join_by(agency, crime_model, threshold, racevar)))
```

### Tables for each agency with multiple statistics for AUC
```{r}
(summarytable.all <- modelresultsummary %>%
  group_by(agency) %>%
  group_map(
    ~ {
      .x %>%
        filter(iteration == "Original") %>%
        group_by(crime_model, threshold, racevar) %>%
        summarize(nreps = n()) %>%
        inner_join({
          .x %>%
            mutate(iteration = factor(iteration),
                                   iteration  =  forcats::fct_recode(iteration,
!!!setNames(names(newlabels),
                                                           newlabels))) %>%
            group_by(crime_model, threshold, racevar, iteration) %>%
            summarize(AUCmean = mean(auc),
````

```
                    AUCmedian = median(auc),
                    AUCmin = min(auc),
                    AUCmax = max(auc),
                    AUCsd = sd(auc),
                    AUCse = sd(auc)/sqrt(length(auc)))
        }, by = join_by(crime_model, threshold, racevar)) %>%
        rename(crime_model = crime_model,
               Threshold = threshold,
               `Race Variable` = racevar,
               Iteration = iteration,
               `# Repetitions` = nreps) %>%
        mutate(Agency = .y$agency[1]) %>%
        relocate(Agency)
    })
)
```

**Supplementary Figures:**
**Agency A gain figures**

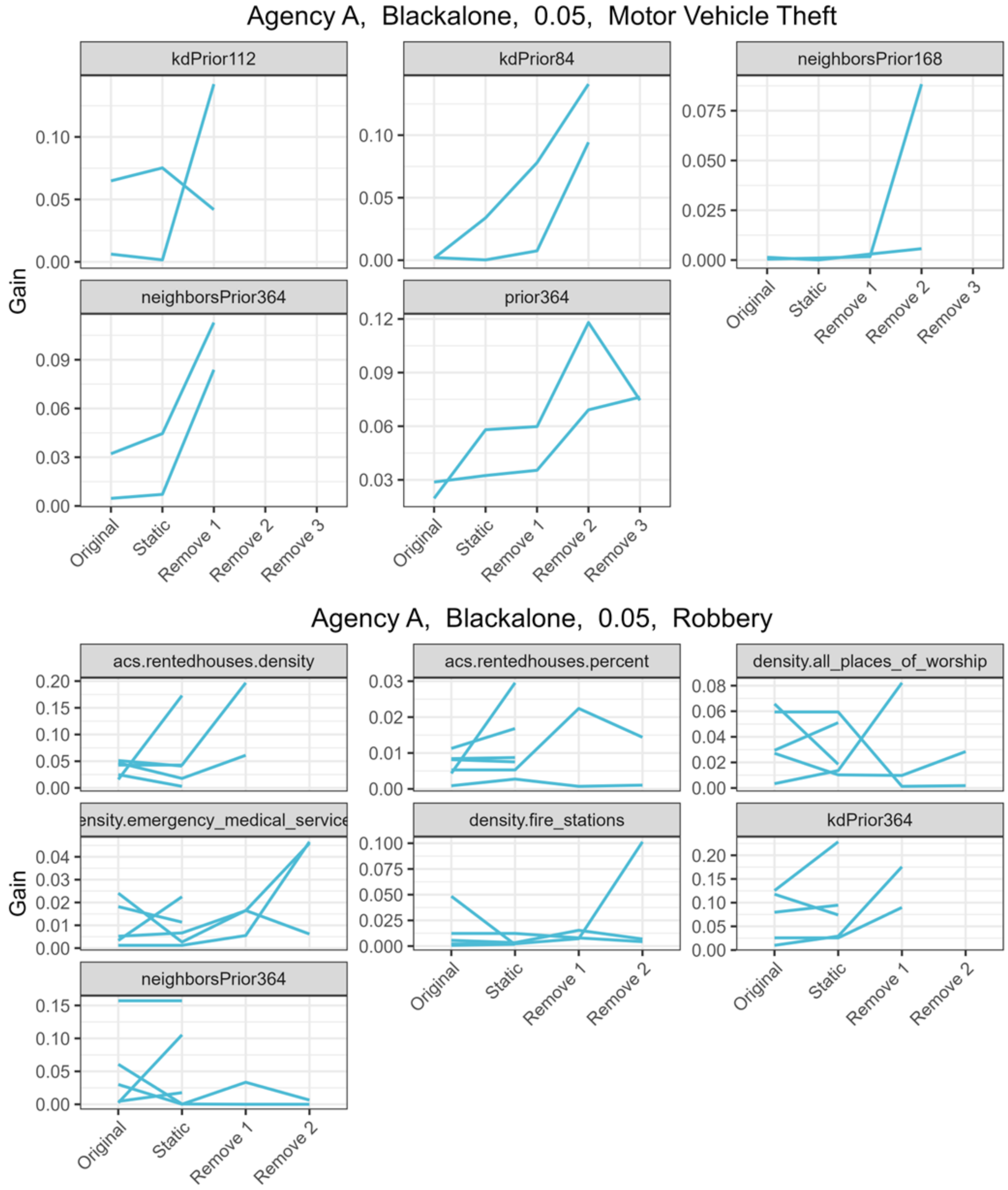

**Supplementary Figure 1.** Difference in gain for features until elimination for the black alone harmful bias variable. Each line is a model run.

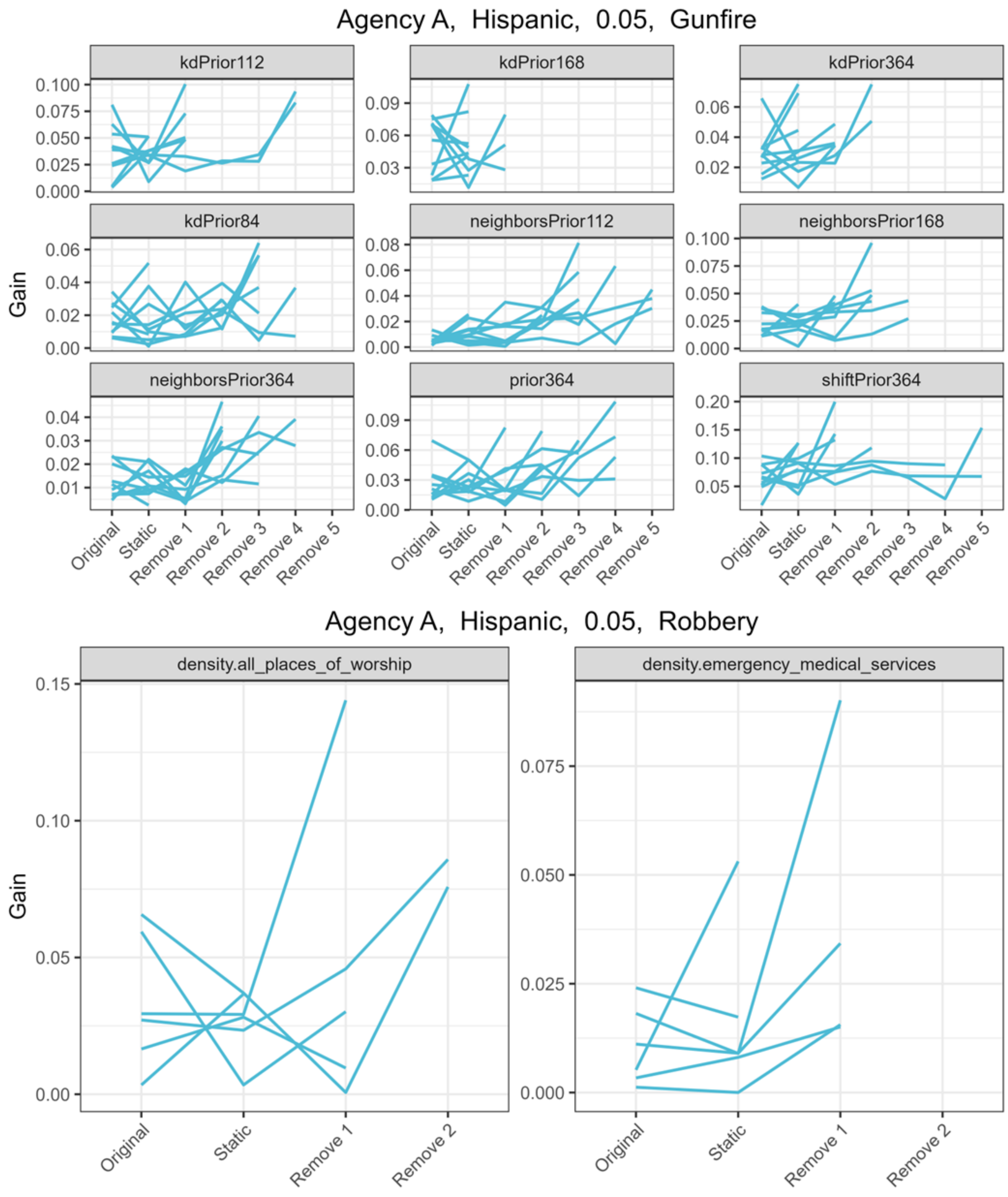

**Supplementary Figure 2.** Difference in gain for features until elimination for the hispanic harmful bias variable. Each line is a model run.

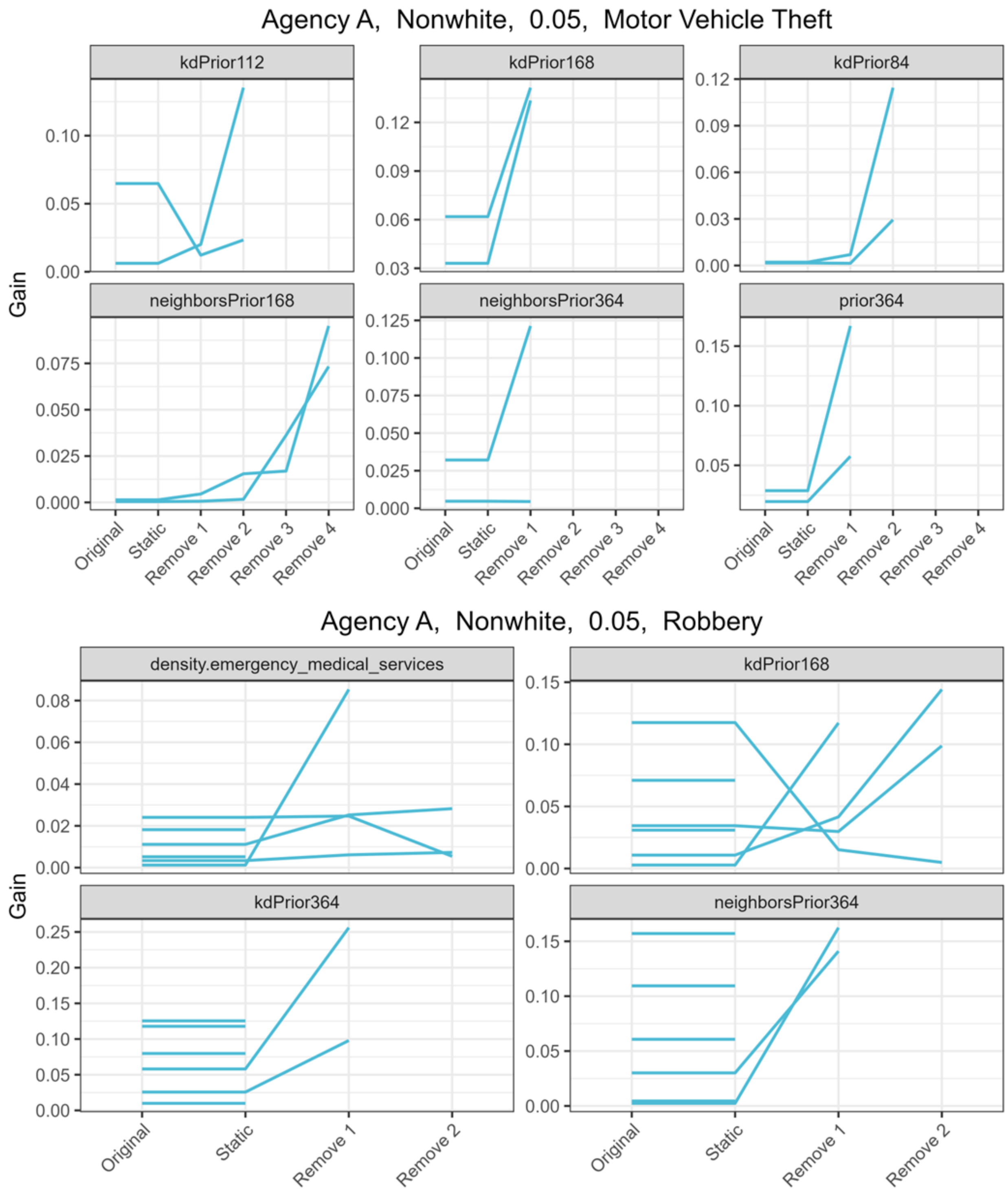

**Supplementary Figure 3.** Difference in gain for features until elimination for the non-white harmful bias variable. Each line is a model run.

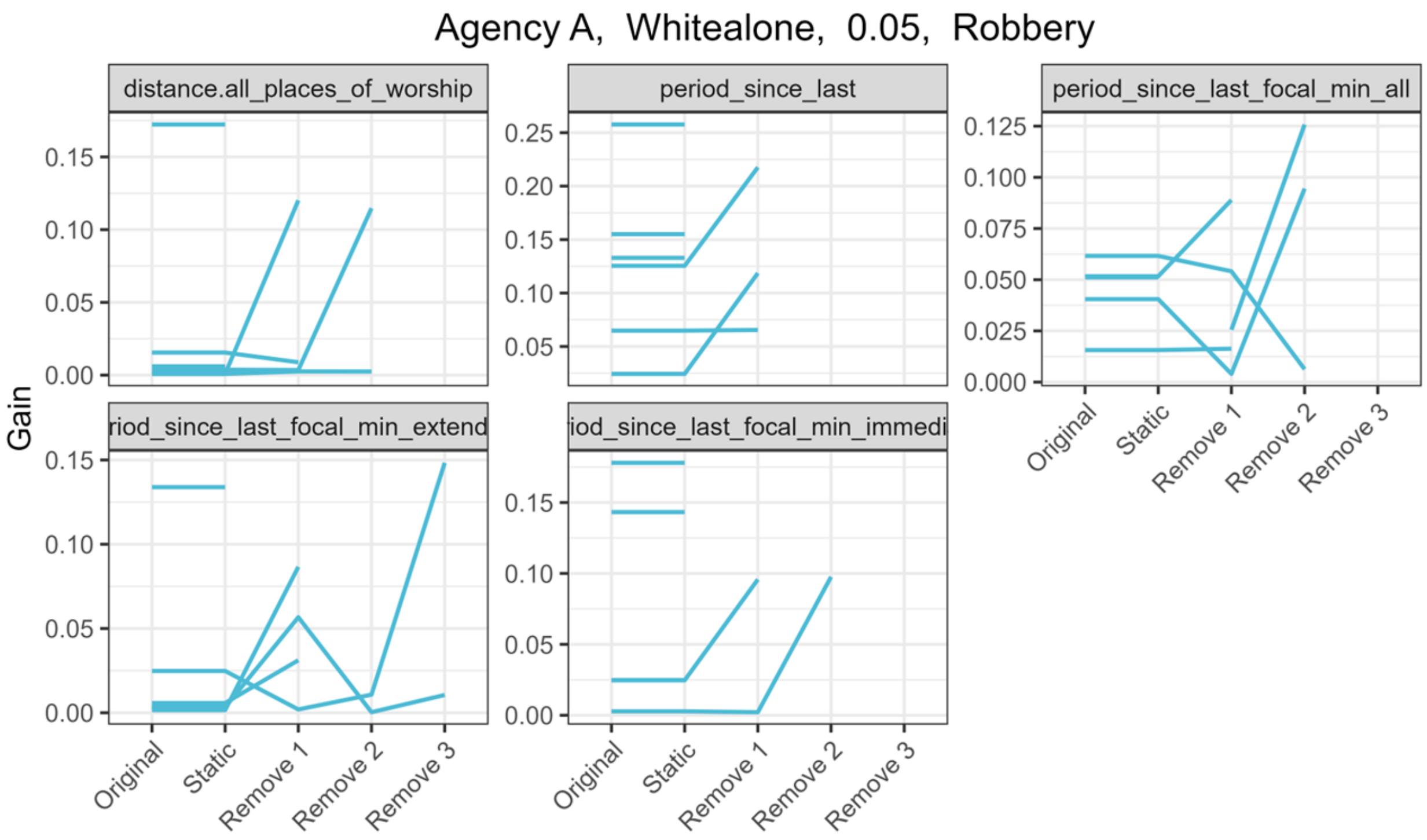

**Supplementary Figure 4.** Difference in gain for features until elimination for white alone harmful bias variable. Each line is a model run.

**Agency B gain figures:**

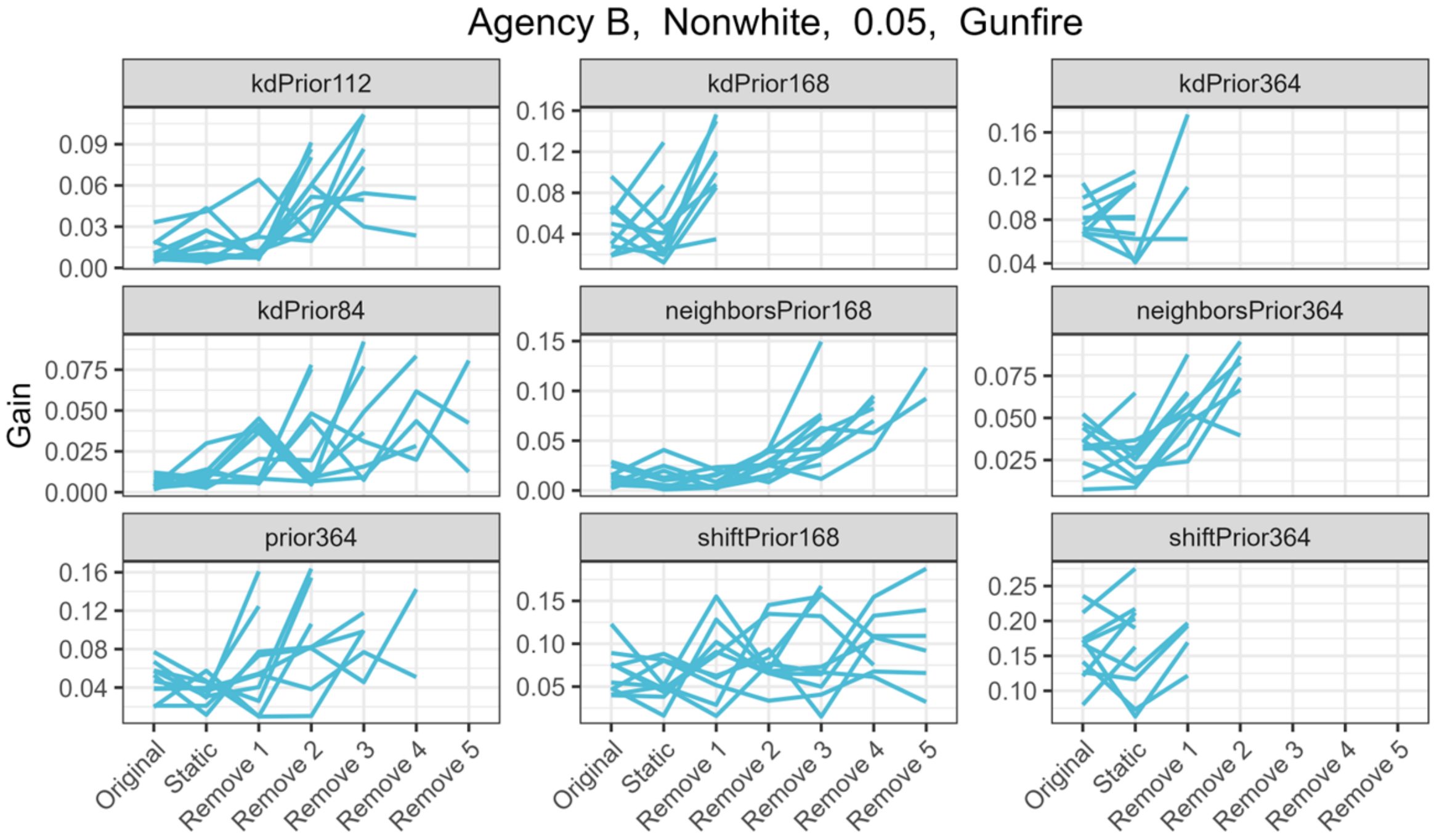

**Supplementary Figure 5.** Difference in gain for features until elimination for the non-white harmful bias variable. Each line is a model run.

**Agency C gain figures:**

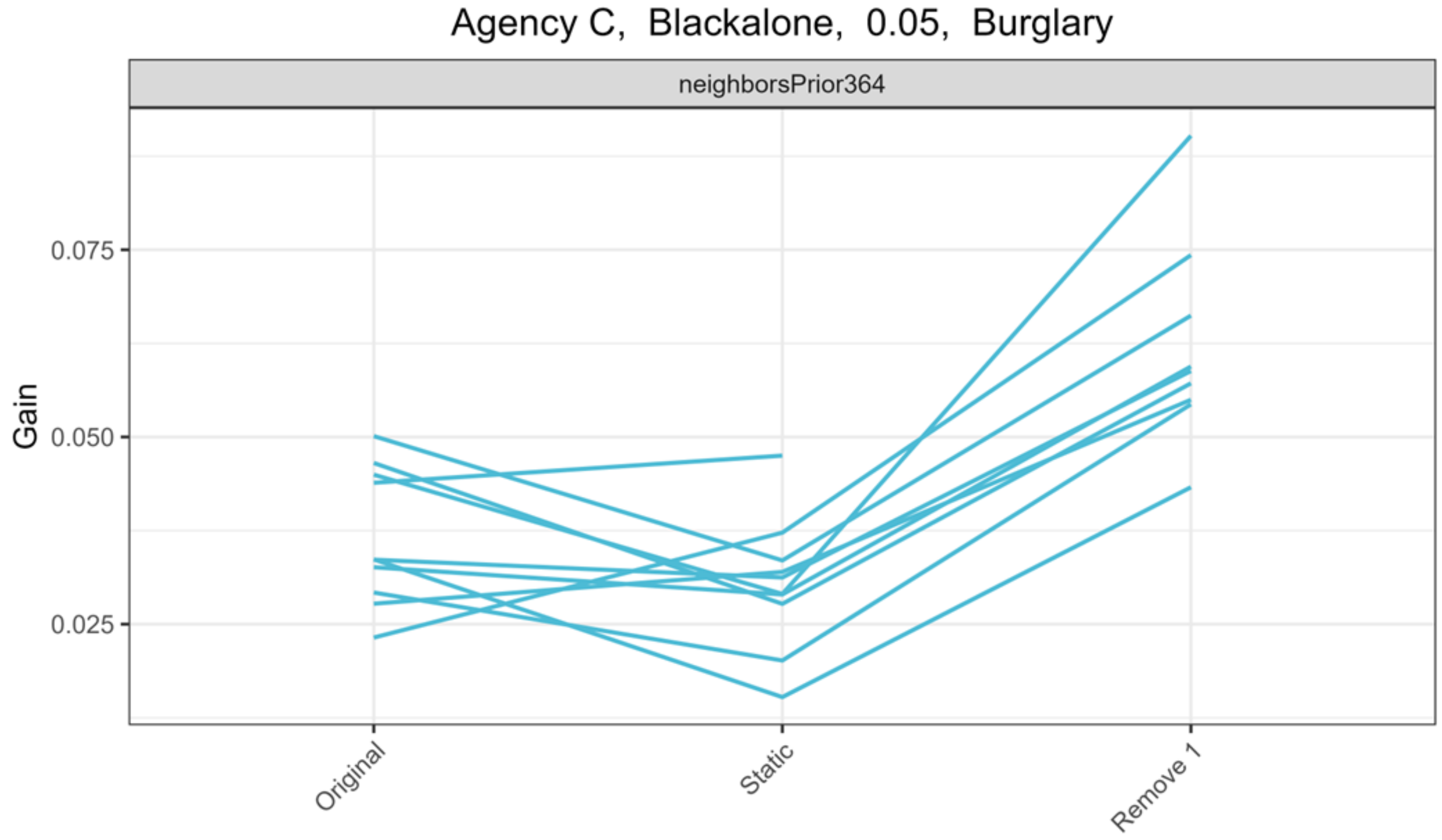

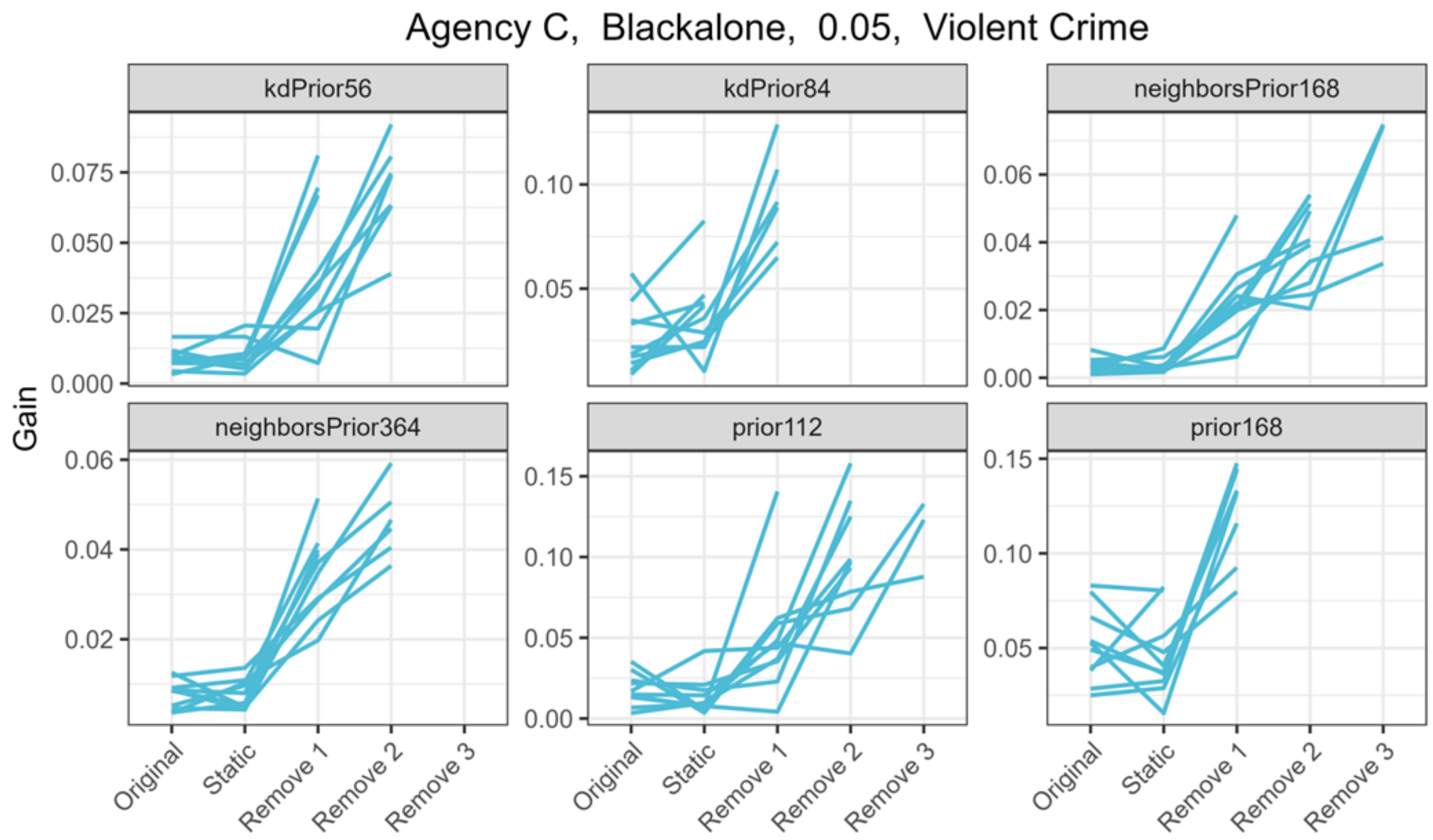

**Supplementary Figure 6.** Difference in gain for features until elimination for the black alone harmful bias variable. Each line is a model run.

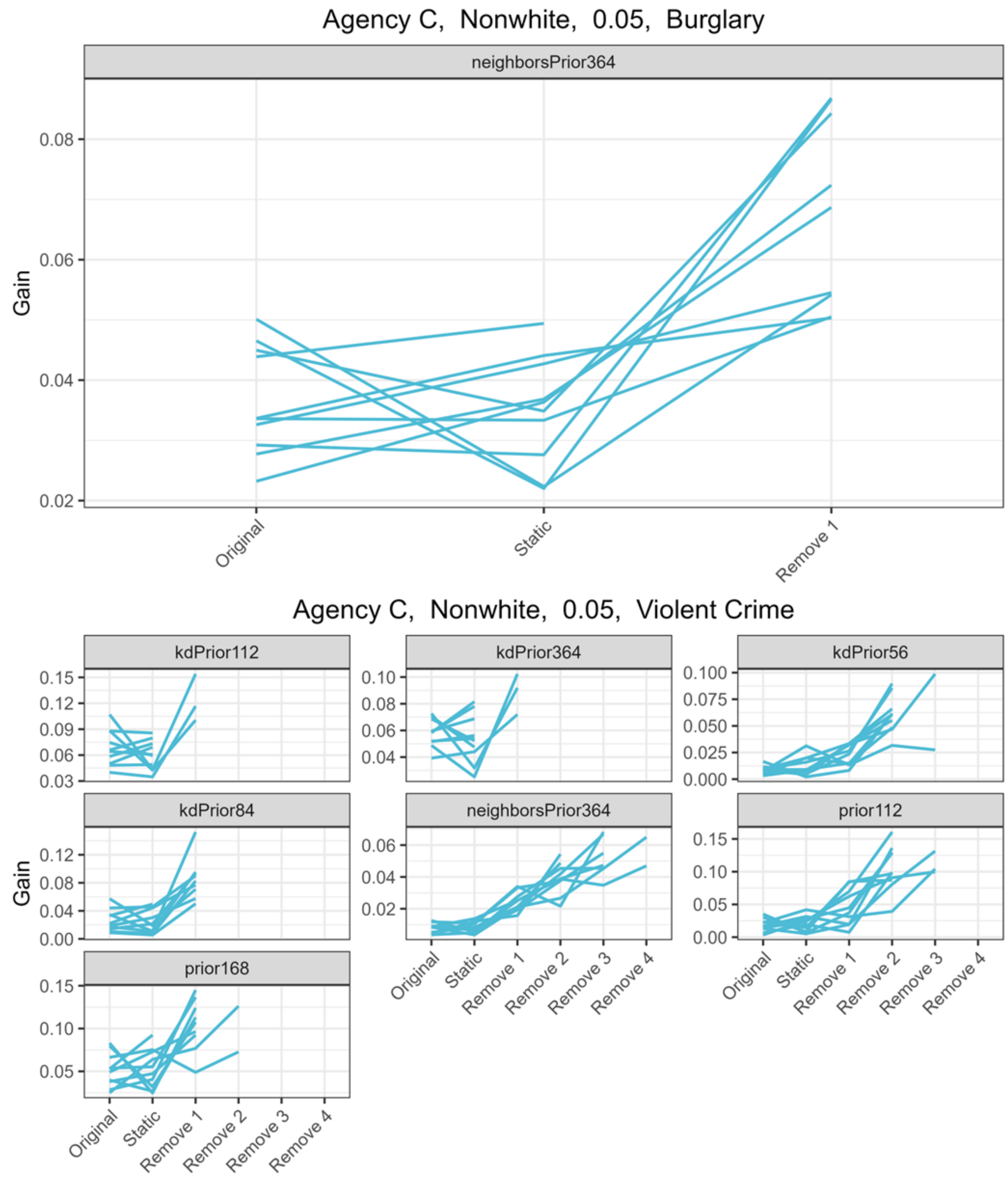

**Supplementary Figure 7.** Difference in gain for features until elimination for the non-white harmful bias variable. Each line is a model run.

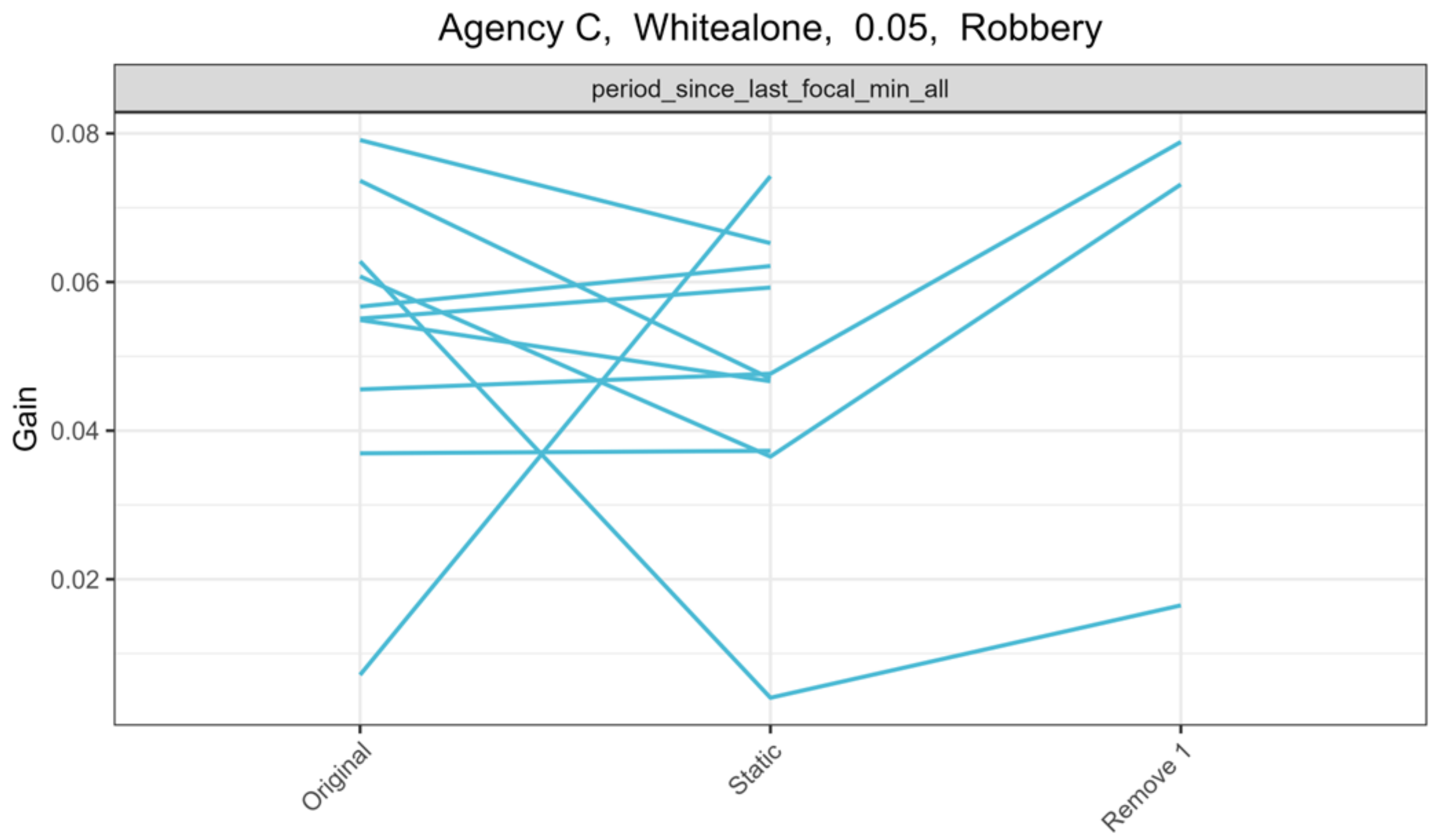

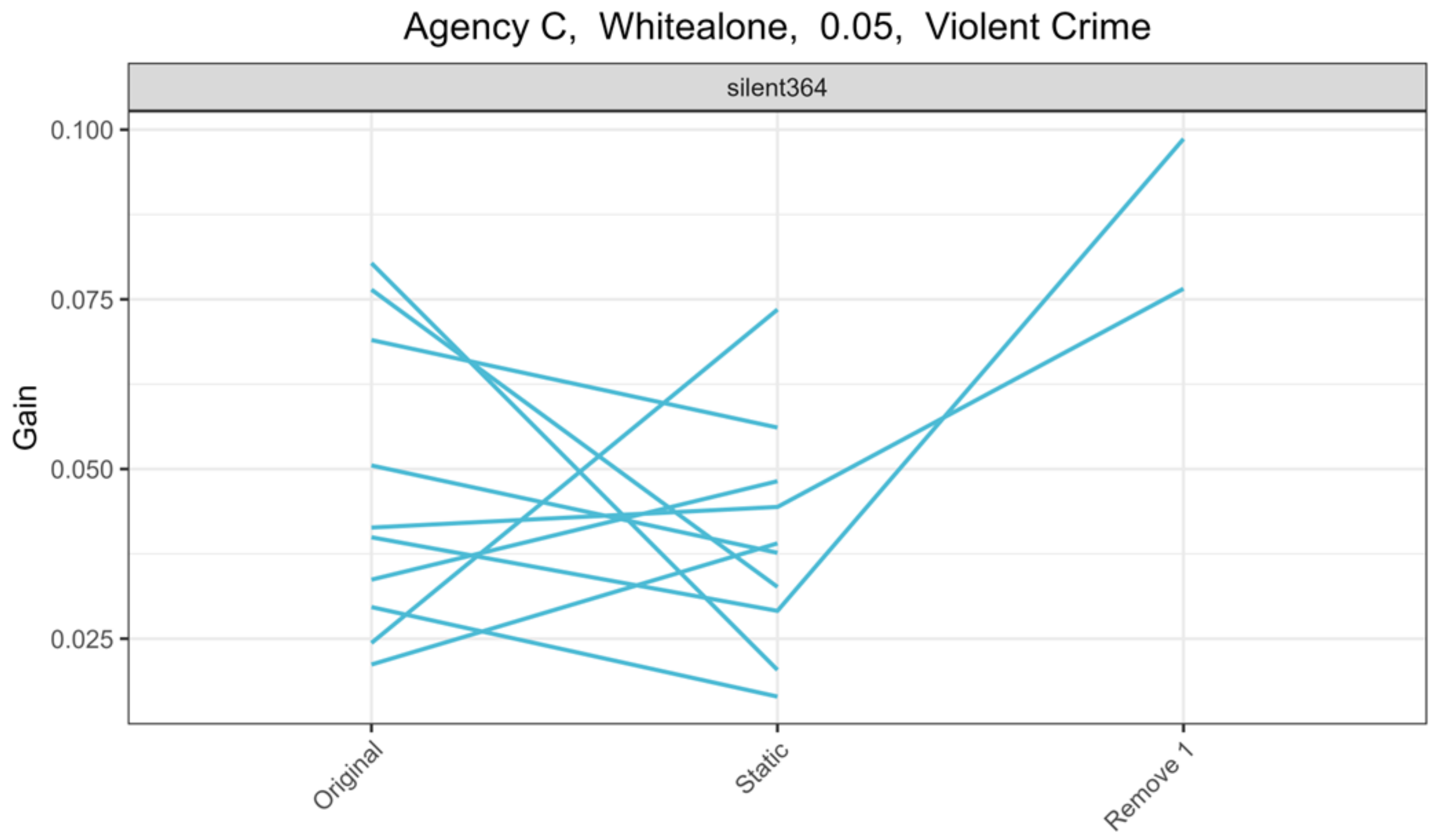

**Supplementary Figure 8.** Difference in gain for features until elimination for the white alone harmful bias variable. Each line is a model run.

## Agency D gain figures:

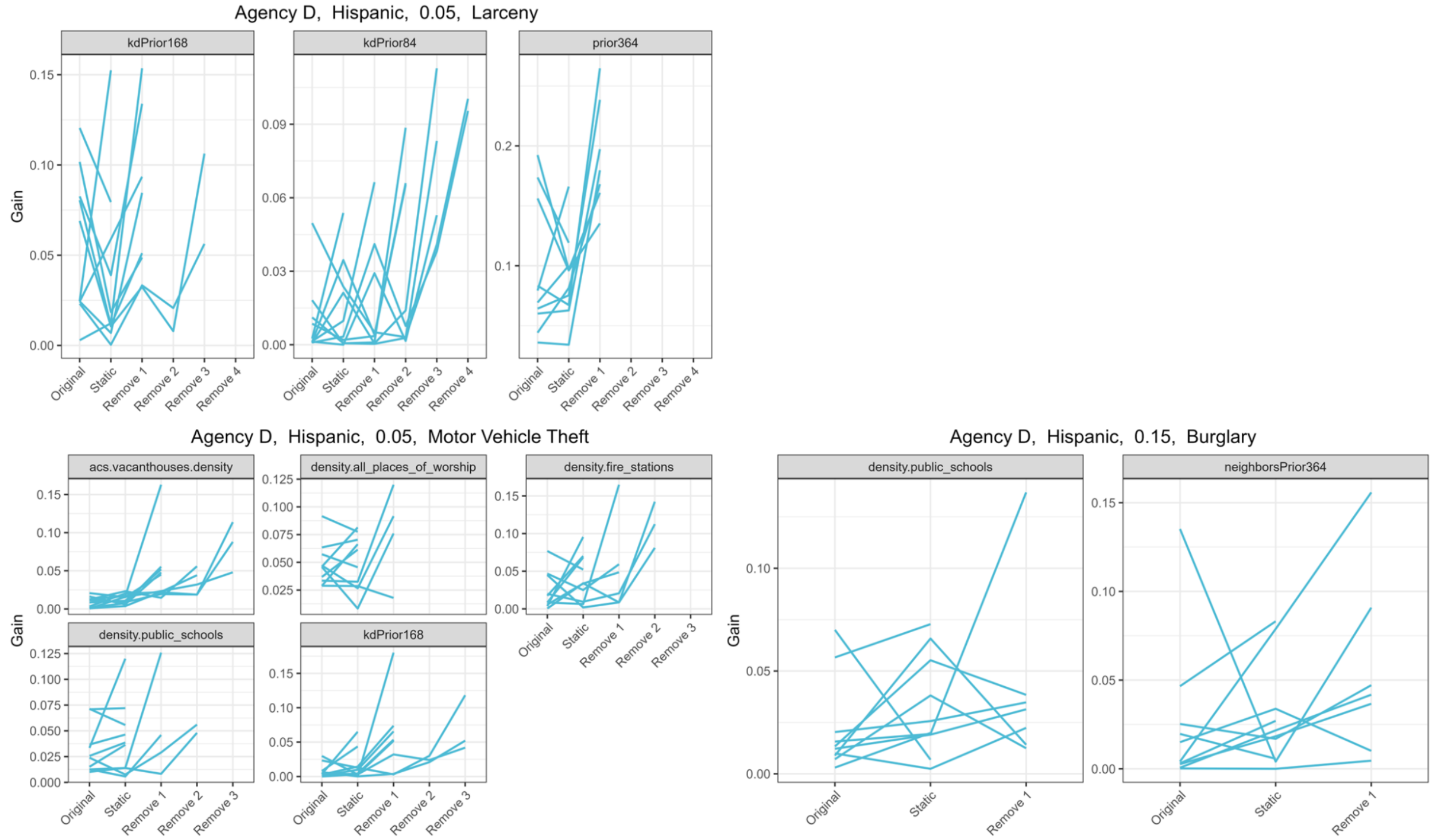

**Supplementary Figure 9.** Difference in gain for features until elimination for the hispanic harmful bias variable. Each line is a model run.

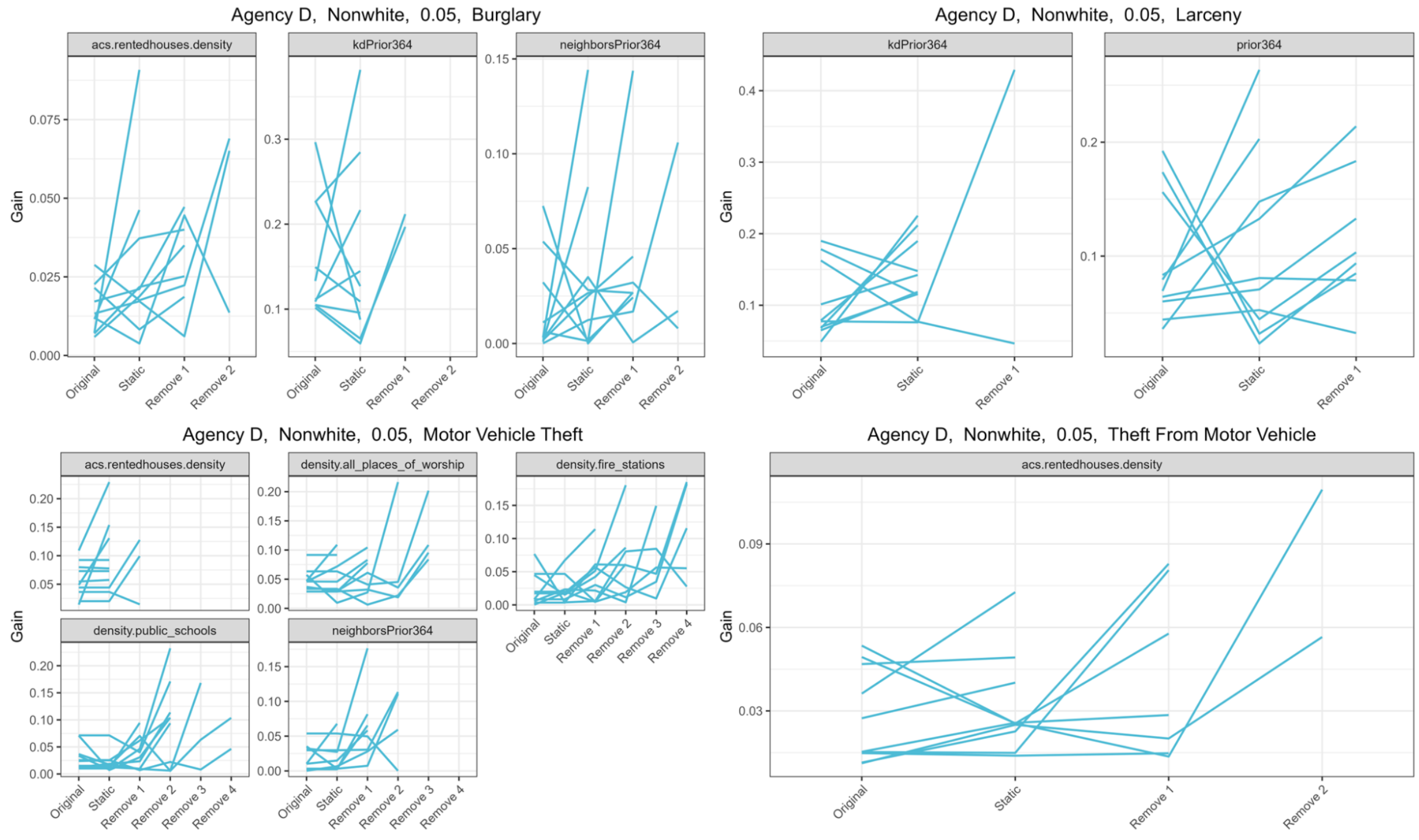

**Supplementary Figure 10.** Difference in gain for features until elimination for the non-white harmful bias variable. Each line is a model run.

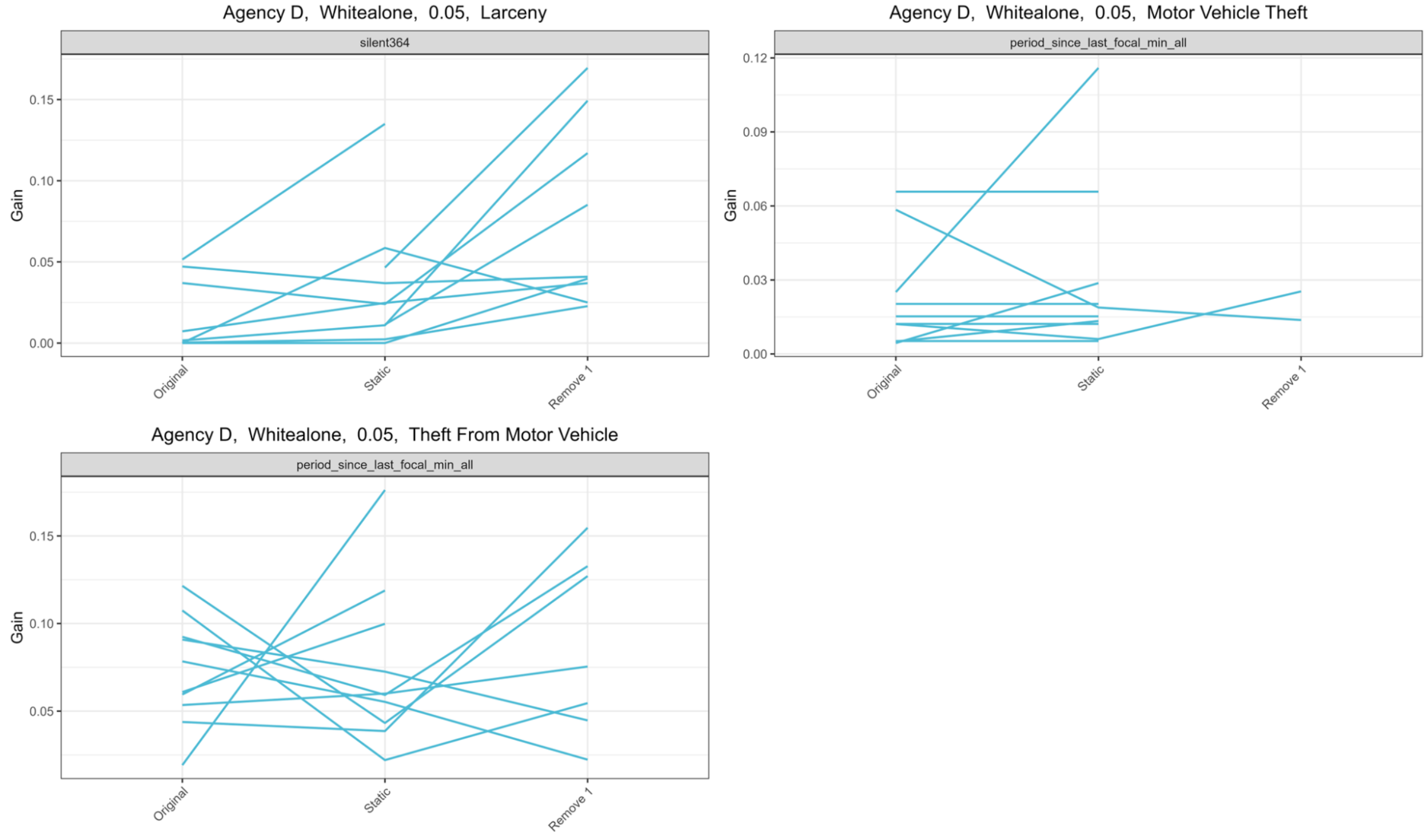

**Supplementary Figure 11.** Difference in gain for features until elimination for the black alone harmful bias variable. Each line is a model run.

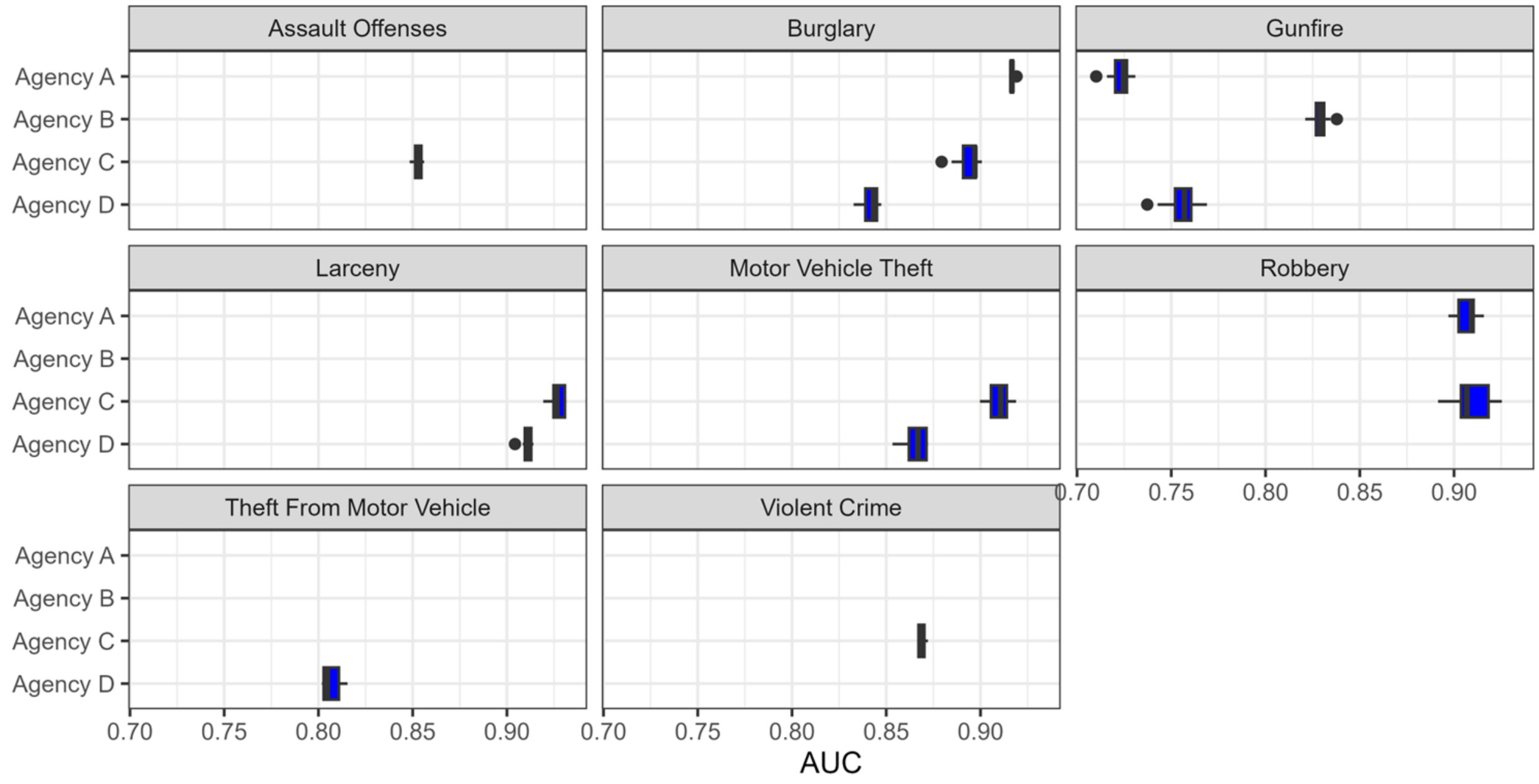

**Supplementary Figure 12.** AUC variance across multiple model runs on the same dataset.

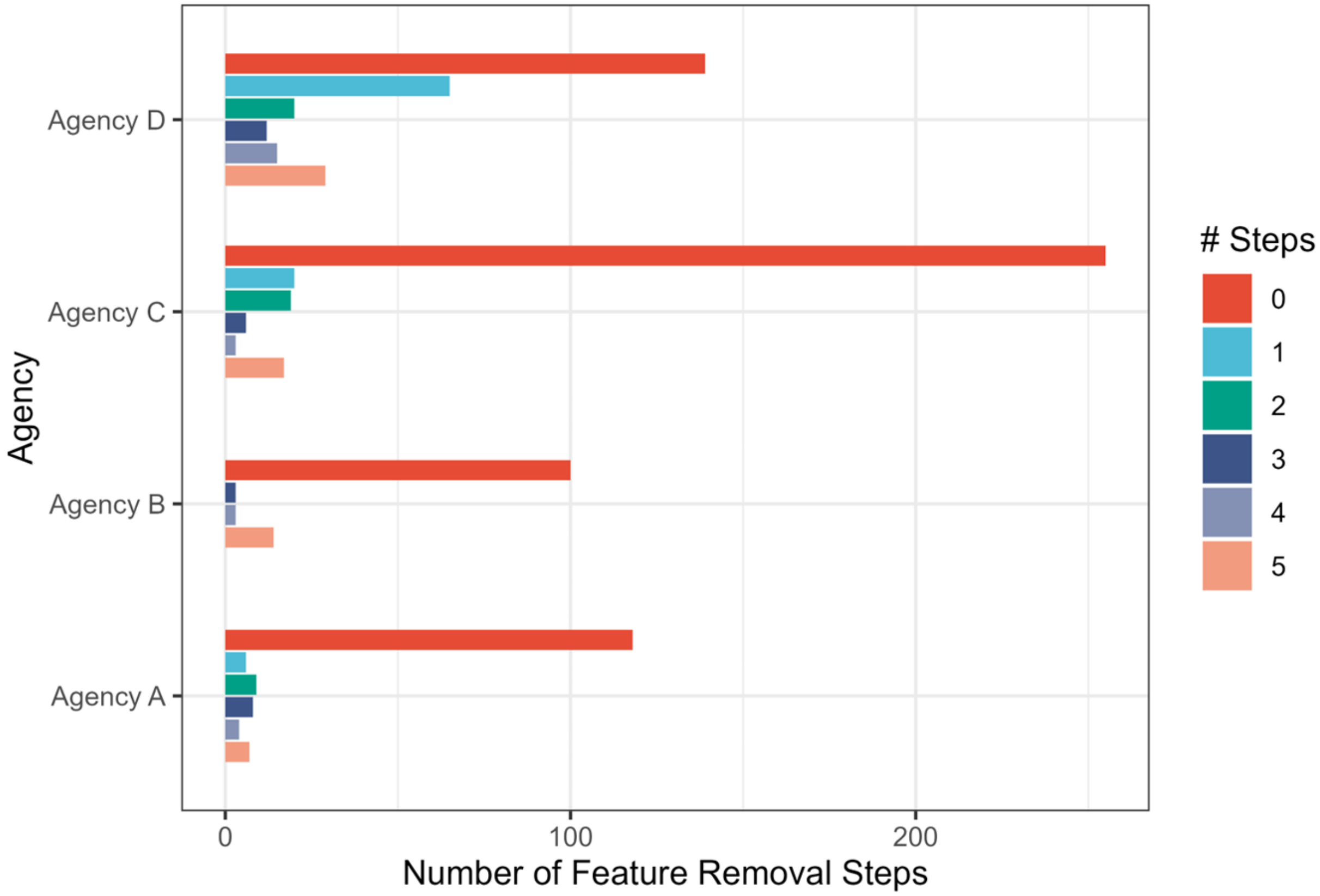

**Supplementary Figure 13.** Count of feature removal iterations for each Agency.

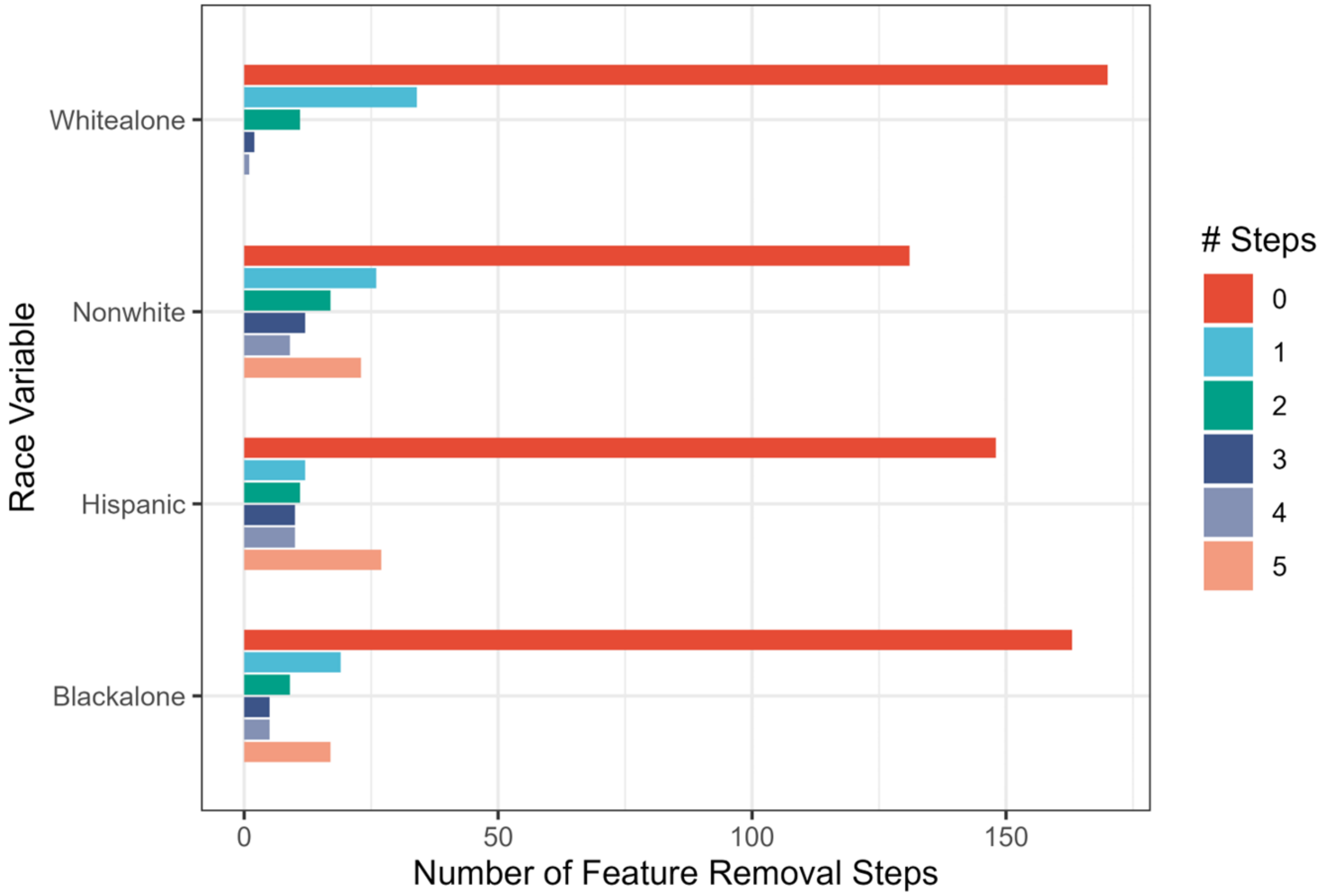

**Supplementary Figure 14.** Count of feature removal iterations for each harmful bias variable.

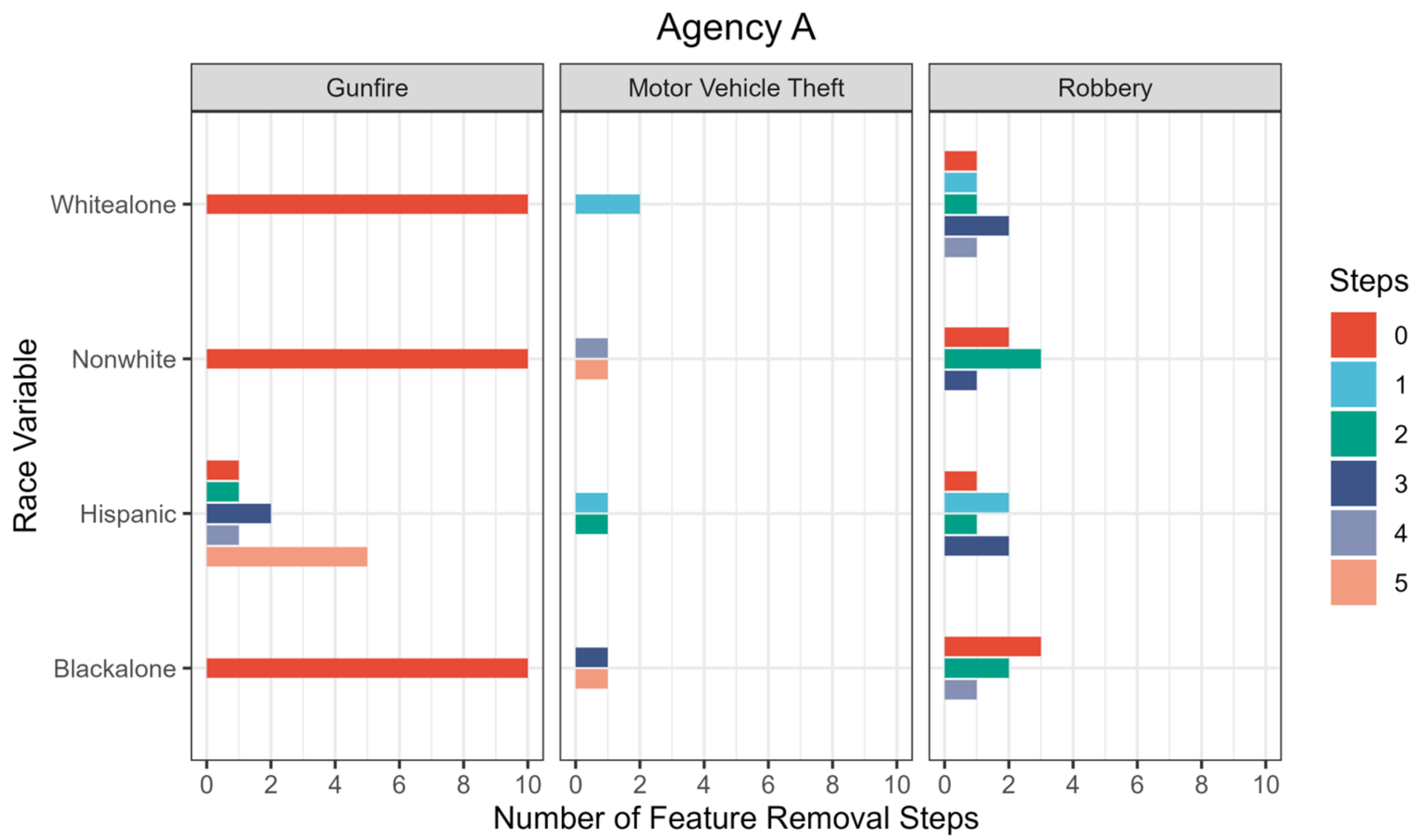

**Supplementary Figure 15.** Count of feature removal iterations for each harmful bias variable by Crime Model in Agency A.

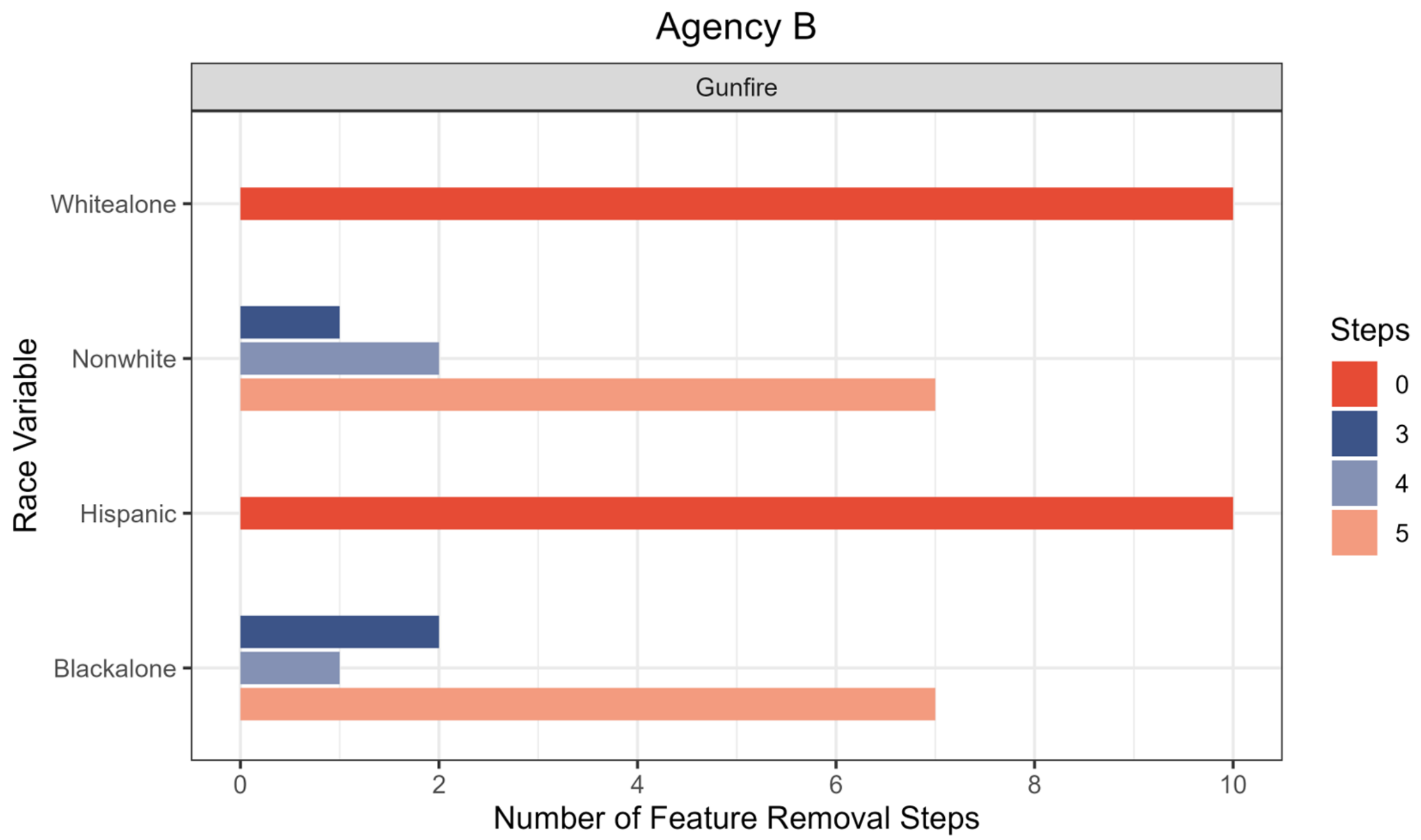

**Supplementary Figure 16.** Count of feature removal iterations for each harmful bias variable by Crime Model in Agency B.

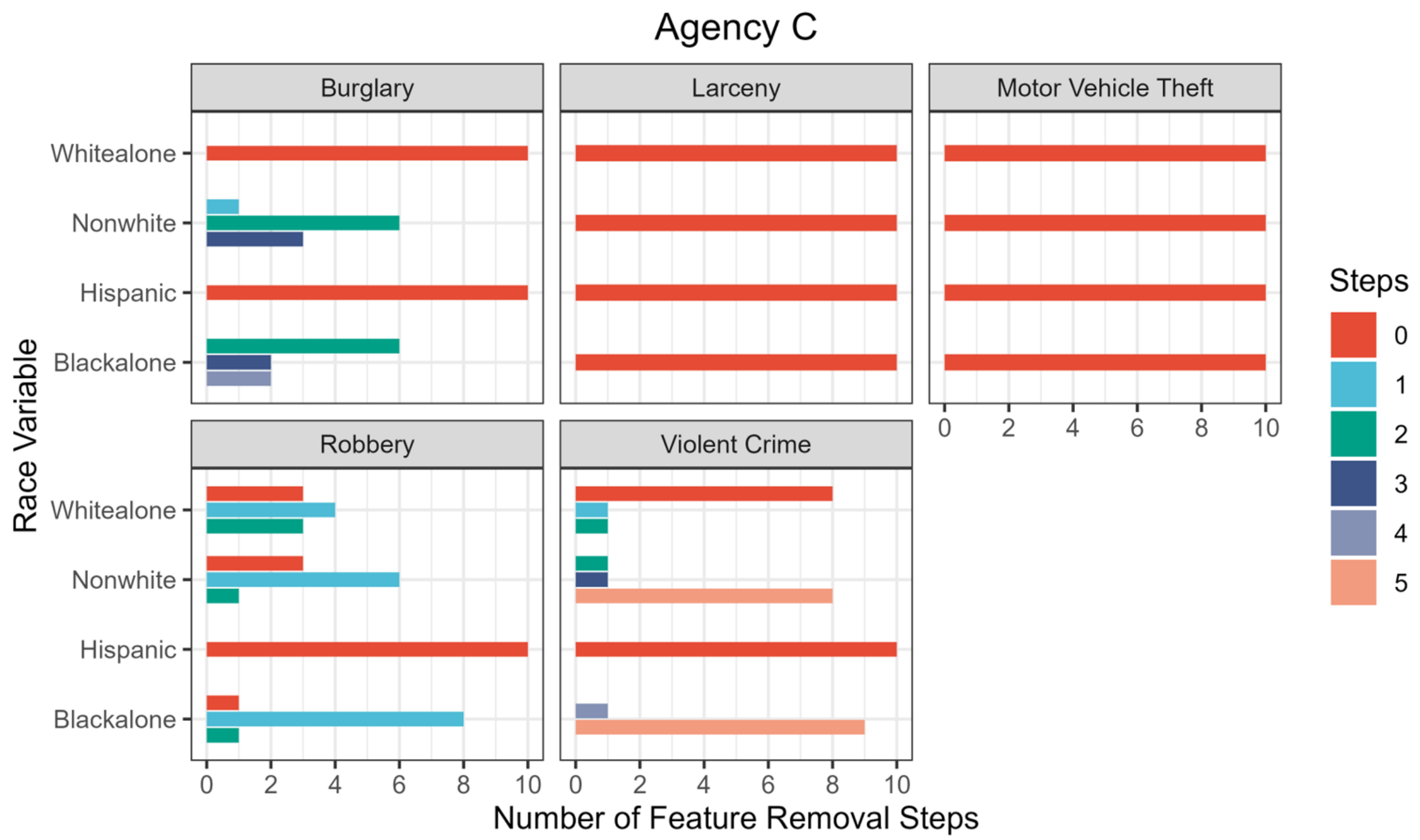

**Supplementary Figure 17.** Count of feature removal iterations for each harmful bias variable by Crime Model in Agency C.

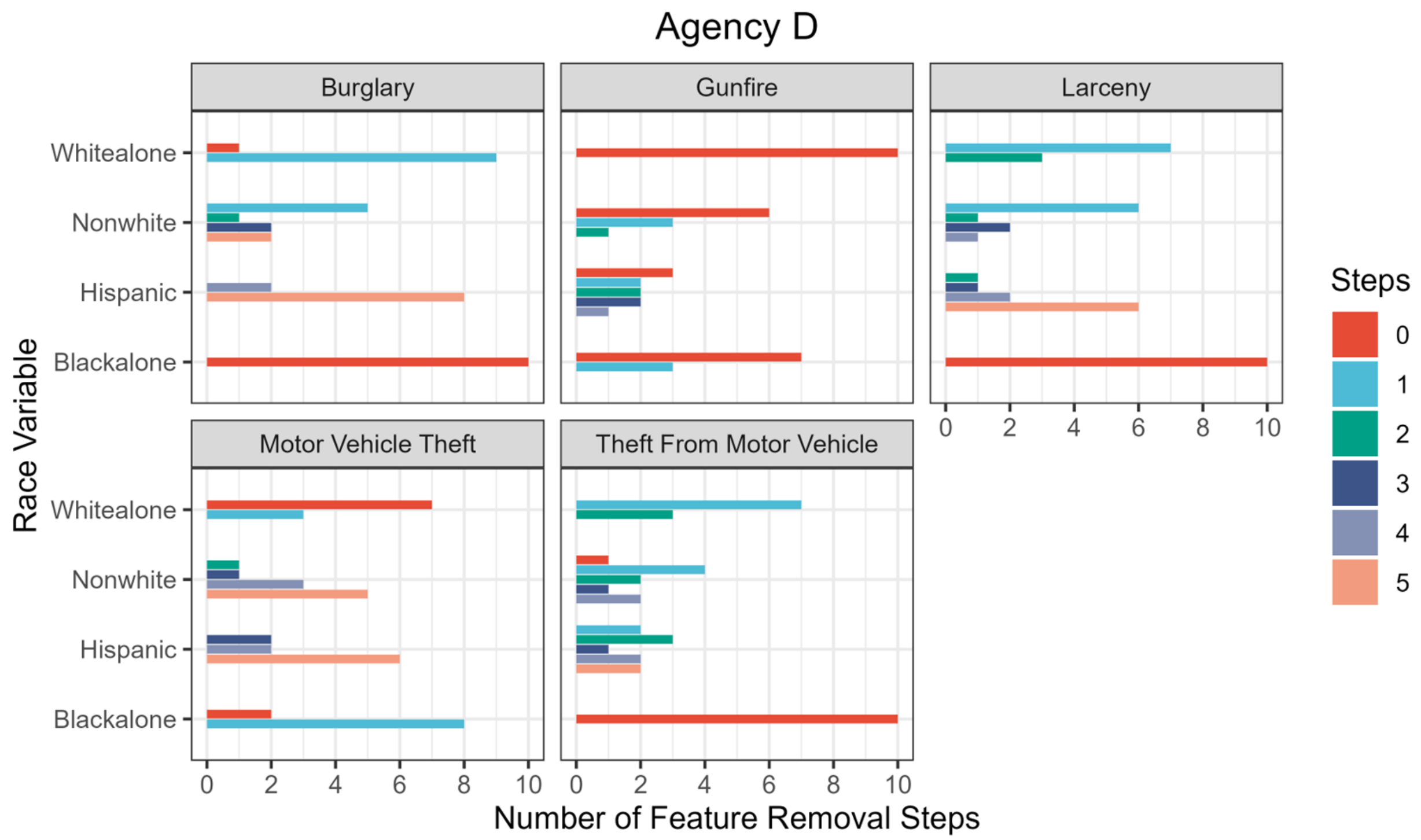

**Supplementary Figure 18.** Count of feature removal iterations for each harmful bias variable by Crime Model in Agency D.

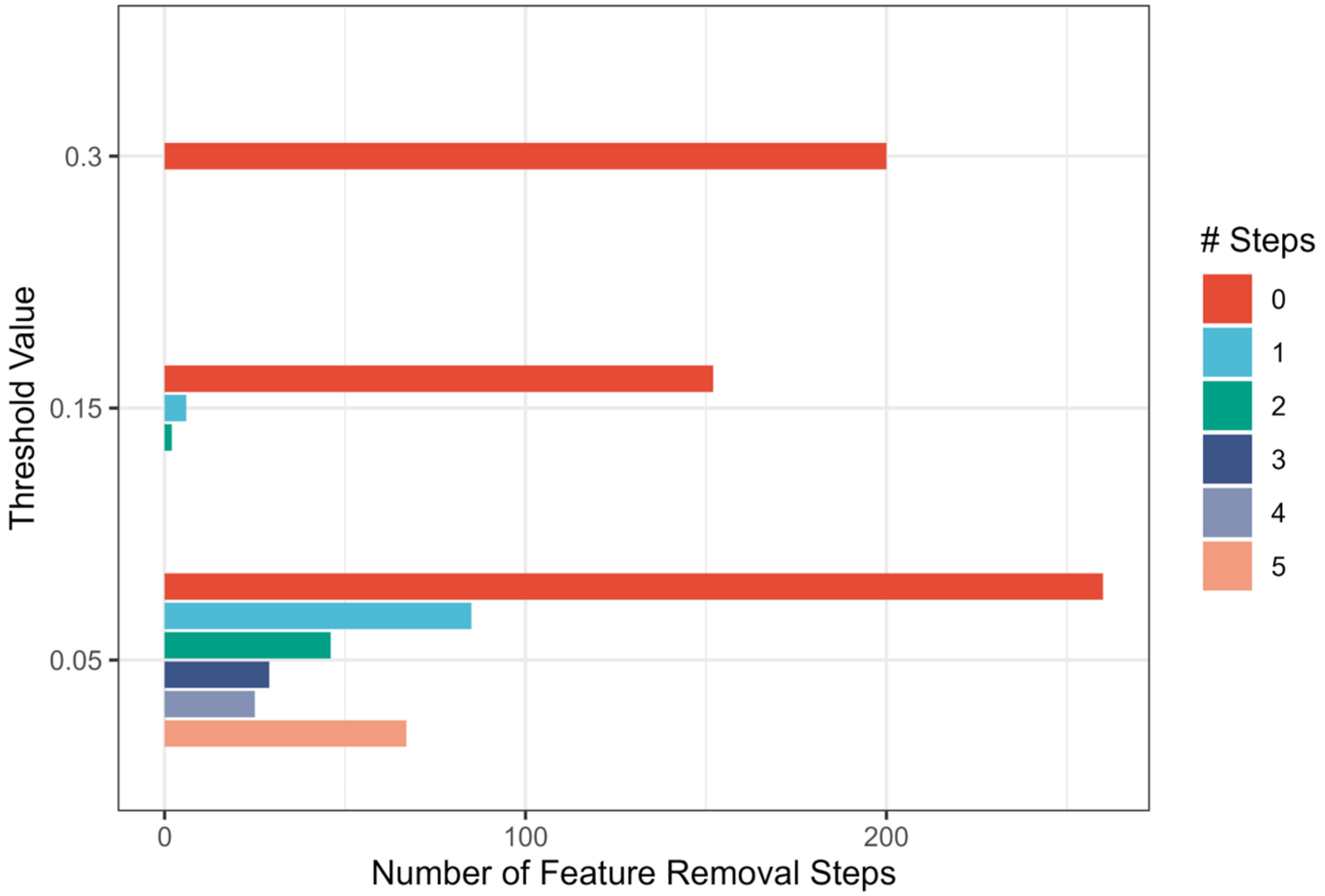

**Supplementary Figure 19.** Count of feature removal iterations for each bias tolerance threshold.

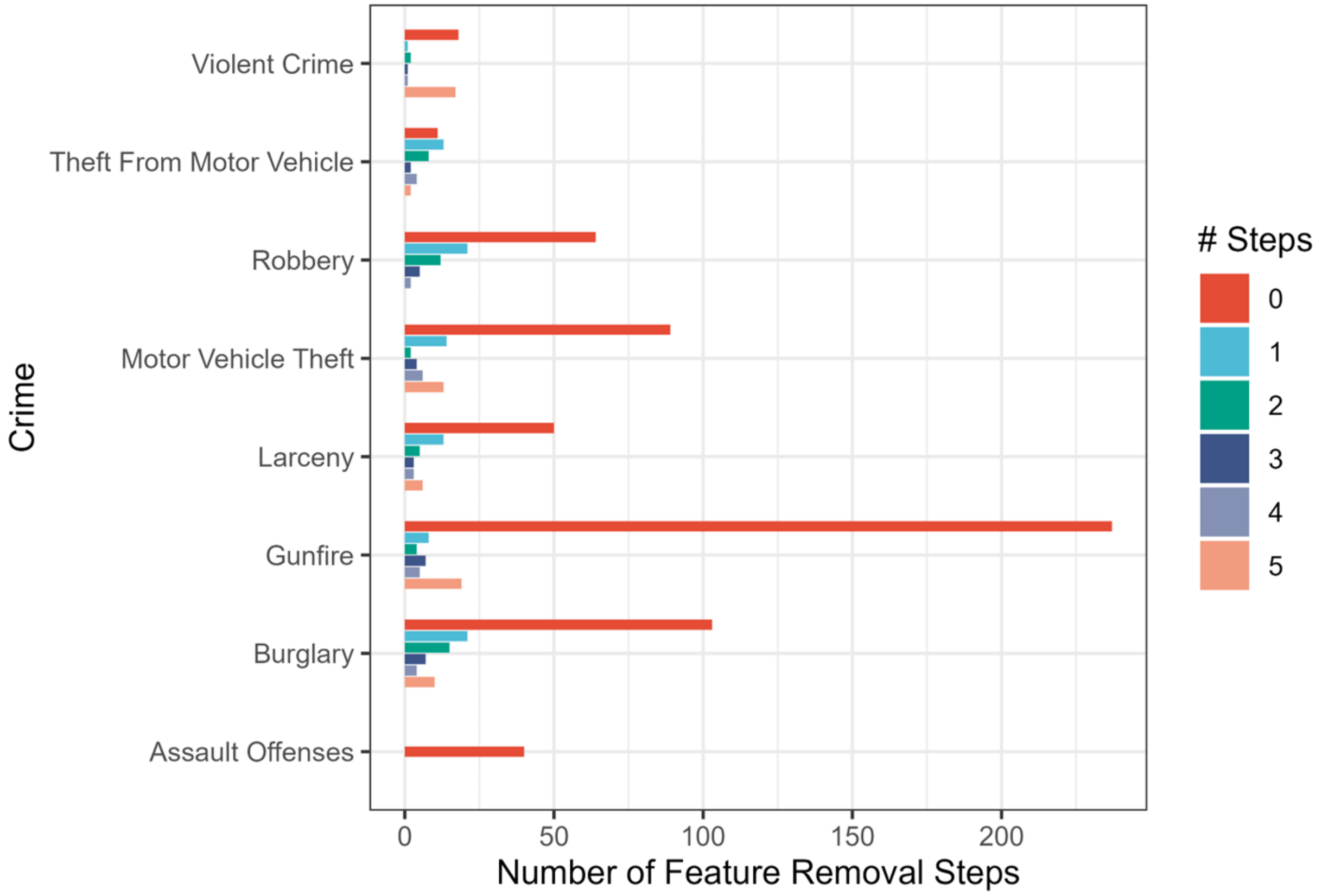

**Supplementary Figure 20.** Count of feature removal iterations for each crime model.

## Supplementary Tables:

**Supplementary Table 1.** Agency A descriptive statistics of AUC score by experiment runs.

| Crime Model | Threshold | Harmful Bias Variable | n= | Iteration | Mean | Median | Min | Max | Std. Dev | Std. Err |
|---|---|---|---|---|---|---|---|---|---|---|
| Burglary | 0.15 | Blackalone | 10 | Original | 0.9169 | 0.9167 | 0.9151 | 0.9191 | 0.0013 | 0.0004 |
| Burglary | 0.15 | Blackalone | 10 | Static | 0.9166 | 0.9164 | 0.9154 | 0.9188 | 0.0009 | 0.0003 |
| Burglary | 0.15 | Hispanic | 10 | Original | 0.9169 | 0.9167 | 0.9151 | 0.9191 | 0.0013 | 0.0004 |
| Burglary | 0.15 | Hispanic | 10 | Static | 0.9175 | 0.9180 | 0.9150 | 0.9197 | 0.0015 | 0.0005 |
| Burglary | 0.15 | Nonwhite | 10 | Original | 0.9169 | 0.9167 | 0.9151 | 0.9191 | 0.0013 | 0.0004 |
| Burglary | 0.15 | Nonwhite | 10 | Static | 0.9169 | 0.9167 | 0.9151 | 0.9191 | 0.0013 | 0.0004 |
| Burglary | 0.15 | Whitealone | 10 | Original | 0.9169 | 0.9167 | 0.9151 | 0.9191 | 0.0013 | 0.0004 |
| Burglary | 0.15 | Whitealone | 10 | Static | 0.9169 | 0.9167 | 0.9151 | 0.9191 | 0.0013 | 0.0004 |
| Gunfire | 0.05 | Blackalone | 10 | Original | 0.7229 | 0.7225 | 0.7159 | 0.7296 | 0.0048 | 0.0015 |
| Gunfire | 0.05 | Blackalone | 10 | Static | 0.7229 | 0.7225 | 0.7159 | 0.7296 | 0.0048 | 0.0015 |
| Gunfire | 0.05 | Hispanic | 10 | Original | 0.7229 | 0.7225 | 0.7159 | 0.7296 | 0.0048 | 0.0015 |
| Gunfire | 0.05 | Hispanic | 10 | Static | 0.7250 | 0.7227 | 0.7127 | 0.7361 | 0.0088 | 0.0028 |
| Gunfire | 0.05 | Hispanic | 10 | Remove 1 | 0.7231 | 0.7237 | 0.6957 | 0.7376 | 0.0121 | 0.0040 |
| Gunfire | 0.05 | Hispanic | 10 | Remove 2 | 0.7191 | 0.7206 | 0.6975 | 0.7271 | 0.0093 | 0.0031 |
| Gunfire | 0.05 | Hispanic | 10 | Remove 3 | 0.7215 | 0.7244 | 0.7082 | 0.7279 | 0.0068 | 0.0024 |
| Gunfire | 0.05 | Hispanic | 10 | Remove 4 | 0.7181 | 0.7167 | 0.7132 | 0.7256 | 0.0051 | 0.0021 |
| Gunfire | 0.05 | Hispanic | 10 | Remove 5 | 0.7188 | 0.7190 | 0.7095 | 0.7275 | 0.0067 | 0.0030 |
| Gunfire | 0.05 | Nonwhite | 10 | Original | 0.7229 | 0.7225 | 0.7159 | 0.7296 | 0.0048 | 0.0015 |
| Gunfire | 0.05 | Nonwhite | 10 | Static | 0.7229 | 0.7225 | 0.7159 | 0.7296 | 0.0048 | 0.0015 |
| Gunfire | 0.05 | Whitealone | 10 | Original | 0.7229 | 0.7225 | 0.7159 | 0.7296 | 0.0048 | 0.0015 |
| Gunfire | 0.05 | Whitealone | 10 | Static | 0.7229 | 0.7225 | 0.7159 | 0.7296 | 0.0048 | 0.0015 |
| Gunfire | 0.3 | Blackalone | 10 | Original | 0.7235 | 0.7252 | 0.7102 | 0.7310 | 0.0060 | 0.0019 |
| Gunfire | 0.3 | Blackalone | 10 | Static | 0.7235 | 0.7252 | 0.7102 | 0.7310 | 0.0060 | 0.0019 |
| Gunfire | 0.3 | Hispanic | 10 | Original | 0.7235 | 0.7252 | 0.7102 | 0.7310 | 0.0060 | 0.0019 |
| Gunfire | 0.3 | Hispanic | 10 | Static | 0.7262 | 0.7246 | 0.7203 | 0.7353 | 0.0049 | 0.0015 |
| Gunfire | 0.3 | Nonwhite | 10 | Original | 0.7235 | 0.7252 | 0.7102 | 0.7310 | 0.0060 | 0.0019 |
| Gunfire | 0.3 | Nonwhite | 10 | Static | 0.7235 | 0.7252 | 0.7102 | 0.7310 | 0.0060 | 0.0019 |
| Gunfire | 0.3 | Whitealone | 10 | Original | 0.7235 | 0.7252 | 0.7102 | 0.7310 | 0.0060 | 0.0019 |
| Gunfire | 0.3 | Whitealone | 10 | Static | 0.7235 | 0.7252 | 0.7102 | 0.7310 | 0.0060 | 0.0019 |
| Motor Vehicle Theft | 0.05 | Blackalone | 2 | Original | 0.9062 | 0.9062 | 0.9061 | 0.9062 | 0.0001 | 0.0001 |
| Motor Vehicle Theft | 0.05 | Blackalone | 2 | Static | 0.9061 | 0.9061 | 0.9060 | 0.9061 | 0.0001 | 0.0001 |
| Motor Vehicle Theft | 0.05 | Blackalone | 2 | Remove 1 | 0.9025 | 0.9025 | 0.8993 | 0.9056 | 0.0045 | 0.0032 |
| Motor Vehicle Theft | 0.05 | Blackalone | 2 | Remove 2 | 0.9052 | 0.9052 | 0.9046 | 0.9059 | 0.0010 | 0.0007 |
| Motor Vehicle Theft | 0.05 | Blackalone | 2 | Remove 3 | 0.9043 | 0.9043 | 0.9033 | 0.9053 | 0.0014 | 0.0010 |
| Motor Vehicle Theft | 0.05 | Blackalone | 2 | Remove 4 | 0.9046 | 0.9046 | 0.9046 | 0.9046 | | |
| Motor Vehicle Theft | 0.05 | Blackalone | 2 | Remove 5 | 0.9071 | 0.9071 | 0.9071 | 0.9071 | | |
| Motor Vehicle Theft | 0.05 | Hispanic | 2 | Original | 0.9062 | 0.9062 | 0.9061 | 0.9062 | 0.0001 | 0.0001 |
| Motor Vehicle Theft | 0.05 | Hispanic | 2 | Static | 0.9057 | 0.9057 | 0.9045 | 0.9069 | 0.0017 | 0.0012 |
| Motor Vehicle Theft | 0.05 | Hispanic | 2 | Remove 1 | 0.9078 | 0.9078 | 0.9073 | 0.9084 | 0.0007 | 0.0005 |
| Motor Vehicle Theft | 0.05 | Hispanic | 2 | Remove 2 | 0.9051 | 0.9051 | 0.9051 | 0.9051 | | |
| Motor Vehicle Theft | 0.05 | Nonwhite | 2 | Original | 0.9062 | 0.9062 | 0.9061 | 0.9062 | 0.0001 | 0.0001 |
| Motor Vehicle Theft | 0.05 | Nonwhite | 2 | Static | 0.9062 | 0.9062 | 0.9061 | 0.9062 | 0.0001 | 0.0001 |
| Motor Vehicle Theft | 0.05 | Nonwhite | 2 | Remove 1 | 0.9064 | 0.9064 | 0.9059 | 0.9069 | 0.0007 | 0.0005 |
| Motor Vehicle Theft | 0.05 | Nonwhite | 2 | Remove 2 | 0.9058 | 0.9058 | 0.9038 | 0.9079 | 0.0029 | 0.0021 |
| Motor Vehicle Theft | 0.05 | Nonwhite | 2 | Remove 3 | 0.9036 | 0.9036 | 0.9019 | 0.9052 | 0.0024 | 0.0017 |
| Motor Vehicle Theft | 0.05 | Nonwhite | 2 | Remove 4 | 0.9037 | 0.9037 | 0.9033 | 0.9041 | 0.0006 | 0.0004 |
| Motor Vehicle Theft | 0.05 | Nonwhite | 2 | Remove 5 | 0.9070 | 0.9070 | 0.9070 | 0.9070 | | |
| Motor Vehicle Theft | 0.05 | Whitealone | 2 | Original | 0.9062 | 0.9062 | 0.9061 | 0.9062 | 0.0001 | 0.0001 |
| Motor Vehicle Theft | 0.05 | Whitealone | 2 | Static | 0.9062 | 0.9062 | 0.9061 | 0.9062 | 0.0001 | 0.0001 |
| Motor Vehicle Theft | 0.05 | Whitealone | 2 | Remove 1 | 0.9051 | 0.9051 | 0.9049 | 0.9052 | 0.0002 | 0.0001 |
| Robbery | 0.05 | Blackalone | 6 | Original | 0.9071 | 0.9093 | 0.8970 | 0.9159 | 0.0070 | 0.0029 |
| Robbery | 0.05 | Blackalone | 6 | Static | 0.8874 | 0.8888 | 0.8593 | 0.9131 | 0.0223 | 0.0091 |
| Robbery | 0.05 | Blackalone | 6 | Remove 1 | 0.8984 | 0.8968 | 0.8962 | 0.9023 | 0.0034 | 0.0019 |
| Robbery | 0.05 | Blackalone | 6 | Remove 2 | 0.8981 | 0.9090 | 0.8741 | 0.9112 | 0.0208 | 0.0120 |
| Robbery | 0.05 | Blackalone | 6 | Remove 3 | 0.8611 | 0.8611 | 0.8611 | 0.8611 | | |
| Robbery | 0.05 | Blackalone | 6 | Remove 4 | 0.9079 | 0.9079 | 0.9079 | 0.9079 | | |
| Robbery | 0.05 | Hispanic | 6 | Original | 0.9071 | 0.9093 | 0.8970 | 0.9159 | 0.0070 | 0.0029 |
| Robbery | 0.05 | Hispanic | 6 | Static | 0.9096 | 0.9114 | 0.8936 | 0.9232 | 0.0099 | 0.0041 |
| Robbery | 0.05 | Hispanic | 6 | Remove 1 | 0.8999 | 0.9065 | 0.8832 | 0.9087 | 0.0111 | 0.0050 |
| Robbery | 0.05 | Hispanic | 6 | Remove 2 | 0.9056 | 0.9053 | 0.9052 | 0.9063 | 0.0007 | 0.0004 |
| Robbery | 0.05 | Hispanic | 6 | Remove 3 | 0.8902 | 0.8902 | 0.8725 | 0.9078 | 0.0250 | 0.0177 |
| Robbery | 0.05 | Nonwhite | 6 | Original | 0.9071 | 0.9093 | 0.8970 | 0.9159 | 0.0070 | 0.0029 |
| Robbery | 0.05 | Nonwhite | 6 | Static | 0.9071 | 0.9093 | 0.8970 | 0.9159 | 0.0070 | 0.0029 |
| Robbery | 0.05 | Nonwhite | 6 | Remove 1 | 0.9062 | 0.9078 | 0.8955 | 0.9135 | 0.0078 | 0.0039 |
| Robbery | 0.05 | Nonwhite | 6 | Remove 2 | 0.9039 | 0.9043 | 0.8941 | 0.9128 | 0.0077 | 0.0039 |
| Robbery | 0.05 | Nonwhite | 6 | Remove 3 | 0.8873 | 0.8873 | 0.8873 | 0.8873 | | |
| Robbery | 0.05 | Whitealone | 6 | Original | 0.9071 | 0.9093 | 0.8970 | 0.9159 | 0.0070 | 0.0029 |
| Robbery | 0.05 | Whitealone | 6 | Static | 0.9071 | 0.9093 | 0.8970 | 0.9159 | 0.0070 | 0.0029 |
| Robbery | 0.05 | Whitealone | 6 | Remove 1 | 0.9038 | 0.9015 | 0.8987 | 0.9119 | 0.0053 | 0.0024 |
| Robbery | 0.05 | Whitealone | 6 | Remove 2 | 0.8896 | 0.8895 | 0.8740 | 0.9055 | 0.0180 | 0.0090 |
| Robbery | 0.05 | Whitealone | 6 | Remove 3 | 0.9033 | 0.9048 | 0.8980 | 0.9072 | 0.0048 | 0.0028 |
| Robbery | 0.05 | Whitealone | 6 | Remove 4 | 0.9050 | 0.9050 | 0.9050 | 0.9050 | | |

**Supplementary Table 2.** Agency B descriptive statistics of AUC score by experiment runs.

| Crime Model | Threshold | Harmful Bias Variable | n= | Iteration | Mean | Median | Min | Max | Std. Dev | Std. Err |
|---|---|---|---|---|---|---|---|---|---|---|
| Gunfire | 0.05 | Blackalone | 10 | Original | 0.8277 | 0.8285 | 0.8237 | 0.8314 | 0.0027 | 0.0008 |
| Gunfire | 0.05 | Blackalone | 10 | Static | 0.8291 | 0.8286 | 0.8245 | 0.8345 | 0.0034 | 0.0011 |
| Gunfire | 0.05 | Blackalone | 10 | Remove 1 | 0.8295 | 0.8314 | 0.8219 | 0.8365 | 0.0053 | 0.0017 |
| Gunfire | 0.05 | Blackalone | 10 | Remove 2 | 0.8287 | 0.8287 | 0.8232 | 0.8329 | 0.0035 | 0.0011 |
| Gunfire | 0.05 | Blackalone | 10 | Remove 3 | 0.8299 | 0.8291 | 0.8257 | 0.8346 | 0.0027 | 0.0009 |
| Gunfire | 0.05 | Blackalone | 10 | Remove 4 | 0.8292 | 0.8294 | 0.8251 | 0.8315 | 0.0021 | 0.0007 |
| Gunfire | 0.05 | Blackalone | 10 | Remove 5 | 0.8274 | 0.8284 | 0.8199 | 0.8318 | 0.0038 | 0.0015 |
| Gunfire | 0.05 | Hispanic | 10 | Original | 0.8277 | 0.8285 | 0.8237 | 0.8314 | 0.0027 | 0.0008 |
| Gunfire | 0.05 | Hispanic | 10 | Static | 0.8282 | 0.8294 | 0.8206 | 0.8327 | 0.0037 | 0.0012 |
| Gunfire | 0.05 | Nonwhite | 10 | Original | 0.8277 | 0.8285 | 0.8237 | 0.8314 | 0.0027 | 0.0008 |
| Gunfire | 0.05 | Nonwhite | 10 | Static | 0.8295 | 0.8282 | 0.8252 | 0.8367 | 0.0037 | 0.0012 |
| Gunfire | 0.05 | Nonwhite | 10 | Remove 1 | 0.8300 | 0.8300 | 0.8249 | 0.8338 | 0.0024 | 0.0008 |
| Gunfire | 0.05 | Nonwhite | 10 | Remove 2 | 0.8293 | 0.8295 | 0.8274 | 0.8312 | 0.0011 | 0.0003 |
| Gunfire | 0.05 | Nonwhite | 10 | Remove 3 | 0.8284 | 0.8301 | 0.8208 | 0.8336 | 0.0042 | 0.0013 |
| Gunfire | 0.05 | Nonwhite | 10 | Remove 4 | 0.8286 | 0.8296 | 0.8208 | 0.8326 | 0.0038 | 0.0013 |
| Gunfire | 0.05 | Nonwhite | 10 | Remove 5 | 0.8277 | 0.8294 | 0.8229 | 0.8319 | 0.0038 | 0.0014 |
| Gunfire | 0.05 | Whitealone | 10 | Original | 0.8277 | 0.8285 | 0.8237 | 0.8314 | 0.0027 | 0.0008 |
| Gunfire | 0.05 | Whitealone | 10 | Static | 0.8297 | 0.8289 | 0.8251 | 0.8351 | 0.0029 | 0.0009 |
| Gunfire | 0.15 | Blackalone | 10 | Original | 0.8302 | 0.8294 | 0.8234 | 0.8378 | 0.0043 | 0.0014 |
| Gunfire | 0.15 | Blackalone | 10 | Static | 0.8282 | 0.8285 | 0.8242 | 0.8320 | 0.0030 | 0.0009 |
| Gunfire | 0.15 | Hispanic | 10 | Original | 0.8302 | 0.8294 | 0.8234 | 0.8378 | 0.0043 | 0.0014 |
| Gunfire | 0.15 | Hispanic | 10 | Static | 0.8304 | 0.8307 | 0.8250 | 0.8342 | 0.0025 | 0.0008 |
| Gunfire | 0.15 | Nonwhite | 10 | Original | 0.8302 | 0.8294 | 0.8234 | 0.8378 | 0.0043 | 0.0014 |
| Gunfire | 0.15 | Nonwhite | 10 | Static | 0.8302 | 0.8299 | 0.8270 | 0.8351 | 0.0029 | 0.0009 |
| Gunfire | 0.15 | Whitealone | 10 | Original | 0.8302 | 0.8294 | 0.8234 | 0.8378 | 0.0043 | 0.0014 |
| Gunfire | 0.15 | Whitealone | 10 | Static | 0.8279 | 0.8274 | 0.8253 | 0.8332 | 0.0022 | 0.0007 |
| Gunfire | 0.3 | Blackalone | 10 | Original | 0.8288 | 0.8300 | 0.8210 | 0.8335 | 0.0039 | 0.0012 |
| Gunfire | 0.3 | Blackalone | 10 | Static | 0.8304 | 0.8311 | 0.8256 | 0.8332 | 0.0027 | 0.0009 |
| Gunfire | 0.3 | Hispanic | 10 | Original | 0.8288 | 0.8300 | 0.8210 | 0.8335 | 0.0039 | 0.0012 |
| Gunfire | 0.3 | Hispanic | 10 | Static | 0.8300 | 0.8299 | 0.8237 | 0.8377 | 0.0041 | 0.0013 |
| Gunfire | 0.3 | Nonwhite | 10 | Original | 0.8288 | 0.8300 | 0.8210 | 0.8335 | 0.0039 | 0.0012 |
| Gunfire | 0.3 | Nonwhite | 10 | Static | 0.8297 | 0.8298 | 0.8267 | 0.8325 | 0.0021 | 0.0007 |
| Gunfire | 0.3 | Whitealone | 10 | Original | 0.8288 | 0.8300 | 0.8210 | 0.8335 | 0.0039 | 0.0012 |
| Gunfire | 0.3 | Whitealone | 10 | Static | 0.8301 | 0.8303 | 0.8275 | 0.8321 | 0.0015 | 0.0005 |

**Supplementary Table 3.** Agency C descriptive statistics of AUC score by experiment runs.

| Crime Model | Threshold | Harmful Bias Variable | n= | Iteration | Mean | Median | Min | Max | Std. Dev | Std. Err |
|---|---|---|---|---|---|---|---|---|---|---|
| Assault Offenses | 0.3 | Blackalone | 10 | Original | 0.8526 | 0.8527 | 0.8484 | 0.8561 | 0.0027 | 0.0009 |
| Assault Offenses | 0.3 | Blackalone | 10 | Static | 0.8532 | 0.8526 | 0.8494 | 0.8586 | 0.0030 | 0.0009 |
| Assault Offenses | 0.3 | Hispanic | 10 | Original | 0.8526 | 0.8527 | 0.8484 | 0.8561 | 0.0027 | 0.0009 |
| Assault Offenses | 0.3 | Hispanic | 10 | Static | 0.8526 | 0.8527 | 0.8484 | 0.8561 | 0.0027 | 0.0009 |
| Assault Offenses | 0.3 | Nonwhite | 10 | Original | 0.8526 | 0.8527 | 0.8484 | 0.8561 | 0.0027 | 0.0009 |
| Assault Offenses | 0.3 | Nonwhite | 10 | Static | 0.8530 | 0.8522 | 0.8499 | 0.8577 | 0.0026 | 0.0008 |
| Assault Offenses | 0.3 | Whitealone | 10 | Original | 0.8526 | 0.8527 | 0.8484 | 0.8561 | 0.0027 | 0.0009 |
| Assault Offenses | 0.3 | Whitealone | 10 | Static | 0.8519 | 0.8521 | 0.8455 | 0.8569 | 0.0031 | 0.0010 |
| Burglary | 0.05 | Blackalone | 10 | Original | 0.8937 | 0.8970 | 0.8794 | 0.9007 | 0.0070 | 0.0022 |
| Burglary | 0.05 | Blackalone | 10 | Static | 0.8960 | 0.8960 | 0.8816 | 0.9061 | 0.0084 | 0.0027 |
| Burglary | 0.05 | Blackalone | 10 | Remove 1 | 0.8928 | 0.8930 | 0.8798 | 0.9054 | 0.0076 | 0.0024 |
| Burglary | 0.05 | Blackalone | 10 | Remove 2 | 0.8899 | 0.8914 | 0.8746 | 0.8964 | 0.0063 | 0.0020 |
| Burglary | 0.05 | Blackalone | 10 | Remove 3 | 0.8979 | 0.8985 | 0.8941 | 0.9006 | 0.0029 | 0.0014 |
| Burglary | 0.05 | Blackalone | 10 | Remove 4 | 0.8958 | 0.8958 | 0.8957 | 0.8958 | 0.0001 | 0.0001 |
| Burglary | 0.05 | Hispanic | 10 | Original | 0.8937 | 0.8970 | 0.8794 | 0.9007 | 0.0070 | 0.0022 |
| Burglary | 0.05 | Hispanic | 10 | Static | 0.8937 | 0.8970 | 0.8794 | 0.9007 | 0.0070 | 0.0022 |
| Burglary | 0.05 | Nonwhite | 10 | Original | 0.8937 | 0.8970 | 0.8794 | 0.9007 | 0.0070 | 0.0022 |
| Burglary | 0.05 | Nonwhite | 10 | Static | 0.8944 | 0.8929 | 0.8919 | 0.9019 | 0.0032 | 0.0010 |
| Burglary | 0.05 | Nonwhite | 10 | Remove 1 | 0.8938 | 0.8935 | 0.8874 | 0.8995 | 0.0044 | 0.0014 |
| Burglary | 0.05 | Nonwhite | 10 | Remove 2 | 0.8922 | 0.8961 | 0.8721 | 0.9054 | 0.0105 | 0.0035 |
| Burglary | 0.05 | Nonwhite | 10 | Remove 3 | 0.8902 | 0.8931 | 0.8805 | 0.8972 | 0.0087 | 0.0050 |
| Burglary | 0.05 | Whitealone | 10 | Original | 0.8937 | 0.8970 | 0.8794 | 0.9007 | 0.0070 | 0.0022 |
| Burglary | 0.05 | Whitealone | 10 | Static | 0.8928 | 0.8912 | 0.8824 | 0.9086 | 0.0075 | 0.0024 |
| Larceny | 0.05 | Blackalone | 10 | Original | 0.9268 | 0.9265 | 0.9193 | 0.9315 | 0.0041 | 0.0013 |
| Larceny | 0.05 | Blackalone | 10 | Static | 0.9253 | 0.9248 | 0.9225 | 0.9299 | 0.0024 | 0.0008 |
| Larceny | 0.05 | Hispanic | 10 | Original | 0.9268 | 0.9265 | 0.9193 | 0.9315 | 0.0041 | 0.0013 |
| Larceny | 0.05 | Hispanic | 10 | Static | 0.9268 | 0.9265 | 0.9193 | 0.9315 | 0.0041 | 0.0013 |
| Larceny | 0.05 | Nonwhite | 10 | Original | 0.9268 | 0.9265 | 0.9193 | 0.9315 | 0.0041 | 0.0013 |
| Larceny | 0.05 | Nonwhite | 10 | Static | 0.9269 | 0.9273 | 0.9236 | 0.9299 | 0.0022 | 0.0007 |
| Larceny | 0.05 | Whitealone | 10 | Original | 0.9268 | 0.9265 | 0.9193 | 0.9315 | 0.0041 | 0.0013 |
| Larceny | 0.05 | Whitealone | 10 | Static | 0.9268 | 0.9264 | 0.9237 | 0.9321 | 0.0023 | 0.0007 |
| Motor Vehicle Theft | 0.05 | Blackalone | 10 | Original | 0.9090 | 0.9096 | 0.8997 | 0.9164 | 0.0052 | 0.0016 |
| Motor Vehicle Theft | 0.05 | Blackalone | 10 | Static | 0.9087 | 0.9083 | 0.9012 | 0.9174 | 0.0051 | 0.0016 |
| Motor Vehicle Theft | 0.05 | Hispanic | 10 | Original | 0.9090 | 0.9096 | 0.8997 | 0.9164 | 0.0052 | 0.0016 |
| Motor Vehicle Theft | 0.05 | Hispanic | 10 | Static | 0.9090 | 0.9096 | 0.8997 | 0.9164 | 0.0052 | 0.0016 |
| Motor Vehicle Theft | 0.05 | Nonwhite | 10 | Original | 0.9090 | 0.9096 | 0.8997 | 0.9164 | 0.0052 | 0.0016 |
| Motor Vehicle Theft | 0.05 | Nonwhite | 10 | Static | 0.9108 | 0.9095 | 0.9051 | 0.9198 | 0.0049 | 0.0016 |
| Motor Vehicle Theft | 0.05 | Whitealone | 10 | Original | 0.9090 | 0.9096 | 0.8997 | 0.9164 | 0.0052 | 0.0016 |
| Motor Vehicle Theft | 0.05 | Whitealone | 10 | Static | 0.9080 | 0.9082 | 0.9017 | 0.9144 | 0.0039 | 0.0012 |
| Motor Vehicle Theft | 0.15 | Blackalone | 10 | Original | 0.9113 | 0.9116 | 0.9024 | 0.9189 | 0.0051 | 0.0016 |
| Motor Vehicle Theft | 0.15 | Blackalone | 10 | Static | 0.9144 | 0.9152 | 0.9059 | 0.9195 | 0.0047 | 0.0015 |
| Motor Vehicle Theft | 0.15 | Hispanic | 10 | Original | 0.9113 | 0.9116 | 0.9024 | 0.9189 | 0.0051 | 0.0016 |
| Motor Vehicle Theft | 0.15 | Hispanic | 10 | Static | 0.9113 | 0.9116 | 0.9024 | 0.9189 | 0.0051 | 0.0016 |
| Motor Vehicle Theft | 0.15 | Nonwhite | 10 | Original | 0.9113 | 0.9116 | 0.9024 | 0.9189 | 0.0051 | 0.0016 |
| Motor Vehicle Theft | 0.15 | Nonwhite | 10 | Static | 0.9134 | 0.9130 | 0.9078 | 0.9211 | 0.0043 | 0.0013 |
| Motor Vehicle Theft | 0.15 | Whitealone | 10 | Original | 0.9113 | 0.9116 | 0.9024 | 0.9189 | 0.0051 | 0.0016 |
| Motor Vehicle Theft | 0.15 | Whitealone | 10 | Static | 0.9104 | 0.9092 | 0.9028 | 0.9217 | 0.0059 | 0.0019 |
| Robbery | 0.05 | Blackalone | 10 | Original | 0.9082 | 0.9068 | 0.8915 | 0.9238 | 0.0105 | 0.0033 |
| Robbery | 0.05 | Blackalone | 10 | Static | 0.9035 | 0.9067 | 0.8777 | 0.9164 | 0.0128 | 0.0040 |
| Robbery | 0.05 | Blackalone | 10 | Remove 1 | 0.9135 | 0.9133 | 0.8985 | 0.9278 | 0.0091 | 0.0030 |
| Robbery | 0.05 | Blackalone | 10 | Remove 2 | 0.9032 | 0.9032 | 0.9032 | 0.9032 | | |
| Robbery | 0.05 | Hispanic | 10 | Original | 0.9082 | 0.9068 | 0.8915 | 0.9238 | 0.0105 | 0.0033 |
| Robbery | 0.05 | Hispanic | 10 | Static | 0.9082 | 0.9068 | 0.8915 | 0.9238 | 0.0105 | 0.0033 |
| Robbery | 0.05 | Nonwhite | 10 | Original | 0.9082 | 0.9068 | 0.8915 | 0.9238 | 0.0105 | 0.0033 |
| Robbery | 0.05 | Nonwhite | 10 | Static | 0.9092 | 0.9088 | 0.8955 | 0.9208 | 0.0081 | 0.0025 |
| Robbery | 0.05 | Nonwhite | 10 | Remove 1 | 0.9137 | 0.9138 | 0.9058 | 0.9262 | 0.0068 | 0.0026 |
| Robbery | 0.05 | Nonwhite | 10 | Remove 2 | 0.9095 | 0.9095 | 0.9095 | 0.9095 | | |
| Robbery | 0.05 | Whitealone | 10 | Original | 0.9082 | 0.9068 | 0.8915 | 0.9238 | 0.0105 | 0.0033 |
| Robbery | 0.05 | Whitealone | 10 | Static | 0.9058 | 0.9085 | 0.8849 | 0.9229 | 0.0124 | 0.0039 |
| Robbery | 0.05 | Whitealone | 10 | Remove 1 | 0.9081 | 0.9104 | 0.8961 | 0.9164 | 0.0064 | 0.0024 |
| Robbery | 0.05 | Whitealone | 10 | Remove 2 | 0.9065 | 0.9079 | 0.8999 | 0.9118 | 0.0061 | 0.0035 |
| Robbery | 0.3 | Blackalone | 10 | Original | 0.9098 | 0.9071 | 0.8969 | 0.9254 | 0.0098 | 0.0031 |
| Robbery | 0.3 | Blackalone | 10 | Static | 0.9051 | 0.9052 | 0.8900 | 0.9162 | 0.0068 | 0.0022 |

| | | | | | | | | | | | |
|---|---|---|---|---|---|---|---|---|---|---|---|
| Robbery | 0.3 | Hispanic | 10 | Original | 0.9098 | 0.9071 | 0.8969 | 0.9254 | 0.0098 | 0.0031 |
| Robbery | 0.3 | Hispanic | 10 | Static | 0.9098 | 0.9071 | 0.8969 | 0.9254 | 0.0098 | 0.0031 |
| Robbery | 0.3 | Nonwhite | 10 | Original | 0.9098 | 0.9071 | 0.8969 | 0.9254 | 0.0098 | 0.0031 |
| Robbery | 0.3 | Nonwhite | 10 | Static | 0.9114 | 0.9124 | 0.8879 | 0.9307 | 0.0146 | 0.0046 |
| Robbery | 0.3 | Whitealone | 10 | Original | 0.9098 | 0.9071 | 0.8969 | 0.9254 | 0.0098 | 0.0031 |
| Robbery | 0.3 | Whitealone | 10 | Static | 0.9162 | 0.9184 | 0.8901 | 0.9272 | 0.0108 | 0.0034 |
| Violent Crime | 0.05 | Blackalone | 10 | Original | 0.8690 | 0.8693 | 0.8664 | 0.8721 | 0.0021 | 0.0007 |
| Violent Crime | 0.05 | Blackalone | 10 | Static | 0.8689 | 0.8698 | 0.8623 | 0.8723 | 0.0030 | 0.0010 |
| Violent Crime | 0.05 | Blackalone | 10 | Remove 1 | 0.8673 | 0.8672 | 0.8647 | 0.8693 | 0.0014 | 0.0005 |
| Violent Crime | 0.05 | Blackalone | 10 | Remove 2 | 0.8687 | 0.8689 | 0.8649 | 0.8718 | 0.0021 | 0.0007 |
| Violent Crime | 0.05 | Blackalone | 10 | Remove 3 | 0.8708 | 0.8713 | 0.8662 | 0.8728 | 0.0018 | 0.0006 |
| Violent Crime | 0.05 | Blackalone | 10 | Remove 4 | 0.8694 | 0.8697 | 0.8670 | 0.8721 | 0.0017 | 0.0006 |
| Violent Crime | 0.05 | Blackalone | 10 | Remove 5 | 0.8698 | 0.8709 | 0.8656 | 0.8726 | 0.0022 | 0.0007 |
| Violent Crime | 0.05 | Hispanic | 10 | Original | 0.8690 | 0.8693 | 0.8664 | 0.8721 | 0.0021 | 0.0007 |
| Violent Crime | 0.05 | Hispanic | 10 | Static | 0.8690 | 0.8693 | 0.8664 | 0.8721 | 0.0021 | 0.0007 |
| Violent Crime | 0.05 | Nonwhite | 10 | Original | 0.8690 | 0.8693 | 0.8664 | 0.8721 | 0.0021 | 0.0007 |
| Violent Crime | 0.05 | Nonwhite | 10 | Static | 0.8684 | 0.8689 | 0.8634 | 0.8710 | 0.0025 | 0.0008 |
| Violent Crime | 0.05 | Nonwhite | 10 | Remove 1 | 0.8680 | 0.8672 | 0.8618 | 0.8745 | 0.0036 | 0.0011 |
| Violent Crime | 0.05 | Nonwhite | 10 | Remove 2 | 0.8677 | 0.8686 | 0.8642 | 0.8703 | 0.0022 | 0.0007 |
| Violent Crime | 0.05 | Nonwhite | 10 | Remove 3 | 0.8703 | 0.8696 | 0.8676 | 0.8750 | 0.0023 | 0.0008 |
| Violent Crime | 0.05 | Nonwhite | 10 | Remove 4 | 0.8676 | 0.8675 | 0.8649 | 0.8694 | 0.0017 | 0.0006 |
| Violent Crime | 0.05 | Nonwhite | 10 | Remove 5 | 0.8670 | 0.8662 | 0.8634 | 0.8731 | 0.0030 | 0.0011 |
| Violent Crime | 0.05 | Whitealone | 10 | Original | 0.8690 | 0.8693 | 0.8664 | 0.8721 | 0.0021 | 0.0007 |
| Violent Crime | 0.05 | Whitealone | 10 | Static | 0.8689 | 0.8685 | 0.8668 | 0.8743 | 0.0023 | 0.0007 |
| Violent Crime | 0.05 | Whitealone | 10 | Remove 1 | 0.8694 | 0.8694 | 0.8692 | 0.8695 | 0.0002 | 0.0001 |
| Violent Crime | 0.05 | Whitealone | 10 | Remove 2 | 0.8680 | 0.8680 | 0.8680 | 0.8680 | | |

**Supplementary Table 4.** Agency D descriptive statistics of AUC score by experiment runs.

| Crime Model | Threshold | Harmful Bias Variable | n= | Iteration | Mean | Median | Min | Max | Std. Dev | Std. Err |
|---|---|---|---|---|---|---|---|---|---|---|
| Burglary | 0.05 | Blackalone | 10 | Original | 0.8420 | 0.8432 | 0.8327 | 0.8474 | 0.0047 | 0.0015 |
| Burglary | 0.05 | Blackalone | 10 | Static | 0.8420 | 0.8432 | 0.8327 | 0.8474 | 0.0047 | 0.0015 |
| Burglary | 0.05 | Hispanic | 10 | Original | 0.8420 | 0.8432 | 0.8327 | 0.8474 | 0.0047 | 0.0015 |
| Burglary | 0.05 | Hispanic | 10 | Static | 0.8340 | 0.8347 | 0.8207 | 0.8450 | 0.0085 | 0.0027 |
| Burglary | 0.05 | Hispanic | 10 | Remove 1 | 0.8306 | 0.8304 | 0.8143 | 0.8417 | 0.0086 | 0.0027 |
| Burglary | 0.05 | Hispanic | 10 | Remove 2 | 0.8348 | 0.8338 | 0.8282 | 0.8466 | 0.0056 | 0.0018 |
| Burglary | 0.05 | Hispanic | 10 | Remove 3 | 0.8331 | 0.8316 | 0.8222 | 0.8469 | 0.0073 | 0.0023 |
| Burglary | 0.05 | Hispanic | 10 | Remove 4 | 0.8369 | 0.8334 | 0.8295 | 0.8527 | 0.0082 | 0.0026 |
| Burglary | 0.05 | Hispanic | 10 | Remove 5 | 0.8357 | 0.8356 | 0.8323 | 0.8399 | 0.0027 | 0.0010 |
| Burglary | 0.05 | Nonwhite | 10 | Original | 0.8420 | 0.8432 | 0.8327 | 0.8474 | 0.0047 | 0.0015 |
| Burglary | 0.05 | Nonwhite | 10 | Static | 0.8453 | 0.8433 | 0.8357 | 0.8566 | 0.0069 | 0.0022 |
| Burglary | 0.05 | Nonwhite | 10 | Remove 1 | 0.8381 | 0.8379 | 0.8164 | 0.8508 | 0.0100 | 0.0032 |
| Burglary | 0.05 | Nonwhite | 10 | Remove 2 | 0.8375 | 0.8387 | 0.8306 | 0.8433 | 0.0047 | 0.0021 |
| Burglary | 0.05 | Nonwhite | 10 | Remove 3 | 0.8427 | 0.8424 | 0.8392 | 0.8469 | 0.0034 | 0.0017 |
| Burglary | 0.05 | Nonwhite | 10 | Remove 4 | 0.8372 | 0.8372 | 0.8287 | 0.8456 | 0.0120 | 0.0085 |
| Burglary | 0.05 | Nonwhite | 10 | Remove 5 | 0.8400 | 0.8400 | 0.8385 | 0.8414 | 0.0020 | 0.0014 |
| Burglary | 0.05 | Whitealone | 10 | Original | 0.8420 | 0.8432 | 0.8327 | 0.8474 | 0.0047 | 0.0015 |
| Burglary | 0.05 | Whitealone | 10 | Static | 0.8436 | 0.8424 | 0.8352 | 0.8579 | 0.0066 | 0.0021 |
| Burglary | 0.05 | Whitealone | 10 | Remove 1 | 0.8421 | 0.8467 | 0.8242 | 0.8541 | 0.0103 | 0.0034 |
| Burglary | 0.15 | Blackalone | 10 | Original | 0.8412 | 0.8427 | 0.8328 | 0.8463 | 0.0050 | 0.0016 |
| Burglary | 0.15 | Blackalone | 10 | Static | 0.8412 | 0.8427 | 0.8328 | 0.8463 | 0.0050 | 0.0016 |
| Burglary | 0.15 | Hispanic | 10 | Original | 0.8412 | 0.8427 | 0.8328 | 0.8463 | 0.0050 | 0.0016 |
| Burglary | 0.15 | Hispanic | 10 | Static | 0.8394 | 0.8409 | 0.8156 | 0.8488 | 0.0095 | 0.0030 |
| Burglary | 0.15 | Hispanic | 10 | Remove 1 | 0.8374 | 0.8391 | 0.8259 | 0.8470 | 0.0078 | 0.0030 |
| Burglary | 0.15 | Hispanic | 10 | Remove 2 | 0.8313 | 0.8313 | 0.8299 | 0.8327 | 0.0020 | 0.0014 |
| Burglary | 0.15 | Nonwhite | 10 | Original | 0.8412 | 0.8427 | 0.8328 | 0.8463 | 0.0050 | 0.0016 |
| Burglary | 0.15 | Nonwhite | 10 | Static | 0.8395 | 0.8398 | 0.8292 | 0.8463 | 0.0051 | 0.0016 |
| Burglary | 0.15 | Nonwhite | 10 | Remove 1 | 0.8374 | 0.8374 | 0.8374 | 0.8374 | | |
| Burglary | 0.15 | Whitealone | 10 | Original | 0.8412 | 0.8427 | 0.8328 | 0.8463 | 0.0050 | 0.0016 |
| Burglary | 0.15 | Whitealone | 10 | Static | 0.8397 | 0.8393 | 0.8280 | 0.8572 | 0.0091 | 0.0029 |
| Gunfire | 0.05 | Blackalone | 10 | Original | 0.7573 | 0.7569 | 0.7477 | 0.7689 | 0.0068 | 0.0021 |
| Gunfire | 0.05 | Blackalone | 10 | Static | 0.7573 | 0.7569 | 0.7477 | 0.7689 | 0.0068 | 0.0021 |
| Gunfire | 0.05 | Blackalone | 10 | Remove 1 | 0.7582 | 0.7608 | 0.7460 | 0.7679 | 0.0112 | 0.0064 |
| Gunfire | 0.05 | Hispanic | 10 | Original | 0.7573 | 0.7569 | 0.7477 | 0.7689 | 0.0068 | 0.0021 |
| Gunfire | 0.05 | Hispanic | 10 | Static | 0.7550 | 0.7575 | 0.7428 | 0.7650 | 0.0079 | 0.0025 |
| Gunfire | 0.05 | Hispanic | 10 | Remove 1 | 0.7508 | 0.7508 | 0.7425 | 0.7578 | 0.0058 | 0.0022 |
| Gunfire | 0.05 | Hispanic | 10 | Remove 2 | 0.7529 | 0.7528 | 0.7408 | 0.7610 | 0.0077 | 0.0035 |
| Gunfire | 0.05 | Hispanic | 10 | Remove 3 | 0.7508 | 0.7528 | 0.7398 | 0.7600 | 0.0102 | 0.0059 |
| Gunfire | 0.05 | Hispanic | 10 | Remove 4 | 0.7447 | 0.7447 | 0.7447 | 0.7447 | | |
| Gunfire | 0.05 | Nonwhite | 10 | Original | 0.7573 | 0.7569 | 0.7477 | 0.7689 | 0.0068 | 0.0021 |
| Gunfire | 0.05 | Nonwhite | 10 | Static | 0.7578 | 0.7571 | 0.7517 | 0.7646 | 0.0050 | 0.0016 |
| Gunfire | 0.05 | Nonwhite | 10 | Remove 1 | 0.7547 | 0.7550 | 0.7508 | 0.7581 | 0.0032 | 0.0016 |
| Gunfire | 0.05 | Nonwhite | 10 | Remove 2 | 0.7596 | 0.7596 | 0.7596 | 0.7596 | | |
| Gunfire | 0.05 | Whitealone | 10 | Original | 0.7573 | 0.7569 | 0.7477 | 0.7689 | 0.0068 | 0.0021 |
| Gunfire | 0.05 | Whitealone | 10 | Static | 0.7534 | 0.7541 | 0.7422 | 0.7630 | 0.0060 | 0.0019 |
| Gunfire | 0.3 | Blackalone | 10 | Original | 0.7548 | 0.7570 | 0.7372 | 0.7667 | 0.0095 | 0.0030 |
| Gunfire | 0.3 | Blackalone | 10 | Static | 0.7548 | 0.7570 | 0.7372 | 0.7667 | 0.0095 | 0.0030 |
| Gunfire | 0.3 | Hispanic | 10 | Original | 0.7548 | 0.7570 | 0.7372 | 0.7667 | 0.0095 | 0.0030 |
| Gunfire | 0.3 | Hispanic | 10 | Static | 0.7581 | 0.7575 | 0.7534 | 0.7633 | 0.0036 | 0.0011 |
| Gunfire | 0.3 | Nonwhite | 10 | Original | 0.7548 | 0.7570 | 0.7372 | 0.7667 | 0.0095 | 0.0030 |
| Gunfire | 0.3 | Nonwhite | 10 | Static | 0.7567 | 0.7592 | 0.7447 | 0.7686 | 0.0085 | 0.0027 |
| Gunfire | 0.3 | Whitealone | 10 | Original | 0.7548 | 0.7570 | 0.7372 | 0.7667 | 0.0095 | 0.0030 |
| Gunfire | 0.3 | Whitealone | 10 | Static | 0.7533 | 0.7527 | 0.7413 | 0.7631 | 0.0063 | 0.0020 |
| Larceny | 0.05 | Blackalone | 10 | Original | 0.9106 | 0.9110 | 0.9043 | 0.9140 | 0.0029 | 0.0009 |
| Larceny | 0.05 | Blackalone | 10 | Static | 0.9106 | 0.9110 | 0.9043 | 0.9140 | 0.0029 | 0.0009 |
| Larceny | 0.05 | Hispanic | 10 | Original | 0.9106 | 0.9110 | 0.9043 | 0.9140 | 0.0029 | 0.0009 |
| Larceny | 0.05 | Hispanic | 10 | Static | 0.9109 | 0.9110 | 0.9090 | 0.9131 | 0.0014 | 0.0004 |
| Larceny | 0.05 | Hispanic | 10 | Remove 1 | 0.9111 | 0.9120 | 0.9036 | 0.9152 | 0.0033 | 0.0010 |
| Larceny | 0.05 | Hispanic | 10 | Remove 2 | 0.9099 | 0.9104 | 0.9072 | 0.9118 | 0.0018 | 0.0006 |
| Larceny | 0.05 | Hispanic | 10 | Remove 3 | 0.9107 | 0.9106 | 0.9068 | 0.9148 | 0.0027 | 0.0009 |
| Larceny | 0.05 | Hispanic | 10 | Remove 4 | 0.9090 | 0.9093 | 0.9037 | 0.9122 | 0.0029 | 0.0010 |
| Larceny | 0.05 | Hispanic | 10 | Remove 5 | 0.9072 | 0.9078 | 0.9038 | 0.9094 | 0.0022 | 0.0009 |
| Larceny | 0.05 | Nonwhite | 10 | Original | 0.9106 | 0.9110 | 0.9043 | 0.9140 | 0.0029 | 0.0009 |
| Larceny | 0.05 | Nonwhite | 10 | Static | 0.9114 | 0.9118 | 0.9055 | 0.9149 | 0.0030 | 0.0010 |
| Larceny | 0.05 | Nonwhite | 10 | Remove 1 | 0.9095 | 0.9092 | 0.9027 | 0.9163 | 0.0039 | 0.0012 |
| Larceny | 0.05 | Nonwhite | 10 | Remove 2 | 0.9102 | 0.9104 | 0.9079 | 0.9120 | 0.0021 | 0.0010 |
| Larceny | 0.05 | Nonwhite | 10 | Remove 3 | 0.9099 | 0.9091 | 0.9077 | 0.9128 | 0.0027 | 0.0015 |
| Larceny | 0.05 | Nonwhite | 10 | Remove 4 | 0.9062 | 0.9062 | 0.9062 | 0.9062 | | |
| Larceny | 0.05 | Whitealone | 10 | Original | 0.9106 | 0.9110 | 0.9043 | 0.9140 | 0.0029 | 0.0009 |
| Larceny | 0.05 | Whitealone | 10 | Static | 0.9115 | 0.9113 | 0.9069 | 0.9147 | 0.0024 | 0.0007 |
| Larceny | 0.05 | Whitealone | 10 | Remove 1 | 0.9105 | 0.9106 | 0.9041 | 0.9134 | 0.0025 | 0.0008 |
| Larceny | 0.05 | Whitealone | 10 | Remove 2 | 0.9097 | 0.9090 | 0.9088 | 0.9114 | 0.0014 | 0.0008 |
| Motor Vehicle Theft | 0.05 | Blackalone | 10 | Original | 0.8656 | 0.8668 | 0.8533 | 0.8722 | 0.0066 | 0.0021 |
| Motor Vehicle Theft | 0.05 | Blackalone | 10 | Static | 0.8656 | 0.8668 | 0.8533 | 0.8722 | 0.0066 | 0.0021 |
| Motor Vehicle Theft | 0.05 | Blackalone | 10 | Remove 1 | 0.8667 | 0.8684 | 0.8587 | 0.8708 | 0.0046 | 0.0016 |
| Motor Vehicle Theft | 0.05 | Hispanic | 10 | Original | 0.8656 | 0.8668 | 0.8533 | 0.8722 | 0.0066 | 0.0021 |
| Motor Vehicle Theft | 0.05 | Hispanic | 10 | Static | 0.8620 | 0.8635 | 0.8525 | 0.8689 | 0.0053 | 0.0017 |
| Motor Vehicle Theft | 0.05 | Hispanic | 10 | Remove 1 | 0.8666 | 0.8663 | 0.8520 | 0.8848 | 0.0098 | 0.0031 |
| Motor Vehicle Theft | 0.05 | Hispanic | 10 | Remove 2 | 0.8661 | 0.8682 | 0.8520 | 0.8745 | 0.0070 | 0.0022 |
| Motor Vehicle Theft | 0.05 | Hispanic | 10 | Remove 3 | 0.8699 | 0.8706 | 0.8614 | 0.8810 | 0.0056 | 0.0018 |
| Motor Vehicle Theft | 0.05 | Hispanic | 10 | Remove 4 | 0.8691 | 0.8712 | 0.8590 | 0.8786 | 0.0062 | 0.0022 |
| Motor Vehicle Theft | 0.05 | Hispanic | 10 | Remove 5 | 0.8712 | 0.8706 | 0.8663 | 0.8793 | 0.0048 | 0.0020 |
| Motor Vehicle Theft | 0.05 | Nonwhite | 10 | Original | 0.8656 | 0.8668 | 0.8533 | 0.8722 | 0.0066 | 0.0021 |
| Motor Vehicle Theft | 0.05 | Nonwhite | 10 | Static | 0.8658 | 0.8652 | 0.8514 | 0.8757 | 0.0078 | 0.0025 |
| Motor Vehicle Theft | 0.05 | Nonwhite | 10 | Remove 1 | 0.8660 | 0.8674 | 0.8451 | 0.8820 | 0.0102 | 0.0032 |
| Motor Vehicle Theft | 0.05 | Nonwhite | 10 | Remove 2 | 0.8662 | 0.8648 | 0.8586 | 0.8792 | 0.0069 | 0.0022 |
| Motor Vehicle Theft | 0.05 | Nonwhite | 10 | Remove 3 | 0.8645 | 0.8650 | 0.8520 | 0.8772 | 0.0069 | 0.0023 |
| Motor Vehicle Theft | 0.05 | Nonwhite | 10 | Remove 4 | 0.8627 | 0.8592 | 0.8549 | 0.8759 | 0.0077 | 0.0027 |
| Motor Vehicle Theft | 0.05 | Nonwhite | 10 | Remove 5 | 0.8690 | 0.8677 | 0.8640 | 0.8747 | 0.0050 | 0.0022 |
| Motor Vehicle Theft | 0.05 | Whitealone | 10 | Original | 0.8656 | 0.8668 | 0.8533 | 0.8722 | 0.0066 | 0.0021 |
| Motor Vehicle Theft | 0.05 | Whitealone | 10 | Static | 0.8636 | 0.8652 | 0.8528 | 0.8722 | 0.0069 | 0.0022 |
| Motor Vehicle Theft | 0.05 | Whitealone | 10 | Remove 1 | 0.8701 | 0.8699 | 0.8671 | 0.8733 | 0.0031 | 0.0018 |
| Theft From Motor Vehicle | 0.05 | Blackalone | 10 | Original | 0.8070 | 0.8047 | 0.8018 | 0.8154 | 0.0053 | 0.0017 |
| Theft From Motor Vehicle | 0.05 | Blackalone | 10 | Static | 0.8070 | 0.8047 | 0.8018 | 0.8154 | 0.0053 | 0.0017 |
| Theft From Motor Vehicle | 0.05 | Hispanic | 10 | Original | 0.8070 | 0.8047 | 0.8018 | 0.8154 | 0.0053 | 0.0017 |
| Theft From Motor Vehicle | 0.05 | Hispanic | 10 | Static | 0.8045 | 0.8042 | 0.7952 | 0.8201 | 0.0076 | 0.0024 |
| Theft From Motor Vehicle | 0.05 | Hispanic | 10 | Remove 1 | 0.8050 | 0.8049 | 0.7999 | 0.8099 | 0.0032 | 0.0010 |
| Theft From Motor Vehicle | 0.05 | Hispanic | 10 | Remove 2 | 0.8034 | 0.8034 | 0.7942 | 0.8086 | 0.0043 | 0.0015 |
| Theft From Motor Vehicle | 0.05 | Hispanic | 10 | Remove 3 | 0.8044 | 0.8080 | 0.7930 | 0.8151 | 0.0094 | 0.0042 |
| Theft From Motor Vehicle | 0.05 | Hispanic | 10 | Remove 4 | 0.8097 | 0.8085 | 0.8061 | 0.8156 | 0.0044 | 0.0022 |
| Theft From Motor Vehicle | 0.05 | Hispanic | 10 | Remove 5 | 0.8127 | 0.8127 | 0.8069 | 0.8185 | 0.0082 | 0.0058 |
| Theft From Motor Vehicle | 0.05 | Nonwhite | 10 | Original | 0.8070 | 0.8047 | 0.8018 | 0.8154 | 0.0053 | 0.0017 |
| Theft From Motor Vehicle | 0.05 | Nonwhite | 10 | Static | 0.8034 | 0.8039 | 0.7926 | 0.8141 | 0.0065 | 0.0020 |
| Theft From Motor Vehicle | 0.05 | Nonwhite | 10 | Remove 1 | 0.8062 | 0.8093 | 0.7965 | 0.8132 | 0.0059 | 0.0020 |
| Theft From Motor Vehicle | 0.05 | Nonwhite | 10 | Remove 2 | 0.8072 | 0.8089 | 0.7969 | 0.8129 | 0.0060 | 0.0027 |
| Theft From Motor Vehicle | 0.05 | Nonwhite | 10 | Remove 3 | 0.8044 | 0.8033 | 0.8017 | 0.8084 | 0.0035 | 0.0020 |
| Theft From Motor Vehicle | 0.05 | Nonwhite | 10 | Remove 4 | 0.8091 | 0.8091 | 0.8015 | 0.8166 | 0.0107 | 0.0076 |
| Theft From Motor Vehicle | 0.05 | Whitealone | 10 | Original | 0.8070 | 0.8047 | 0.8018 | 0.8154 | 0.0053 | 0.0017 |
| Theft From Motor Vehicle | 0.05 | Whitealone | 10 | Static | 0.8071 | 0.8073 | 0.7943 | 0.8174 | 0.0079 | 0.0025 |
| Theft From Motor Vehicle | 0.05 | Whitealone | 10 | Remove 1 | 0.8022 | 0.8008 | 0.7949 | 0.8103 | 0.0058 | 0.0018 |
| Theft From Motor Vehicle | 0.05 | Whitealone | 10 | Remove 2 | 0.7995 | 0.8031 | 0.7882 | 0.8073 | 0.0100 | 0.0058 |

**Supplementary Table 5.** Harmful bias variables and count of features above 0.5 correlation threshold across all model runs.

| ***Harmful Bias Variables and Features*** | ***Assault Offenses*** | ***Burglary*** | ***Gunfire*** | ***Larceny*** | ***Motor Vehicle Theft*** | ***Robbery*** | ***Theft From Motor Vehicle*** | ***Violent Crime*** | ***Grand Total*** |
|---|---|---|---|---|---|---|---|---|---|
| **Nonwhite** | **8** | **9** | **9** | **9** | **9** | **8** | **1** | **8** | **61** |
| acs.hhincome.median | 1 | 1 | | 1 | 1 | 1 | | 1 | 6 |
| acs.popbelowpovertylevel.density | 1 | 1 | | 1 | 1 | 1 | | 1 | 6 |
| acs.popbelowpovertylevel.percent | 1 | 1 | 1 | 1 | 1 | 1 | | 1 | 7 |
| acs.rent.median | 1 | 1 | | 1 | 1 | 1 | | 1 | 6 |
| acs.rentedhouses.percent | 1 | 2 | 1 | 2 | 2 | 1 | 1 | 1 | 11 |
| col | 1 | 1 | 1 | 1 | 1 | 1 | | 1 | 7 |
| distance.cycleways | | | 1 | | | | | | 1 |
| distance.hospitals | | | 1 | | | | | | 1 |
| distance.motorway | 1 | 1 | | 1 | 1 | 1 | | 1 | 6 |
| distance.motorway_link | 1 | 1 | | 1 | 1 | 1 | | 1 | 6 |
| distance.street_lamp | | | 1 | | | | | | 1 |
| terrain-elevation | | | 1 | | | | | | 1 |
| terrain-roughness | | | 1 | | | | | | 1 |
| terrain-slope | | | 1 | | | | | | 1 |
| **Whitealone** | **8** | **9** | **9** | **9** | **9** | **8** | **1** | **8** | **61** |
| acs.hhincome.median | 1 | 1 | | 1 | 1 | 1 | | 1 | 6 |
| acs.popbelowpovertylevel.density | 1 | 1 | | 1 | 1 | 1 | | 1 | 6 |
| acs.popbelowpovertylevel.percent | 1 | 1 | 1 | 1 | 1 | 1 | | 1 | 7 |
| acs.rent.median | 1 | 1 | | 1 | 1 | 1 | | 1 | 6 |
| acs.rentedhouses.percent | 1 | 2 | 1 | 2 | 2 | 1 | 1 | 1 | 11 |
| col | 1 | 1 | 1 | 1 | 1 | 1 | | 1 | 7 |
| distance.cycleways | | | 1 | | | | | | 1 |
| distance.hospitals | | | 1 | | | | | | 1 |
| distance.motorway | 1 | 1 | | 1 | 1 | 1 | | 1 | 6 |
| distance.motorway_link | 1 | 1 | | 1 | 1 | 1 | | 1 | 6 |

| | | | | | | | | |
|---|---|---|---|---|---|---|---|---|
| distance.street_lamp | | 1 | | | | | | 1 |
| terrain-elevation | | 1 | | | | | | 1 |
| terrain-roughness | | 1 | | | | | | 1 |
| terrain-slope | | 1 | | | | | | 1 |
| **Hispanic** | **10** | **20** | **6** | **8** | | **6** | | **50** |
| acs.age.median | 1 | | | | | | | 1 |
| acs.hhincome.median | 2 | 1 | 1 | 1 | | 1 | | 6 |
| acs.house.density | 1 | | 1 | 1 | | 1 | | 4 |
| acs.popbelowpovertylevel.density | 1 | 1 | 1 | 1 | | 1 | | 5 |
| acs.popbelowpovertylevel.percent | 1 | 1 | | | | | | 2 |
| acs.population.density | 1 | | 1 | 1 | | 1 | | 4 |
| acs.rentedhouses.density | 1 | | 1 | 1 | | 1 | | 4 |
| acs.rentedhouses.percent | 1 | | 1 | 1 | | 1 | | 4 |
| col | | 2 | | | | | | 2 |
| density.ncua_insured_credit_unions | | 1 | | | | | | 1 |
| density.railroad_bridges | | 1 | | | | | | 1 |
| density.turning_loops | | 2 | | | | | | 2 |
| distance.bus_stops | | 1 | | | | | | 1 |
| distance.cycleways | | 2 | | | | | | 2 |
| distance.FDIC_insured_banks | | 1 | | | | | | 1 |
| distance.hospitals | | 1 | | | | | | 1 |
| distance.paths | | 1 | | | | | | 1 |
| distance.railroad_bridges | | 1 | | | | | | 1 |
| distance.trunk_links | | 1 | | | | | | 1 |

| | | | | | | | | | |
|---|---|---|---|---|---|---|---|---|---|
| kdPrior364 | | | | | 1 | | | | 1 |
| neighborsPrior364 | | | | | 1 | | | | 1 |
| row | | 1 | 1 | | | | | | 2 |
| terrain-roughness | | | 1 | | | | | | 1 |
| terrain-slope | | | 1 | | | | | | 1 |
| **Blackalone** | **6** | **7** | **3** | **6** | **6** | **6** | | **6** | **40** |
| acs.popbelowpovertylevel.density | 1 | 1 | | 1 | 1 | 1 | | 1 | 6 |
| acs.popbelowpovertylevel.percent | 1 | 2 | 1 | 1 | 1 | 1 | | 1 | 8 |
| acs.rent.median | 1 | 1 | | 1 | 1 | 1 | | 1 | 6 |
| acs.rentedhouses.percent | 1 | 1 | | 1 | 1 | 1 | | 1 | 6 |
| distance.crossings | | | 1 | | | | | | 1 |
| distance.motorway | 1 | 1 | | 1 | 1 | 1 | | 1 | 6 |
| distance.motorway_link | 1 | 1 | | 1 | 1 | 1 | | 1 | 6 |
| terrain-elevation | | | 1 | | | | | | 1 |
| **Grand Total** | **22** | **35** | **41** | **30** | **32** | **22** | **8** | **22** | **212** |

**Supplementary Table 6.** Harmful bias variables used in this study.

| Harmful Bias Variable | Description |
|---|---|
| Black Alone | From ACS variable ID: B02001_003 divided by Variable ID: B02001_001 (Total) |
| All Non-White | From ACS variable IDs: B02001_003, B02001_004, B02001_005, B02001_006, B02001_007, B02001_008, B02001_009, B02001_010, divided by variable ID: B02001_001 (Total) |
| White Alone | From ACS variable ID: B02001_002 divided by variable ID: B02001_001 (Total) |
| Hispanic | From ACS variable ID: B03002_012 divided by variable ID: B02001_001 (Total) |

**Supplementary Table 7.** Model hyperparameters.

| Hyperparameters | Value |
|---|---|
| eta | 0.1 |
| gamma | 2 |
| max_depth | 5 |
| min_child_weight | 6 |
| max_delta_step | 2 |
| subsample | 0.86 |
| colsample_bytree | 0.42 |
| lambda | 4 |
| alpha | 4 |

**Supplementary Table 8.** List of all American Community Survey (ACS) features used in machine learning experiments.

| Feature | Purpose | Description | Agency A | Agency B | Agency C | Agency D | ResourceRouter Production |
|---|---|---|---|---|---|---|---|
| acs.age.median_ACS2021 | Target Desirability | Median age | TRUE | TRUE | TRUE | TRUE | TRUE |
| acs.belowhsedu.density_ACS2021 | Collective Efficacy | Density of people with education below high-school level | TRUE | TRUE | TRUE | TRUE | FALSE |
| acs.belowhsedu.percent_ACS2021 | Collective Efficacy | Percent people with education below high-school level | TRUE | TRUE | TRUE | TRUE | FALSE |
| acs.hhincome.median_ACS2021 | Collective Efficacy, Target Desirability | Median household income | TRUE | TRUE | TRUE | TRUE | TRUE |
| acs.hhnoincome.density_ACS2021 | Collective Efficacy | Density of households with no income | TRUE | TRUE | TRUE | TRUE | FALSE |
| acs.hhnoincome.percent_ACS2021 | Collective Efficacy | Percent households with no income | TRUE | TRUE | TRUE | TRUE | TRUE |
| acs.hhsize.mean_ACS2021 | Target Distribution | Mean household size | TRUE | TRUE | TRUE | TRUE | TRUE |
| acs.house.density_ACS2021 | Target Distribution | Density of housing units | TRUE | TRUE | TRUE | TRUE | TRUE |
| acs.popbelowpovertylevel.density_ACS2021 | Collective Efficacy | Density of population below poverty level | TRUE | TRUE | TRUE | TRUE | FALSE |
| acs.popbelowpovertylevel.percent_ACS2021 | Collective Efficacy | Percent population below poverty level | TRUE | TRUE | TRUE | TRUE | FALSE |
| acs.population.density_ACS2021 | Target Distribution | Density of population | TRUE | TRUE | TRUE | TRUE | TRUE |
| acs.rent.median_ACS2021 | Collective Efficacy, Target Desirability | Median rent | TRUE | TRUE | TRUE | TRUE | TRUE |
| acs.rentedhouses.density_ACS2021 | Collective Efficacy, Target Desirability | Density of rented houses | TRUE | TRUE | TRUE | TRUE | FALSE |
| acs.rentedhouses.percent_ACS2021 | Collective Efficacy, Target Desirability | Percent rented houses | TRUE | TRUE | TRUE | TRUE | TRUE |
| acs.unemployment.density_ACS2021 | Collective Efficacy | Density of unemployment | TRUE | TRUE | TRUE | TRUE | FALSE |
| acs.unemployment.percent_ACS2021 | Collective Efficacy | Percent unemployment | TRUE | TRUE | TRUE | TRUE | FALSE |
| acs.vacanthouses.density_ACS2021 | Risk Terrain Modeling | Density of vacant houses | TRUE | TRUE | TRUE | TRUE | FALSE |
| acs.vacanthouses.percent_ACS2021 | Risk Terrain Modeling | Percent vacant houses | TRUE | TRUE | TRUE | TRUE | TRUE |
| acs.vehicles.density_ACS2021 | Target Distribution | Density of vehicles | TRUE | TRUE | TRUE | TRUE | TRUE |

**Supplementary Table 9.** List of all Risk Terrain Modeling (RTM) features used in machine learning experiments.

| Feature | Purpose | Description | Agency A | Agency B | Agency C | Agency D | ResourceRouter Production |
|---|---|---|---|---|---|---|---|
| coverage.gas_stations | RTM | The percent of the cell covered by a gas station | FALSE | FALSE | TRUE | FALSE | TRUE |
| coverage.parking | RTM | The percent of the cell covered by a parking lot | FALSE | FALSE | TRUE | FALSE | TRUE |
| coverage.pharmacies | RTM | The percent of the cell covered by a pharmacy | FALSE | FALSE | TRUE | FALSE | TRUE |
| coverage.prison_boundaries | RTM | The percent of the cell covered by a prison | TRUE | TRUE | FALSE | FALSE | TRUE |
| density.all_places_of_worship | RTM | Density of features, bandwidth chosen automatically based upon the data, normalized from 0 to 1 | TRUE | TRUE | TRUE | TRUE | TRUE |
| density.bus_stops | RTM | Density of features, bandwidth chosen automatically based upon the data, normalized from 0 to 1 | TRUE | FALSE | TRUE | TRUE | TRUE |
| density.colleges_and_universities | RTM | Density of features, bandwidth chosen automatically based upon the data, | FALSE | FALSE | FALSE | TRUE | TRUE |
| | | normalized from 0 to 1 | | | | | |
| density.crossings | RTM | Density of features, bandwidth chosen automatically based upon the data, normalized from 0 to 1 | TRUE | TRUE | TRUE | TRUE | TRUE |
| density.emergency_medical_services | RTM | Density of features, bandwidth chosen automatically based upon the data, normalized from 0 to 1 | TRUE | TRUE | TRUE | FALSE | TRUE |
| density.fdic_insured_banks | RTM | Density of features, bandwidth chosen automatically based upon the data, normalized from 0 to 1 | TRUE | TRUE | FALSE | TRUE | TRUE |
| density.fire_stations | RTM | Density of features, bandwidth chosen automatically based upon the data, normalized from 0 to 1 | TRUE | TRUE | TRUE | TRUE | TRUE |
| density.mobile_home_parks | RTM | Density of features, bandwidth chosen automatically based upon the data, normalized from 0 to 1 | TRUE | FALSE | TRUE | FALSE | TRUE |
| density.motorway_junctions | RTM | Density of features, bandwidth chosen automatically based upon the data, | TRUE | TRUE | TRUE | FALSE | TRUE |
| | | normalized from 0 to 1 | | | | | |
| density.ncua_insured_credit_unions | RTM | Density of features, bandwidth chosen automatically based upon the data, normalized from 0 to 1 | TRUE | TRUE | TRUE | TRUE | TRUE |
| density.private_schools | RTM | Density of features, bandwidth chosen automatically based upon the data, normalized from 0 to 1 | TRUE | TRUE | FALSE | TRUE | TRUE |
| density.public_schools | RTM | Density of features, bandwidth chosen automatically based upon the data, normalized from 0 to 1 | TRUE | TRUE | FALSE | TRUE | TRUE |
| density.railroad_bridges | RTM | Density of features, bandwidth chosen automatically based upon the data, normalized from 0 to 1 | FALSE | FALSE | TRUE | FALSE | TRUE |
| density.trunk_road_circles | RTM | Density of features, bandwidth chosen automatically based upon the data, normalized from 0 to 1 | FALSE | FALSE | FALSE | TRUE | TRUE |
| density.turning_circles | RTM | Density of features, bandwidth chosen automatically based upon the data, | TRUE | TRUE | TRUE | FALSE | TRUE |
| | | normalized from 0 to 1 | | | | | |
| density.turning_loops | RTM | Density of features, bandwidth chosen automatically based upon the data, normalized from 0 to 1 | TRUE | TRUE | TRUE | FALSE | TRUE |
| density.urgent_care_facilities | RTM | Density of features, bandwidth chosen automatically based upon the data, normalized from 0 to 1 | FALSE | TRUE | FALSE | FALSE | TRUE |
| distance.place_of_worship | RTM | Distance to nearest feature, measured in map units along raster, capped to 2500 units | FALSE | FALSE | TRUE | FALSE | TRUE |
| distance.bus_stop | RTM | Distance to nearest feature, measured in map units along raster, capped to 2500 units | TRUE | TRUE | TRUE | FALSE | TRUE |
| distance.college_and_university | RTM | Distance to nearest feature, measured in map units along raster, capped to 2500 units | TRUE | TRUE | TRUE | FALSE | TRUE |
| distance.convention_center_fairground | RTM | Distance to nearest feature, measured in map units along raster, capped to 2500 units | TRUE | TRUE | TRUE | TRUE | TRUE |
| distance.crossing | RTM | Distance to nearest feature, measured in map units along raster, capped to 2500 units | TRUE | TRUE | TRUE | FALSE | TRUE |
| distance.cycleway | RTM | Distance to nearest feature, measured in map units along raster, capped to 2500 units | TRUE | TRUE | TRUE | TRUE | TRUE |
| distance.emergency_medical_service | RTM | Distance to nearest feature, measured in map units along raster, capped to 2500 units | TRUE | TRUE | TRUE | FALSE | TRUE |
| distance.fdic_insured_bank | RTM | Distance to nearest feature, measured in map units along raster, capped to 2500 units | TRUE | TRUE | TRUE | FALSE | TRUE |
| distance.fire_station | RTM | Distance to nearest feature, measured in map units along raster, capped to 2500 units | TRUE | TRUE | TRUE | FALSE | TRUE |
| distance.footway | RTM | Distance to nearest feature, measured in map units along raster, capped to 2500 units | TRUE | TRUE | TRUE | TRUE | TRUE |
| distance.gas_station | RTM | Distance to nearest feature, measured in map units along raster, | FALSE | FALSE | TRUE | FALSE | TRUE |
| | | capped to 2500 units | | | | | |
| distance.hospital | RTM | Distance to nearest feature, measured in map units along raster, capped to 2500 units | TRUE | TRUE | TRUE | TRUE | TRUE |
| distance.local_law_enforcement | RTM | Distance to nearest feature, measured in map units along raster, capped to 2500 units | TRUE | TRUE | TRUE | TRUE | TRUE |
| distance.major_sport_venue | RTM | Distance to nearest feature, measured in map units along raster, capped to 2500 units | TRUE | FALSE | TRUE | TRUE | TRUE |
| distance.mobile_home_park | RTM | Distance to nearest feature, measured in map units along raster, capped to 2500 units | TRUE | TRUE | TRUE | FALSE | TRUE |
| distance.motorway | RTM | Distance to nearest feature, measured in map units along raster, capped to 2500 units | TRUE | TRUE | TRUE | FALSE | TRUE |
| distance.motorway_junction | RTM | Distance to nearest feature, measured in map units along raster, capped to 2500 units | TRUE | TRUE | TRUE | FALSE | TRUE |
| distance.motorway_link | RTM | Distance to nearest feature, | TRUE | TRUE | TRUE | TRUE | TRUE |
| | | measured in map units along raster, capped to 2500 units | | | | | |
| distance.ncua_insured_credit_union | RTM | Distance to nearest feature, measured in map units along raster, capped to 2500 units | TRUE | TRUE | TRUE | FALSE | TRUE |
| distance.parking | RTM | Distance to nearest feature, measured in map units along raster, capped to 2500 units | FALSE | FALSE | TRUE | FALSE | TRUE |
| distance.path | RTM | Distance to nearest feature, measured in map units along raster, capped to 2500 units | TRUE | TRUE | TRUE | TRUE | TRUE |
| distance.pharmacy | RTM | Distance to nearest feature, measured in map units along raster, capped to 2500 units | FALSE | FALSE | TRUE | FALSE | TRUE |
| distance.place_of_worship | RTM | Distance to nearest feature, measured in map units along raster, capped to 2500 units | TRUE | TRUE | FALSE | FALSE | TRUE |
| distance.primary_street | RTM | Distance to nearest feature, measured in map units along raster, capped to 2500 units | TRUE | TRUE | TRUE | TRUE | TRUE |
| distance.primary_street_link | RTM | Distance to nearest feature, measured in map units along raster, capped to 2500 units | TRUE | TRUE | TRUE | TRUE | TRUE |
| distance.prison_boundary | RTM | Distance to nearest feature, measured in map units along raster, capped to 2500 units | TRUE | TRUE | FALSE | FALSE | TRUE |
| distance.private_school | RTM | Distance to nearest feature, measured in map units along raster, capped to 2500 units | TRUE | TRUE | FALSE | FALSE | TRUE |
| distance.public_school | RTM | Distance to nearest feature, measured in map units along raster, capped to 2500 units | TRUE | TRUE | FALSE | FALSE | TRUE |
| distance.public_transit_route | RTM | Distance to nearest feature, measured in map units along raster, capped to 2500 units | FALSE | FALSE | FALSE | TRUE | TRUE |
| distance.public_transit_station | RTM | Distance to nearest feature, measured in map units along raster, capped to 2500 units | FALSE | FALSE | FALSE | TRUE | TRUE |
| distance.raceway | RTM | Distance to nearest feature, measured in map units along raster, | FALSE | FALSE | TRUE | FALSE | TRUE |
| | | capped to 2500 units | | | | | |
| distance.railroad_bridge | RTM | Distance to nearest feature, measured in map units along raster, capped to 2500 units | TRUE | FALSE | TRUE | TRUE | TRUE |
| distance.residential_street | RTM | Distance to nearest feature, measured in map units along raster, capped to 2500 units | TRUE | TRUE | TRUE | TRUE | TRUE |
| distance.secondary_street | RTM | Distance to nearest feature, measured in map units along raster, capped to 2500 units | TRUE | TRUE | TRUE | TRUE | TRUE |
| distance.secondary_street_link | RTM | Distance to nearest feature, measured in map units along raster, capped to 2500 units | TRUE | TRUE | TRUE | TRUE | TRUE |
| distance.service_road | RTM | Distance to nearest feature, measured in map units along raster, capped to 2500 units | TRUE | TRUE | TRUE | TRUE | TRUE |
| distance.street_lamp | RTM | Distance to nearest feature, measured in map units along raster, capped to 2500 units | FALSE | TRUE | FALSE | FALSE | TRUE |
| distance.supplemental_college | RTM | Distance to nearest feature, | FALSE | TRUE | FALSE | TRUE | TRUE |
| | | measured in map units along raster, capped to 2500 units | | | | | |
| distance.tertiary_street | RTM | Distance to nearest feature, measured in map units along raster, capped to 2500 units | FALSE | TRUE | TRUE | TRUE | TRUE |
| distance.tertiary_street_link | RTM | Distance to nearest feature, measured in map units along raster, capped to 2500 units | TRUE | TRUE | TRUE | TRUE | TRUE |
| distance.track | RTM | Distance to nearest feature, measured in map units along raster, capped to 2500 units | TRUE | TRUE | TRUE | TRUE | TRUE |
| distance.trunk_road | RTM | Distance to nearest feature, measured in map units along raster, capped to 2500 units | TRUE | TRUE | TRUE | TRUE | TRUE |
| distance.trunk_road_circle | RTM | Distance to nearest feature, measured in map units along raster, capped to 2500 units | FALSE | FALSE | FALSE | TRUE | TRUE |
| distance.trunk_road_link | RTM | Distance to nearest feature, measured in map units along raster, capped to 2500 units | TRUE | FALSE | TRUE | TRUE | TRUE |
| distance.turning_circle | RTM | Distance to nearest feature, measured in map units along raster, capped to 2500 units | TRUE | TRUE | TRUE | FALSE | TRUE |
| distance.turning_loop | RTM | Distance to nearest feature, measured in map units along raster, capped to 2500 units | TRUE | TRUE | TRUE | TRUE | TRUE |
| distance.unclassified_street | RTM | Distance to nearest feature, measured in map units along raster, capped to 2500 units | TRUE | TRUE | TRUE | TRUE | TRUE |
| distance.urgent_care_facility | RTM | Distance to nearest feature, measured in map units along raster, capped to 2500 units | TRUE | TRUE | FALSE | TRUE | TRUE |

**Supplementary Table 10.** List of all crime event features used in machine learning experiments.

| **Feature** | **Purpose** | **Description** | **Agency A** | **Agency B** | **Agency C** | **Agency D** | **ResourceRouter Production** |
|---|---|---|---|---|---|---|---|
| kdPrior3 | Near Repeats | Kernel density smoothed mean count of past events over last X days | TRUE | TRUE | TRUE | TRUE | TRUE |
| kdPrior7 | Near Repeats | Kernel density smoothed mean count of past events over last X days | TRUE | TRUE | TRUE | TRUE | TRUE |
| kdPrior14 | Near Repeats | Kernel density smoothed mean count of past events over last X days | TRUE | TRUE | TRUE | TRUE | TRUE |
| kdPrior28 | Hotspot | Kernel density smoothed mean count of past events over last X days | TRUE | TRUE | TRUE | TRUE | TRUE |
| kdPrior56 | Hotspot | Kernel density smoothed mean count of past events over last X days | TRUE | TRUE | TRUE | TRUE | TRUE |
| kdPrior84 | Hotspot | Kernel density smoothed mean count of past events over last X days | TRUE | TRUE | TRUE | TRUE | TRUE |
| kdPrior112 | Hotspot | Kernel density smoothed mean count of past events over last X days | TRUE | TRUE | TRUE | TRUE | TRUE |
| kdPrior168 | Hotspot | Kernel density smoothed mean count of past events over last X days | TRUE | TRUE | TRUE | TRUE | TRUE |
| kdPrior364 | Hotspot | Kernel density smoothed mean count of past events over last X days | TRUE | TRUE | TRUE | TRUE | TRUE |
| neighborsPrior3 | Near Repeats | Mean count of past events over last X days in | TRUE | TRUE | TRUE | TRUE | TRUE |
| | | neighboring cells | | | | | |
| neighborsPrior7 | Near Repeats | Mean count of past events over last X days in neighboring cells | TRUE | TRUE | TRUE | TRUE | TRUE |
| neighborsPrior14 | Near Repeats | Mean count of past events over last X days in neighboring cells | TRUE | TRUE | TRUE | TRUE | TRUE |
| neighborsPrior28 | Hotspot | Mean count of past events over last X days in neighboring cells | TRUE | TRUE | TRUE | TRUE | TRUE |
| neighborsPrior56 | Hotspot | Mean count of past events over last X days in neighboring cells | TRUE | TRUE | TRUE | TRUE | TRUE |
| neighborsPrior84 | Hotspot | Mean count of past events over last X days in neighboring cells | TRUE | TRUE | TRUE | TRUE | TRUE |
| neighborsPrior112 | Hotspot | Mean count of past events over last X days in neighboring cells | TRUE | TRUE | TRUE | TRUE | TRUE |
| neighborsPrior168 | Hotspot | Mean count of past events over last X days in neighboring cells | TRUE | TRUE | TRUE | TRUE | TRUE |
| neighborsPrior364 | Hotspot | Mean count of past events over last X days in neighboring cells | TRUE | TRUE | TRUE | TRUE | TRUE |
| period_since_last | Near Repeats | Number of time periods since last event in that cell | TRUE | TRUE | TRUE | TRUE | TRUE |
| period_since_last_focal_min_all | Near Repeats | Minimum number of time periods since last event across all nearby cells | TRUE | TRUE | TRUE | TRUE | TRUE |
| period_since_last_focal_min_extended | Near Repeats | Minimum number of time periods since last event across cells near but not adjacent to the cell | TRUE | TRUE | TRUE | TRUE | TRUE |
| period_since_last_focal_min_immediate | Near Repeats | Minimum number of time periods since last event across adjacent cells | TRUE | TRUE | TRUE | TRUE | TRUE |
| prior3 | Near Repeats | Mean count of past events in the cell over last X days | TRUE | TRUE | TRUE | TRUE | TRUE |
| prior7 | Near Repeats | Mean count of past events in the cell over last X days | TRUE | TRUE | TRUE | TRUE | TRUE |
| prior14 | Near Repeats | Mean count of past events in the cell over last X days | TRUE | TRUE | TRUE | TRUE | TRUE |
| prior28 | Hotspot | Mean count of past events in the cell over last X days | TRUE | TRUE | TRUE | TRUE | TRUE |
| prior56 | Hotspot | Mean count of past events in the cell over last X days | TRUE | TRUE | TRUE | TRUE | TRUE |

**Supplementary Table 11.** List of all temporal features used in machine learning Experiments.

| **Feature** | **Purpose** | **Description** | **Agency A** | **Agency B** | **Agency C** | **Agency D** | **ResourceRouter Production** |
|---|---|---|---|---|---|---|---|
| dow | Cyclical Patterns | Day of week, 0 to 6 index | TRUE | TRUE | TRUE | TRUE | TRUE |
| dow0 | Cyclical Patterns | 1 if day of week is this day | TRUE | TRUE | TRUE | TRUE | TRUE |
| dow1 | Cyclical Patterns | 2 if day of week is this day | TRUE | TRUE | TRUE | TRUE | TRUE |
| dow2 | Cyclical Patterns | 3 if day of week is this day | TRUE | TRUE | TRUE | TRUE | TRUE |
| dow3 | Cyclical Patterns | 4 if day of week is this day | TRUE | TRUE | TRUE | TRUE | TRUE |
| dow4 | Cyclical Patterns | 5 if day of week is this day | TRUE | TRUE | TRUE | TRUE | TRUE |
| dow5 | Cyclical Patterns | 6 if day of week is this day | TRUE | TRUE | TRUE | TRUE | TRUE |
| dow6 | Cyclical Patterns | 7 if day of week is this day | TRUE | TRUE | TRUE | TRUE | TRUE |
| dowShfPrior112 | Cyclical Patterns | Mean count of past events in the day of the week and in the shift of the day across X Days | TRUE | TRUE | TRUE | TRUE | TRUE |
| dowShfPrior14 | Cyclical Patterns | Mean count of past events in the day of the week and in the shift of the day across X Days | TRUE | TRUE | TRUE | TRUE | TRUE |
| dowShfPrior168 | Cyclical Patterns | Mean count of past events in the day of the week and in the shift of the day across X Days | TRUE | TRUE | TRUE | TRUE | TRUE |
| dowShfPrior28 | Cyclical Patterns | Mean count of past events in the day of the week and in the shift of the day across X Days | TRUE | TRUE | TRUE | TRUE | TRUE |
| dowShfPrior3 | Cyclical Patterns | Mean count of past events in the day of the week and in the shift of the day across X Days | TRUE | TRUE | TRUE | TRUE | TRUE |
| dowShfPrior364 | Cyclical Patterns | Mean count of past events in the day of the week and in the shift of the day across X Days | TRUE | TRUE | TRUE | TRUE | TRUE |
| dowShfPrior56 | Cyclical Patterns | Mean count of past events in the day of the week and in the shift of the day across X Days | TRUE | TRUE | TRUE | TRUE | TRUE |
| dowShfPrior7 | Cyclical Patterns | Mean count of past events in the day of the week and in the shift of the day across X Days | TRUE | TRUE | TRUE | TRUE | TRUE |
| dowShfPrior84 | Cyclical Patterns | Mean count of past events in the day of the week and in the shift of the day across X Days | TRUE | TRUE | TRUE | TRUE | TRUE |
| shift | Cyclical Patterns | The shift of the day, 0 indexed | TRUE | TRUE | TRUE | TRUE | TRUE |
| shift1 | Cyclical Patterns | 1 if the shift is this value | TRUE | TRUE | TRUE | TRUE | TRUE |
| shift2 | Cyclical Patterns | 2 if the shift is this value | TRUE | TRUE | TRUE | TRUE | TRUE |
| shift3 | Cyclical Patterns | 3 if the shift is this value | TRUE | TRUE | TRUE | TRUE | TRUE |
| shift4 | Cyclical Patterns | 4 if the shift is this value | TRUE | FALSE | FALSE | FALSE | TRUE |
| shift5 | Cyclical Patterns | 5 if the shift is this value | TRUE | FALSE | FALSE | FALSE | TRUE |
| shift6 | Cyclical Patterns | 6 if the shift is this value | TRUE | FALSE | FALSE | FALSE | TRUE |
| shift7 | Cyclical Patterns | 7 if the shift is this value | TRUE | FALSE | FALSE | FALSE | TRUE |
| shift8 | Cyclical Patterns | 8 if the shift is this value | TRUE | FALSE | FALSE | FALSE | TRUE |
| shift9 | Cyclical Patterns | 9 if the shift is this value | TRUE | FALSE | FALSE | FALSE | TRUE |
| shift10 | Cyclical Patterns | 10 if the shift is this value | TRUE | FALSE | FALSE | FALSE | TRUE |
| shift11 | Cyclical Patterns | 11 if the shift is this value | TRUE | FALSE | FALSE | FALSE | TRUE |
| shift12 | Cyclical Patterns | 12 if the shift is this value | TRUE | FALSE | FALSE | FALSE | TRUE |
| shiftPrior112 | Cyclical Patterns | Mean count of past events in the shift of the day across X Days | TRUE | TRUE | TRUE | TRUE | TRUE |
| shiftPrior14 | Cyclical Patterns | Mean count of past events in the shift of the day across X Days | TRUE | TRUE | TRUE | TRUE | TRUE |
| shiftPrior168 | Cyclical Patterns | Mean count of past events in the shift of the day across X Days | TRUE | TRUE | TRUE | TRUE | TRUE |
| shiftPrior28 | Cyclical Patterns | Mean count of past events in the shift of the day across X Days | TRUE | TRUE | TRUE | TRUE | TRUE |
| shiftPrior3 | Cyclical Patterns | Mean count of past | TRUE | TRUE | TRUE | TRUE | TRUE |
| | | events in the shift of the day across X Days | | | | | |
| shiftPrior364 | Cyclical Patterns | Mean count of past events in the shift of the day across X Days | TRUE | TRUE | TRUE | TRUE | TRUE |
| shiftPrior56 | Cyclical Patterns | Mean count of past events in the shift of the day across X Days | TRUE | TRUE | TRUE | TRUE | TRUE |
| shiftPrior7 | Cyclical Patterns | Mean count of past events in the shift of the day across X Days | TRUE | TRUE | TRUE | TRUE | TRUE |
| shiftPrior84 | Cyclical Patterns | Mean count of past events in the shift of the day across X Days | TRUE | TRUE | TRUE | TRUE | TRUE |
| weekOfYear | Cyclical Patterns | The week number within the year | TRUE | TRUE | TRUE | TRUE | TRUE |
| monthOfYear | Cyclical Patterns | The month of the year, 1 to 12 | TRUE | TRUE | TRUE | TRUE | TRUE |
| dayOfMonth | Cyclical Patterns | The day of the month | TRUE | TRUE | TRUE | TRUE | TRUE |
| dates-isCalmHoliday-max | Cyclical Patterns | 1 If day is calm federal holiday | TRUE | TRUE | TRUE | TRUE | TRUE |
| dates-isNoisyHoliday-max | Cyclical Patterns | 1 If day is noisy federal holiday | TRUE | TRUE | TRUE | TRUE | TRUE |
| calculated-lunarillumination-mean | Moon cycle | Mean illumination of the moon across a shift | TRUE | TRUE | TRUE | TRUE | TRUE |

**Supplementary Table 12.** List of all weather, geographic and terrain features used in machine learning experiments.

| Feature | Purpose | Description | Agency A | Agency B | Agency C | Agency D | ResourceRouter Production |
|---|---|---|---|---|---|---|---|
| col | Unknown Geographic Factors | Column of the cell within the raster; should not be used very much unless we have missing geographic factors | TRUE | TRUE | TRUE | TRUE | TRUE |
| row | Unknown Geographic Factors | Row of the cell within the raster; should not be used very much unless we have missing geographic factors | TRUE | TRUE | TRUE | TRUE | TRUE |
| terrain.aspect | Physical Terrain | Mean aspect in raster cell | TRUE | TRUE | TRUE | TRUE | TRUE |
| terrain.elevation | Physical Terrain | Mean elevation in raster cell | TRUE | TRUE | TRUE | TRUE | TRUE |
| terrain.roughness | Physical Terrain | Mean roughness in raster cell | TRUE | TRUE | TRUE | TRUE | TRUE |
| terrain.slope | Physical Terrain | Mean slope in raster cell | TRUE | TRUE | TRUE | TRUE | TRUE |
| weather-humidity-mean | Weather patterns | Mean precipitation intenstiy across a time period (shift) | TRUE | TRUE | TRUE | TRUE | TRUE |
| weather-precipIntensity-mean | Weather patterns | Mean temperature intenstiy across a time period (shift) | TRUE | TRUE | TRUE | TRUE | TRUE |
| weather-pressure-mean | Weather patterns | Mean windspeed intenstiy across a time period (shift) | TRUE | TRUE | TRUE | TRUE | TRUE |
| weather-temperature-mean | Weather patterns | Mean pressure intenstiy across a time period (shift) | TRUE | TRUE | TRUE | TRUE | TRUE |
| weather-windSpeed-mean | Weather patterns | Mean humity intenstiy across a time period (shift) | TRUE | TRUE | TRUE | TRUE | TRUE |

**Supplementary Table 13.** City Demographics.

| | | | **Race** | | **Ethnicity** |
|---|---|---|---|---|---|
| **City ID** | **Region** | **Black Alone** | **All Non-White** | **White Alone** | **Hispanic** |
| A | West | 5.10% | 57.00% | 44.90% | 53.00% |
| B | Midwest | 22.50% | 47.70% | 54.00% | 16.60% |
| C | Southeast | 9.00% | 26.90% | 75.90% | 6.00% |
| D | Northeast | 15.30% | 52.90% | 50.30% | 26.60% |

* https://data.census.gov/table, 2021 1-year ACS estimates

**Supplementary Table 14.** Features not being considerate in threshold function because correlation is Null with harmful bias variables.

| **Feature** | **Description** |
|---|---|
| calculated-lunarillumination-mean | Mean illumination of the moon across a time period (shift) |
| dates-isCalmHoliday-max | 1 If day is federal holiday |
| dates-isNoisyHoliday-max | 1 If day is federal holiday |
| dayOfMonth | The day of the month |
| dow | Day of week, 0 to 6 index |
| dow0 | 1 if day of week  is this day |
| dow1 | 2 if day of week  is this day |
| dow2 | 3 if day of week  is this day |
| dow3 | 4 if day of week  is this day |
| dow4 | 5 if day of week  is this day |
| dow5 | 6 if day of week  is this day |
| dow6 | 7 if day of week  is this day |
| monthOfYear | The month of the year, 1 to 12 |
| shift | The shift of the day, 0 indexed |
| shift1 | 1 if the shift is this value |
| shift2 | 2 if the shift is this value |
| shift3 | 3 if the shift is this value |
| weather-humidity-mean | Mean precipitation intensity across a time period (shift) |
| weather-precipIntensity-mean | Mean temperature intensity across a time period (shift) |
| weather-pressure-mean | Mean wind speed intensity across a time period (shift) |
| weather-temperature-mean | Mean pressure intensity across a time period (shift) |
| weather-windSpeed-mean | Mean humidity intensity across a time period (shift) |
| weekOfYear | The week number within the year |